\documentclass{emulateapj}

\usepackage{natbib}
\slugcomment{Accepted by ApJ}
\shorttitle{Birth and Evolution of Isolated Radio Pulsars}
\shortauthors{Faucher-Gigu\`ere and Kaspi}

\begin{document}

\title{Birth and Evolution of Isolated Radio Pulsars}

\author{Claude-Andr\'e Faucher-Gigu\`ere\altaffilmark{1, 2} and Victoria M. Kaspi\altaffilmark{1}}
\altaffiltext{1}{Department of Physics, McGill University, 3600 University St, Montr\'eal, QC, H3A-2T8, Canada}
\altaffiltext{2}{Now at the Harvard-Smithsonian Center for Astrophysics; 60 Garden St, MS-10, Cambridge, MA, 02138, USA; cgiguere@cfa.harvard.edu}

\begin{abstract}
We investigate the birth and evolution of Galactic isolated radio pulsars.
We begin by estimating their birth space velocity distribution from proper motion measurements of \cite{2002ApJ...571..906B, 2003AJ....126.3090B}.
We find no evidence for multimodality of the distribution and favor one in which the absolute one-dimensional velocity components are exponentially distributed and with a three-dimensional mean velocity of \mbox{$380^{+40}_{-60}$ km s$^{-1}$}.
We then proceed with a Monte Carlo-based population synthesis, modelling the birth properties of the pulsars, their time evolution, and their detection in the Parkes and Swinburne Multibeam surveys. 
We present a population model that appears generally consistent with the observations.
Our results suggest that pulsars are born in the spiral arms, with a Galactocentric radial distribution that is well described by the functional form proposed by \cite{2004A&A...422..545Y}, in which the pulsar surface density peaks at radius \mbox{$\sim3$ kpc}.
The birth spin period distribution extends to several hundred milliseconds, with no evidence of multimodality.
Models which assume the radio luminosities of pulsars to be independent of the spin periods and period derivatives are inadequate, as they lead to the detection of too many old simulated pulsars in our simulations.
Dithered radio luminosities proportional to the square root of the spin-down luminosity accommodate the observations well and provide a natural mechanism for the pulsars to dim uniformly as they approach the death line, avoiding an observed pile-up on the latter.
There is no evidence for significant torque decay (due to magnetic field decay or otherwise) over the lifetime of the pulsars as radio sources (\mbox{$\sim100$ Myr}). 
Finally, we estimate the pulsar birthrate and total number of pulsars in the Galaxy.

\end{abstract}


\section{INTRODUCTION}
\label{intro}
The birth and evolution of pulsars are of considerable interest.
The spatial distribution of pulsars at birth may be used to associate them with their progenitors.
Their initial spin periods may also be related to those of the progenitors' cores, which are not well predicted by theory due to differential rotation \citep{1978ApJ...220..279E}.
The birth properties of neutron stars are also intimately related to the physics of core-collapse supernovae in which most are thought to be formed.
For example, their birth space velocities, spin periods, and magnetic fields put severe constraints on the mechanisms that may account for their high observed velocities \citep{lai03}.
The comparison of the Galactic pulsar birth and supernova rates may provide further insight into supernovae by quantifying the fraction which leaves behind a neutron star.
It may also be possible to clarify whether all neutron stars unaffected by peculiar conditions (such as accretion from a binary companion or an anomalously high magnetic field) are in fact radio pulsars, although not always beamed toward us.
Knowledge of the spin evolution of pulsars is particularly valuable in elucidating whether magnetic field decay occurs in isolated neutron stars.
A correlation between the spin down of pulsars and the evolution of their radio luminosity may finally shed light on the long-standing problem of the pulsar radio emission mechanism.

One way to probe the birth and evolution of pulsars is to study the population as a whole.
While the earlier efforts in this field relied on simplified analytical or semi-analytical treatments \citep[e.g.,][]{1970ApJ...160..979G, 1971IAUS...46..165L, 1977ApJ...215..885T, 1977MNRAS.179..635D, 1985MNRAS.213..613L}, the trend over the last two decades has been to attempt detailed computational modelling of the pulsar population and the selection effects affecting the observed sample, often including the evolution of the synthetic pulsars from birth to detection \citep[e.g.,][]{1987A&A...178..143S, 1989ApJ...345..931E, 1992A&A...254..198B, 1997A&A...322..477H, 1997MNRAS.289..592L, 2002ApJ...568..289A, 2002ApJ...565..482G, 2004ApJ...604..775G}.
The broad goal of these ``Monte Carlo" (MC) simulation studies is to statistically reproduce the pulsar sample observed in actual surveys and test whether a given population model is consistent with the data. 

Unfortunately, the conclusions of pulsar population investigations have often been conflicting.

One long standing issue is that of magnetic field decay.
For example, \cite{1970ApJ...160..979G}, \cite{1985MNRAS.213..613L}, \cite{1990ApJ...352..222N}, \cite{2002ApJ...565..482G}, and \cite{2004ApJ...604..775G} have suggested that field decay occurs on short time scales \mbox{$\sim2.5-5$ Myr}.
Similar statistical studies, e.g. those of \cite{1987A&A...178..143S}, \cite{1992A&A...254..198B}, and \cite{1997MNRAS.289..592L}, reached the opposite conclusion that the magnetic field of pulsars does not decay significantly during their lifetime as radio sources, implying decay time constants \mbox{$\gtrsim100$ Myr}.
Meanwhile, theoretical arguments have mostly supported that field decay is unimportant for typical neutron stars \citep[][and references therein]{1969Natur.224..673B, 1992ApJ...395..250G}.
Since the observed kinetic age versus characteristic age diagram (see, e.g., Lyne, Anderson, \& Salter 1982\nocite{1982MNRAS.201..503L} and Harrison, Lyne, \& Anderson 1993\nocite{1993MNRAS.261..113H}, who used it to argue in favor of field decay) has been shown to be an artefact of selection effects \citep{1997MNRAS.289..592L}, direct empirical evidence for field decay is also lacking.
However, evolution of the inclination angle between the magnetic and spin axes of a pulsar can also produce torque variations (see section \ref{evolution rotational}) that are difficult to disentangle from evolution of the magnetic field itself on the basis of spin kinematics.
We will thus henceforth speak more generally of ``torque decay"\footnote{The magnetic torque on a rotating neutron star decreases with time even if the magnetic field and its orientation are constant, owing to its decreasing rotational frequency.
Here, we are referring to any potential decay in addition to this expected decrease.}.

Another matter of debate, introduced by \cite{1981JApA....2..315V}, concerns the ``injection" of a subpopulation of pulsars with birth spin period \mbox{$\sim$0.5 s}, in addition to a population of Crab-like pulsars born with short periods \mbox{$\lesssim$ 100 ms}.
Evidence for injection was found by \cite{1981JApA....2..315V} in an extension of the ``pulsar current" analysis proposed by \cite{1981MNRAS.194..137P}, in which one considers the flow of pulsars in the period-period derivative plane.
Although the pulsar current analysis is model-free in the sense that it does not require explicit assumptions regarding the luminosity and spin-down laws of the pulsars, \cite{1985MNRAS.213..613L} argued that the conclusion of \cite{1981JApA....2..315V} was subject to considerable statistical uncertainty and suffered from an incomplete treatment of the selection effects plaguing pulsar surveys.
Nevertheless, injection received further support from a number of independent analyses \citep{1986ApJ...304..140C, 1987ApJ...319..162N, 1990ApJ...352..222N}, while similar studies maintained that the data did not require it \citep{1987A&A...178..143S, 1992A&A...254..198B}.
Refining the pulsar current analysis by restricting attention to pulsars above a luminosity cut-off of \mbox{10 mJy kpc$^{2}$} at 400 MHz, for which significant statistics were available, \cite{1993MNRAS.263..403L} found no evidence for injection, although they could not exclude that it may affect the fainter end of the pulsar population.
Injection appears to have since lost popularity.
Most recently, \cite{2004ApJ...617L.139V} presented a pulsar current analysis of the Parkes Multibeam pulsar survey \citep{2001MNRAS.328...17M} data which, although it suggests that many, perhaps \mbox{40\%}, of pulsars are born with periods in the range \mbox{0.1-0.5 s}, did not show evidence of distinct subpopulations.

Lately, several authors have attempted to characterize the distribution of the birth space velocities of pulsars.
Different analyses initially led to quantitative disagreements regarding the mean velocity \citep{1994Natur.369..127L, 1997MNRAS.291..569H, 1997MNRAS.289..592L}.
Somewhat surprisingly, other workers have found evidence in favor of a bimodal velocity distribution \citep{1998ApJ...505..315C, 2002ApJ...568..289A, 2003AJ....126.3090B}, so that the issue of subpopulations of ordinary radio pulsars with different birth properties is not completely resolved.
The origin of the two distribution components has not been conclusively identified.
In fact, the two components themselves remain to be directly exhibited in the proper-motion data and the different authors disagree on their modes and relative weight.
Analysing a large sample of pulsar proper motion measurements, \cite{2005MNRAS.tmp..475H} directly challenged the multimodality of the distribution.

Multiple reasons motivate us to reconsider the birth and evolution of isolated radio pulsars in this paper.

First, pulsar astronomy has in the recent years seen a number of major advances.
The Parkes (PM) and Swinburne (SM) Multibeam pulsar surveys \citep{2001MNRAS.328...17M, 2001MNRAS.326..358E} at \mbox{1.4 GHz} have approximately doubled the number of known pulsars.
Over 1500 objects are now cataloged in the Australia Telescope National Facility (ATNF) Pulsar Database\footnote{http://www.atnf.csiro.au/research/pulsar/psrcat/} \citep{2005AJ....129.1993M}.
\cite{NE2001} have introduced a new model of the Galactic free electron density (NE2001) to supersede the previously standard model of \cite{1993ApJ...411..674T} (TC93), providing an updated dispersion measure distance scale.
New astrometric measurements using interferometry provide a sample relatively free of brightness and distance biases from which to estimate the space velocity distribution of pulsars \citep{2002ApJ...571..906B, 2003AJ....126.3090B}.

Second, in spite of poorly understood underlying physics, the recent simulations have become increasingly sophisticated.
Departing from the common practice of modelling the pulsar pseudo-luminosity (where pseudo-luminosity $\times$ distance$^{2} \equiv$ flux density), \cite{2002ApJ...568..289A} have for instance introduced a standard-candle geometrical model of physical luminosity.
While this luminosity model is certainly a valuable step toward more realistic simulations, it is somewhat speculative in several respects (e.g., Gaussian shape and angular size of the radio beams, and the relative radiated power contribution of each).
An important feature of the analysis of \cite{2002ApJ...568..289A} is the use of indirect information (i.e. other than proper motions) contained in the observed pulsar sample in the inferrence of their birth velocity distribution of pulsars, relying on a detailed modelling of the selection effects of major radio surveys at 400 MHz.
Since the accuracy of this modelling is limited by the validity of the assumptions made regarding the intrinsic properties of the pulsars, it is a fair question to ask whether their conclusion that the birth space velocity distribution of isolated pulsars has two well defined components, corresponding to low (\mbox{$\sim90$ km s$^{-1}$}) and high velocities (\mbox{$\sim500$ km s$^{-1}$}) is really required by the data.
A previous analysis by \cite{1998ApJ...505..315C} omitting a detailed treatment of the selection effects suggest that evidence for two components is indeed present in the data, but this claim has been disputed by \cite{2005MNRAS.tmp..475H}.   
Adopting a slight modification of the \cite{2002ApJ...568..289A} luminosity model to include spectral dependence, \cite{2004ApJ...604..775G} claimed evidence for magnetic field decay on a time scale \mbox{$\sim2.8$ Myr}.
Because pulsars are ultimately detected through their electromagnetic fluxes, it appears that the conclusions of pulsar population simulations are highly dependent on the assumed luminosity model.
Since there is no strong independent support for any particular luminosity model, we may also consider whether the complexity of some of the simulations is absolutely needed.

Throughout, we follow two guiding principles.
First, because of the important uncertainties in modelling pulsars from birth to detection, we rely on independent results obtained by more direct means whenever possible.
Second, inspired by Ockham's razor, we strive for simplicity over sophistication.
We further attempt to be as self-consistent as possible.
While \cite{2004ApJ...604..775G} implemented the NE2001 model, for instance, they based their luminosity model and their birth velocity distribution on those of \cite{2002ApJ...568..289A}, who used TC93.

We begin by estimating the pulsar birth space velocity distribution directly from proper motion measurements in section \ref{kick vel}.
In section \ref{pop synth}, we fix this distribution and proceed with Monte Carlo simulations of pulsar birth, evolution, and detection.
We discuss our results and their implications in section \ref{discussion} and conclude in section \ref{conclusion}.

\section{THE PULSAR BIRTH SPACE VELOCITY DISTRIBUTION FROM PROPER MOTION MEASUREMENTS}
\label{kick vel}
Pulsars have long been recognized as a high-velocity stellar population.
Based on the correlation between the ages of pulsars and their distances from the Galactic plane, and on the transverse velocity of the Crab pulsar, \cite{1970ApJ...160..979G} proposed that most pulsars are born with a space velocity \mbox{$\sim$100 km s$^{-1}$}.
We now know that, in fact, some pulsars may have velocity exceeding \mbox{1000 km s$^{-1}$}.
PSR \mbox{B2224+65} in the Guitar Nebula, for instance, has inferred tranverse velocity \mbox{$1640$ km s$^{-1}$} \citep{1993Natur.362..133C, 2002ApJ...575..407C, 2004ApJ...600L..51C}.
Although this estimate assumes the NE2001 dispersion-measure distance of \mbox{1.9 kpc}, the observed bow shock is a strong indication of unusually rapid motion.
PSR \mbox{B1508+55} has a velocity determined directly from proper-motion and parallax measurements of \mbox{$1083^{+103}_{-90}$ km s$^{-1}$} \citep{2005ApJ...630L..61C}.
These velocities are to be contrasted with those \mbox{$\sim$15 km s$^{-1}$} typical of their purported progenitors, massive main sequence stars \citep[e.g.,][]{1979A&A....80...35L}.

The identified mechanisms from which neutron stars may acquire their high initial velocities fall in two categories.
The original proposal is that the velocities originate from recoil in binaries that are disrupted by a symmetric supernova explosion \citep{1961BAN....15..265B, 1970ApJ...160L..91G, 1996ApJ...456..738I}.
The other involves supernova asymmetries, which may induce ``kicks" to nascent neutron stars.
There is strong evidence that this second scenario plays an important role.
The most direct evidence comes from the observed characteristics of binary systems containing a neutron star.
In these systems, the spin and orbital angular momentum vectors are expected to be aligned prior to the last supernova, owing to tidal and mass transfer effects \citep{1991PhR...203....1B}.
A post-supernova misalignment then indicates that the supernova must have imparted a kick to the neutron star out of the orbital plane, which can result only from an asymmetric explosion.
Spin-orbit misalignment has in fact been detected in several binary systems: in the \mbox{PSR J0045$-$7319}/B-star system \citep{1995ApJ...452..819L, kbm+96}, in the \mbox{PSR J1740$-$3052}/massive companion system \citep{2003ASPC..302...85S}, and in the double neutron star systems \mbox{PSR B1913+16} \citep{1998ApJ...509..856K, 2000ApJ...528..401W}, \mbox{PSR J0737$-$3039}, and \mbox{PSR B1534+12} \citep{2004ApJ...616..414W, 2005ApJ...619.1036T}.
Direct polarization observations of supernovae also suggest that that most are asymmetric \citep[e.g.,][]{2001ApJ...550.1030W}.
In general, both binary breakup and kicks due to supernova asymmetries will contribute to the observed pulsar velocities.

As already mentioned, several authors \citep[e.g.][]{1994Natur.369..127L, 1997MNRAS.291..569H, 1997A&A...322..127H, 1997MNRAS.289..592L, 1998ApJ...505..315C, 2002ApJ...568..289A, 2003AJ....126.3090B} have attempted to constrain the birth velocity distribution of pulsars. 
They have found mean birth velocities \mbox{$\sim300-500$ km s$^{-1}$}.
All of these studies (except for Lorimer et al. 1997, who used the older model of Cordes et al. 1991\nocite{1991Natur.354..121C}) relied on the TC93 model.
As was demonstrated by \cite{1994Natur.369..127L}, adoption of a new free electron density model can have a considerable impact on derived velocities.
In the absence of pulsar velocity studies based on the NE2001 model in the literature when this part of the present work was carried out, we reanalyze the sample of proper motions of \cite{2002ApJ...571..906B, 2003AJ....126.3090B}.
This sample consists of 34 pulsars, 8 of which have parallax measurements.
In general, the pulsars in the sample are fainter and more distant than in previous proper motion surveys, reducing the bias against the high-velocity objects which rapidly escape the solar neighborhood.
Moreover, the measurements are typically much more precise than in most earlier works, which also appear to have reported overly optimistic error bars, most likely because of a failure to fully account for ionospheric effects \citep{2003AJ....126.3090B}.
For these reasons, we choose not to include other catalogued proper motion measurements.
A caveat, however, is that the small sample size provides limited statistics.

The basis of our formalism is identical to that of \cite{2003AJ....126.3090B}.
The Bayesian maximum likelihood framework has the benefit of allowing a detailed treatment of the uncertainties on the measured and derived quantities.
This is important, because distance uncertainties, for example, are in general neither Gaussian nor symmetric, and ignoring this would lead to biased derived velocities.
The consideration of non-Gaussian models and the calculation of credibility ranges are our additions.

In section \ref{kick vel method}, we describe the maximum likelihood formalism used.
The models investigated are presented section \ref{kick vel models}.
In section \ref{kick vel results}, we present and discuss the results.

\subsection{Method}
\label{kick vel method}
In this paper, we define the term birth velocity as the post-supernova velocity of the pulsar relative to its local standard of rest (LSR), including any contribution due to pre-supernova binary orbital motion or motion of the progenitor system relative to the LSR.

Let \mbox{$\mu_{l}=\dot{l}$} (where $l$ and $b$ are the Galactic longitude and latitude, respectively) be the (observed) component of proper motion parallel to the Galactic plane, and $D$ be the distance to the pulsar.
Then \mbox{$v_{l}=D \mu_{l} \cos{b} - \Delta v_{l}$} is the component of the pulsar's transverse velocity parallel to the plane, relative to its LSR, where \mbox{$\Delta v_{l}(D, l, b)$} is the contribution to the observed velocity due to differential Galactic rotation and the motion of the Sun relative to its own LSR \citep[see, e.g.,][]{1998gaas.book.....B}.
We assume a flat rotation curve with circular velocity \mbox{225 km s$^{-1}$} and a solar motion of \mbox{16.5 km s$^{-1}$} in the direction \mbox{$(l,b)=(53^{\circ}, 25^{\circ})$} \citep{1981gask.book.....M}.
These values are consistent with those obtained from a more recent analysis of the kinematics of Galactic Cepheids from \emph{Hipparcos} proper motions \citep{1997MNRAS.291..683F}.
The probability density function (PDF) for $v_l$ can be computed as
\begin{eqnarray}
\label{v_l pdf}
p(v_{l})=
\int_{0}^{\infty}
\!\!
\int_{-\infty}^{\infty}
\delta \{v_{l} - [D \cos{b}~\mu_{l} 
-
\Delta_{v_{l}}]\}
p(\mu_{l})
p(D)
\textrm{d}\mu_{l}
\textrm{d}D
.
\end{eqnarray}
The functions $p(\mu_{l})$ and $p(D)$ are the PDF for the component of the proper motion parallel to the plane and the distance to the pulsar, respectively, and reflect the uncertainties on their measurements.

For $\mu_{l}$, the uncertainty is Gaussian, as reported by the observers.
For $D$, there are two cases.
If a parallax ($\varpi$) measurement is available for the pulsar, we consider the parallax distance and the reported Gaussian uncertainty on the parallax is transformed to the corresponding (non-Gaussian) uncertainty on the derived distance:
\begin{equation}
p(D)=
\frac{1}
{\sqrt{2 \pi} \sigma_{\varpi} D^{2}}
\exp{
\left[
-\frac{(1/D - \varpi)^{2}}
{2 \sigma_{\varpi}^{2}}
\right]
}.
\end{equation}
For pulsars for which no parallax measurement is available, we use the distance derived from the dispersion measure ($DM$) using the NE2001 model.
To model the uncertainty on the distance due to imperfections of the free electron density model, a Gaussian uncertainty with variance $0.4DM$ is assumed on the measured $DM$.
Then 
\begin{equation}
p(D)=
p(DM)
\left|
\frac{\textrm{d}(DM)}
{\textrm{d}D}
\right|.
\end{equation}
The derivative \mbox{d($DM$)/d$D$} is numerically approximated by interpolation from a table of dispersion measures as a function of distance computed using the NE2001 model. 
The distance step is set to \mbox{10 pc}.
In reality, the error on the measured $DM$ is negligible.
The 40\% figure is approximately equal to the variance of the fractional errors of the model dispersion measures, with respect to the measured ones, for the pulsars in ATNF Pulsar Database with parallax measurements, assuming the NE2001 model and that the parallax distance is exact.
The rationale behind this rough model introduced by \cite{2002ApJ...571..906B} is that the estimated uncertainties tend to be greater for pulsars in regions where the free electron density fluctuates rapidly, as expected. 
We note that out of 34 pulsars in the sample, 20 have $v_{l}<0$ versus 14 with $v_{l}>0$. 
Under the assumption of isotropic velocities, an asymmetry of this magnitude has 20\% probability of occurring for our sample size and is not worrisome.

We quantify the likelihood of models based on the distribution of $v_{l}$ for the pulsars in the sample.
Although the sample is not restricted to young objects, dynamical evolution and the potential bias toward low-velocity objects that remain in the solar neighborhood for longer periods are not expected to significantly affect our analysis.
In fact, we verified that the velocity components parallel to the Galactic plane of the old pulsars are, on average, close to their birth values.
Specifically, we used the computer model described in section \ref{pop synth} to compare the averages of the Cartesian velocity components parallel to the Galactic plane after evolution of the pulsars to the averages of their birth values.
The differences were $\sim5$\%, which is less than the uncertainties on the model parameters derived here (c.f. section \ref{kick vel results}).
Because younger pulsars appear generally easier to detect (see section \ref{pulsar luminosity}), the actual effect in the observed sample is presumably even smaller. 
The bias due to high-velocity pulsars rapidly moving away from the Galactic plane (and hence from the sampled volume) is mitigated because pulsars with a large velocity component perpendicular to the plane do not necessarily have correspondingly large components parallel to it.
It is implicitly assumed that the birth velocities of the pulsars are independent of their other characteristics.

The likelihood for a model $\mathcal{M}$, depending on the (possibly vectorial) parameter $\theta$, is 
\begin{equation}
\mathcal{L}(\mathcal{M}(\theta))=
\prod_{psr} \int_{-\infty}^{\infty}
p_{psr}(v_{l})p_{\mathcal{M}(\theta)}(v_{l})
\textrm{d}v_{l}.
\end{equation}
We maximize this function with respect to $\theta$ to find the maximum likelihood value of the parameter.
The evidence for the model is
 \begin{equation}
\mathcal{E(\mathcal{M})}=
\int
\mathcal{L}(\mathcal{M}(\theta))p_{0}(\theta)
\textrm{d}\theta
,
\end{equation}
where $p_{0}(\theta)$ is the prior PDF for the parameter and the integration is over the entire parameter space, i.e. the set of values that $\theta$ may take.
The evidences allow comparison of different models.
The ``odds ratios" $\mathcal{E}_{i}/\mathcal{E}_{j}$ may be thought of as relative probabilities of models $i$ and $j$ \citep{DAG03}.
In all cases, flat priors ($p_{0} \equiv constant$) are assumed over a reasonable parameter space.
The exact evidence for each model depends on this (somewhat arbitrary) choice of priors, though this dependence vanishes in the limit of a large data set.
We quote (normalized) evidences to two digits to illustrate the differences between models with similar evidences, although only the first digit is really significant and the numerical values should not be overinterpreted.

For each model, we provide an uncertainty estimate for the maximum likelihood parameter.
Let $\theta_{ML}$ be the parameter of maximum likelihood.
We solve for \mbox{$\theta_{min}<\theta_{ML}<\theta_{max}$} so that
\begin{equation}
\frac{
\int _{\theta_{min}} ^{\theta_{ML}} \mathcal{L}(\mathcal{M}(\theta)) \textrm{d} \theta
}
{
\int _{-\infty} ^{\infty} \mathcal{L}(\mathcal{M}(\theta)) \textrm{d} \theta
}
=
\frac{
\int_{\theta_{ML}}^{\theta_{max}} \mathcal{L}(\mathcal{M}(\theta)) \textrm{d}\theta
}
{
\int_{-\infty}^{\infty} \mathcal{L}(\mathcal{M}(\theta)) \textrm{d}\theta
} 
=
C,
\end{equation}
where $C$ is the ``probability content" \citep{1992ApJ...398..146G}.
We consider $1\sigma$ errors, for which $C$=34.15\%.
If $\theta$ is a vector, as in the case of the Gaussian model below, we treat each component as above, fixing the others to the maximum likelihood values.
The intervals $[\theta_{min}, \theta_{max}]$ are known as ``credibility ranges".

\subsection{Models}
\label{kick vel models}
We investigate several functional forms for the pulsar birth velocity distribution.
We describe them in this section.

\subsubsection{Gaussian Model}
The first model that we consider is a two-component Gaussian: 
\begin{equation}
p(v_{l})=
\frac{w_{1}}
{
\sqrt{2\pi}
\sigma_{1}
}
\exp{
\left(
-
\frac
{
v_{l}^{2}
}
{
2\sigma_{1}^{2}
}
\right)
}
+
\frac{1-w_{1}}{\sqrt{2\pi}\sigma_{2}}
\exp{
\left(
-
\frac
{
v_{l}^{2}
}
{
2\sigma_{2}^{2}
}
\right)
}
\end{equation}
where $w_{1}$ is the fraction of pulsars in the low-velocity component, and $\sigma_{1}$ and $\sigma_{2}$ are the 1-D velocity dispersions for the low- and high-velocity components, respectively.
Gaussian models are natural to investigate, as they frequently arise as a consequence of the central limit theorem.
The possibility of two components is motivated by the studies of \cite{1998ApJ...505..315C}, \cite{2002ApJ...568..289A}, and \cite{2003AJ....126.3090B}, who all considered such models.
In the case $w_{1}=1$, this model reduces to an ordinary Gaussian.
The corresponding three-dimensional velocity distribution is Maxwellian and has mean \mbox{$\langle v_{3D} \rangle = \sqrt{8 / \pi} [w_{1} \sigma_{1} + (1-w_{1}) \sigma_{2}]$}. 

\begin{figure}[ht]
\begin{center}
\includegraphics[width=0.3\textwidth, angle=-90]{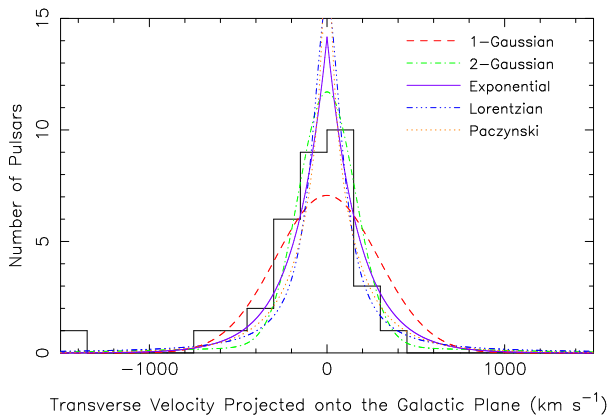}
\caption{
Histogram of the tranverse velocity components parallel to the Galactic plane for the pulsars used in our analysis of the birth velocity distribution.
For this histogram, the velocity is calculated directly using the most probable distance (from parallax if available; otherwise derived from the $DM$ using the NE2001 model) and proper motion of each pulsar, without taking uncertainties into account.
The overlaid curves illustrate the maximum likelihood probability density function, fitted to the data with uncertainties as described in section \ref{kick vel method}, for each of the investigated models: single-component Gaussian (dashed), two-component Gaussian (dashed-dotted), exponential (solid), Lorentzian (dashed-triple-dotted), and \cite{1990ApJ...348..485P} (dotted).}
\label{v_l hist}
\end{center}
\end{figure}

As illustrated by Figure \ref{v_l hist}, while 27 of the 34 pulsars in the sample have \mbox{$|v_{l}|<300$ km s$^{-1}$}, the distribution extends beyond \mbox{$|v_{l}|=1000$ km s$^{-1}$}, with \mbox{$|v_{l}|=1340$ km s$^{-1}$} for \mbox{PSR B2011+38}.
Here, the velocity components are calculated using the most probable distance and proper motion for each pulsar, without taking the uncertainties into account.
Since Gaussian functions have rapidly decaying tails, we do not expect a single-component Gaussian to describe the data well.
Rather, we expect a two-component Gaussian with extended tails to be favored in order to accommodate the highest-velocity objects. 
Such a two-component Gaussian model requires three free parameters.
It also results in a bimodal three-dimensional velocity distribution \citep[see, e.g., Figure 3 of][]{2002ApJ...568..289A}, a feature which, if real, requires an astrophysical explanation.
To investigate whether the complexity of the two-component Gaussian model is justified, we consider a number of alternative single-parameter models, with heavier tails than Gaussian functions.

\subsubsection{Alternative Single-Parameter Models}
The (double-sided) exponential model has functional form
\begin{equation}
p(v_{l})=\frac{1}{2 \langle v_{l} \rangle}\exp{\left( -\frac{|v_{l}|}{\langle v_{l} \rangle} \right)},
\end{equation}
where $\langle v_{l} \rangle$ is the mean absolute value of the velocity component.
The corresponding three-dimensional distribution is difficult to derive analytically, but its mean is easily estimated numerically by simulation of velocity vectors.
The Lorentzian model is defined by
\begin{equation}
p(v_{l})=\frac{HWHM}{\pi(v_{l}^2 + HWHM^2)},
\end{equation}
where $HWHM$ is the half width at half maximum.
The first moment of the corresponding three-dimensional distribution in this case does not exist and the mean is thus undefined.
Finally, we consider a model introduced by \cite{1990ApJ...348..485P}:
\begin{equation}
p(v_{l})=\frac{[1+(v_{l}/v_{*})] \ln{\left[ (1+(v_{l}/v_{*})^2)/(v_{l}/v_{*})^2 \right]} - 1}{\pi [1+(v_{l}/v_{*})^2]}
.
\end{equation}
This $v_{l}$ distribution corresponds to the more mathematically appealing three-dimensional distribution
\begin{equation}
p(v_{3D})=
\frac{4}{\pi v_{*} [1+(v_{3D}/v_{*})^2]^2}
.
\end{equation} 
The parameter $v_{*}$ is related to the three-dimensional mean by $\langle v_{3D} \rangle =2 v_{*} / \pi$.

\begin{deluxetable*}{cccccccc}
\tablecaption{Maximum Likelihood (ML) Birth Velocity Models \label{kick vel results table}}
\tablewidth{0pt}
\tablehead{
\colhead{Model} & 
\colhead{ML Parameters} &
\colhead{$\mathcal{E}$\tablenotemark{a}} &
\colhead{Searched Range} &
\colhead{$\langle v_{3D} \rangle$} &
\colhead{$P_{\textrm{K-S}}$} &
\colhead{$f_{60}$\tablenotemark{b}} 
\\
\colhead{} & 
\colhead{(km s$^{-1}$)\tablenotemark{c}} &
\colhead{} &
\colhead{(km s$^{-1}$)} &
\colhead{(km s$^{-1}$)} &
\colhead{} &
\colhead{\%}
}
\startdata

1-Gaussian & $\sigma=290^{+30}_{-30}$ & $9.7\times10^{-4}$ & [100, 500] & 460$^{+50}_{-50}$ & 0.045 & 0.2 \\
2-Gaussian & $w_{1}=0.90^{+0.10}_{-0.03}$ & 1.0 & [0.05, 1.00] & 350$^{+90}_{-140}$ & 0.442 & 1.2 \\
 & $\sigma_{1}=160^{+20}_{-30}$ & & [50, 300] & & & \\
 & $\sigma_{2}=780^{+150}_{-140}$ & & [300, 1000] & & & \\  
Exponential & $\mu=180^{+20}_{-30}$ & 1.0 & [50, 300] & 380$^{+40}_{-60}$ & 0.377 & 1.4 \\
Lorentzian & $HWHM=100^{+20}_{-20}$ & 1.3 & [50, 350] & $\ldots$\tablenotemark{d} & 0.245 & 2.0 & \\
Paczy\'nski & $v_{*}=560^{+110}_{-120}$ & 1.5 & [200, 900] & 360$^{+70}_{-80}$ & 0.153 & 13.5 \\

\enddata

\tablenotetext{a}{The quoted evidences have been normalized so that \mbox{$\mathcal{E} \equiv 1$} for the exponential model.}
\tablenotetext{b}{Fraction of pulsars with birth velocity \mbox{$\leq 60$ km s$^{-1}$}, the typical central escape velocity of the most massive Galactic globular clusters, such as 47 Tucanae \citep{1985IAUS..113..541W}.
It is 2.8\% for the model of \cite{2002ApJ...568..289A}.}
\tablenotetext{c}{Except for the dimensionless fraction $w_{1}$.}
\tablenotetext{d}{The expectation value of $v_{3D}$ is undefined.}

\end{deluxetable*}

\subsection{Results and Discussion}
\label{kick vel results}
The results of the maximum likelihood optimization are summarized in Table \ref{kick vel results table}.
We also list the derived mean three-dimensional velocity and the fraction of pulsars with birth velocity \mbox{$\leq 60$ km s$^{-1}$}, the typical central escape velocity for the most massive Galactic globular clusters, such as \mbox{Tucanae 47} \citep{1985IAUS..113..541W}, for each model.
As a check on our calculations, we also give the Kolmogorov-Smirnov (K-S) $P$-values $P_{\textrm{K-S}}$ (see, e.g., Press et al. 1992\nocite{1992nrca.book.....P}  and section \ref{model assessment}) to compare the observed distribution of $v_{l}$ and the maximum likelihood distribution for each model.
For the K-S $P$-values, only the most probable values of the proper motion and distance are used to calculate the velocity component $v_{l}$, i.e. the uncertainties are ignored, and we only expect the results to roughly validate those of the more detailed maximum likelihood analysis.
A further benefit of the K-S $P$-values is that they provide an absolute measure of the goodness of fit, whereas the maximum likelihood analysis only provides relative figures.

With an evidence $\sim10^{-3}$ that of the other models, and a definitely lower K-S $P$-value, the single-component Gaussian model is clearly inadequate.
The four other models all have identical evidence, up to the significant digit, and we cannot distinguish among them.
None of them can be rejected at a significant level of confidence based on the K-S test either.
In particular, we cannot justify the complexity associated with the two-component Gaussian model.
Nonetheless, it is interesting to note that the maximum likelihood parameters for the two-component Gaussian model place $w_{1}$=0.90$^{+0.10}_{-0.03}$ of pulsars in the low-velocity component, with 1-D dispersion \mbox{$\sigma_{1}$=160$^{+20}_{-30}$ km s$^{-1}$}, and the rest in the high-velocity component, with dispersion \mbox{$\sigma_{2}$=780$^{+150}_{-140}$ km s$^{-1}$} (implying a mean three-dimensional velocity \mbox{$\langle v_{3D} \rangle$=350$^{+90}_{-140}$ km s$^{-1}$}).
\cite{1998ApJ...505..315C} arrived at a similar result, with best-fit two-component Gaussian model parameters \mbox{$w_{1}=$0.86$^{+0.03}_{-0.12}$}, \mbox{$\sigma_{1}$=175$^{+19}_{-24}$ km s$^{-1}$}, and \mbox{$\sigma_{2}$=700$^{+295}_{-132}$ km s$^{-1}$} (with corresponding \mbox{$\langle v_{3D} \rangle$=395$^{+245}_{-80}$ km$^{-1}$}).
\cite{2002ApJ...568..289A} found maximum likelihood for \mbox{$w_{1}$=0.4$^{+0.2}_{-0.2}$}, \mbox{$\sigma_{1}$=90$^{+20}_{-15}$ km s$^{-1}$}, and \mbox{$\sigma_{2}$=500$^{+250}_{-150}$ km s$^{-1}$} (with corresponding \mbox{$\langle v_{3D} \rangle$=540$^{+460}_{-240}$ km s$^{-1}$}).
Finally, \cite{2003AJ....126.3090B} obtained $w_{1}$=0.20, \mbox{$\sigma_{1}$=99 km s$^{-1}$}, and \mbox{$\sigma_{2}$=294 km s$^{-1}$} (with corresponding \mbox{$\langle v_{3D} \rangle$=407 km s$^{-1}$}).
Here, the quoted uncertainties on the model parameters $w_{1}$, $\sigma_{1}$, and $\sigma_{2}$ are at the $1\sigma$ level in each case.
We estimated the uncertainties on the derived $\langle v_{3D} \rangle$ by calculating the minimum and maximum values allowed when each of the parameters is confined in its $1\sigma$ region.
While our derived mean three-dimensional velocity is consistent with each of these results within $1\sigma$, the uncertainties are large due to the multiple model parameters.
Moreover, the fact that the values of the three model parameters describing the two components of the velocity distribution vary wildly (especially the fraction of low-velocity pulsars) suggests that the true velocity distribution does not have two well defined components.
This supports the hypothesis that previous authors were pushed toward a more complex two-component Gaussian model because of the non-Gaussianity of the true distribution, for instance to accommodate the more extreme objects, rather than because of a genuine bi-Gaussianity.

Except for the single-component Gaussian model, for which the derived mean three-dimensional velocity is overestimated because of the highest-velocity pulsars, the maximum likelihood mean three-dimensional velocity for each of the models we have investigated, which varies from \mbox{350 km s$^{-1}$} to \mbox{380 km s$^{-1}$}, is smaller than most previous estimates (e.g., \mbox{450 km s$^{-1}$} for Lyne \& Lorimer 1994\nocite{1994Natur.369..127L}, \mbox{$\sim$500 km s$^{-1}$} for Lorimer, Bailes, \& Harrison 1997\nocite{1997MNRAS.289..592L}, and the results discussed in the previous paragraph).
This is consistent with the distances estimated using the NE2001 model being in general smaller than those obtained using TC93, which is the case for all the pulsars with distances derived from dispersion measures retained in our analysis, except for PSRs \mbox{B1534+12} and \mbox{B1541+09}.
A notable exception is \cite{1997MNRAS.291..569H}, who found a mean birth velocity \mbox{$\sim250-300$ km s$^{-1}$}.
The origin of the discrepancy is not clear, but is possibly attributable to pulsar weight adjustments made by \cite{1997MNRAS.291..569H} to attempt to account for Malmquist-like biases (B.~Hansen 2005, private communication).

There are two main differences between our conclusions and those of \cite{2003AJ....126.3090B}, who used the same pulsar sample and performed the same basic analysis. 
First, our best-fit velocity models, excluding the single-component Gaussian one, have lower (albeit consistent within $1\sigma$) mean three-dimensional velocities (\mbox{$\sim350-380$ km s$^{-1}$} compared to \mbox{407 km s$^{-1}$}). 
This is expected, as just discussed, since we have used the NE2001 model for the Galactic free electron density, while \cite{2003AJ....126.3090B} used TC93.
Second, \cite{2003AJ....126.3090B} favor a two-component Gaussian functional form, while we prefer alternative single-parameter models. 
We understand this to be simply a result of \cite{2003AJ....126.3090B} not investigating non-Gaussian single-parameters models, for had we only considered Gaussian models as they did, we too would have been led to favor a distribution with two components (see Table \ref{kick vel results table}).

After we had completed our analysis, \cite{2005MNRAS.tmp..475H} published a pulsar birth velocity distribution inferred from a sample of 73 pulsars with measured proper motions (most of which from pulse timing) and characteristic ages \mbox{$\tau_{c} \equiv P/2\dot{P} \leq$3 Myr} using NE2001.
They did not, however, treat measurement uncertainties.
They found that the data are well described by a three-dimensional Maxwellian distribution with one-dimensional standard deviation \mbox{$\sigma=265$ km s$^{-1}$}.
This standard deviation is consistent with the one we obtained for our single-parameter Gaussian model (\mbox{$\sigma=290^{+30}_{-30}$ km s$^{-1}$}), but the result is at odds with the fact that we disfavor a single-parameter Gaussian functional form.
\cite{2005MNRAS.tmp..475H} did not investigate alternative single-parameter models and we note that PSRs \mbox{B2011+38} and \mbox{B2224+65} (both with inferred transverse velocity \mbox{$\geq 1600$ km s$^{-1}$}), which they have included in their analysis, have vanishingly small probability of occurring under the hypothesis that the velocities are distributed according to their best-fit Maxwellian distribution\footnote{Consider $n$ three-dimensional velocities independently drawn from a Maxwellian distribution with standard deviation $\sigma$ for the corresponding one-dimensional Gaussian distribution: $p(v_{3D})=\sqrt{\frac{2}{\pi}}\frac{1}{\sigma^{3}}v_{3D}^{2}\exp{\left(-v_{3D}^{2}/2\sigma^2\right)}$ (\mbox{$v_{3D}\geq0$}).
Each velocity has probability \mbox{$p\equiv\int_{v}^{\infty}p(v_{3D})\textrm{d}v_{3D}$} of exceeding the value $v$.
For independently drawn velocities, the total number $E$ of velocities exceeding $v$ is binomially distributed with parameters $n$ and $p$.
The probability that $E$ is at least 2 is thus given by \mbox{$1-(1-p)^{n-1}(1+(n-1)p)$.}
For \mbox{$\sigma=265$ km s$^{-1}$}, $n=73$, and \mbox{$v=1600$ km s$^{-1}$}, we obtain the value $\sim10^{-11}$.}.
Rather than directly fitting a model to the observed one- or two-dimensional distributions of velocities projected on the plane of the sky, \cite{2005MNRAS.tmp..475H} used a deconvolution algorithm to estimate the corresponding three-dimensional velocity distribution, under the assumption of isotropy of the velocity vectors.
They then fitted a model to the deconvolved data.
The fact that PSRs \mbox{B2011+38} and \mbox{B2224+65} are inconsistent with their best-fit model, despite the fit to the deconvolved data being statistically good (reduced \mbox{$\chi^2=0.6$}) suggests that some information was lost in the deconvolution process and that the technique may not be sensitive to subtle tail behavior.
It is true, though, that the authors tested their deconvolution algorithm and verified that it reproduced mock three-dimensional distributions well, at least to a degree sufficient to distinguish between unimodal and bimodal distributions.
We do not necessarily consider our results discrepant with those of \cite{2005MNRAS.tmp..475H}, as the difference appears to lie mainly in whether a few extreme objects in the sample are given weight.
The resolution of this dilemma will require reliable confirmation or rebuttal of the measurements of the high velocities and will depend on whether such objects continue to be discovered.

\cite{1997MNRAS.291..569H} have remarked that there is significant uncertainty regarding the form of the pulsar birth velocity distribution. 
They demonstrated that, due to projection effects, a three-dimensional kick velocity distribution consisting of two delta functions (located at \mbox{250 km s$^{-1}$} and \mbox{1000 km s$^{-1}$}, with weight 0.8 and 0.2, respectively), was equally consistent with the observed proper motions as their preferred, qualitatively very different, Maxwellian with one-dimensional dispersion \mbox{$\sigma=$190 km s$^{-1}$}.
\cite{1997MNRAS.289..592L} also showed that a variety of birth velocity distributions were consistent with the data.
We continue to find that the exact shape of the birth velocity distribution is poorly constrained, as illustrated by our inability to quantitatively discriminate between four different models.
Reliable determination of the correct shape of the kick velocity from proper motion measurements will require study of a considerably larger sample of pulsars.
Perhaps our non-detection of two well defined components is due to the size of the analyzed sample, but given the lack of compelling evidence for a multimodal distribution, we follow Ockham's razor in favoring unimodal, single-parameter models.

\begin{figure}[ht]
\begin{center}
\includegraphics[angle=-90, width=0.45\textwidth]{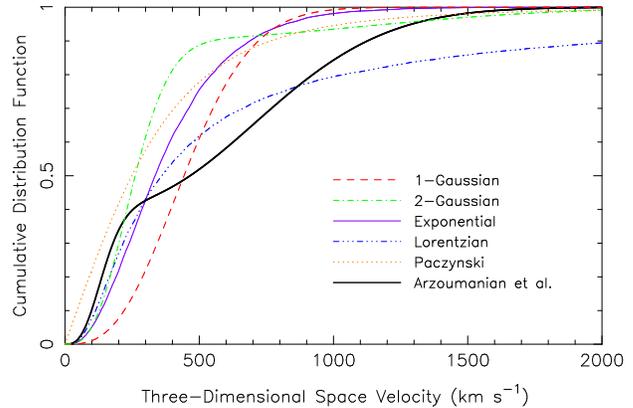}
\caption{Cumulative distribution function for the three-dimensional birth velocity for each of the maximum likelihood models investigated: single-component Gaussian (dashed), two-component Gaussian (dashed-dotted), exponential (solid), Lorentzian (dashed-triple-dotted), and \cite{1990ApJ...348..485P} (dotted).
The distribution inferred by \cite{2002ApJ...568..289A} is indicated by the thick solid curve.
}
\label{cum v plot}
\end{center}
\end{figure}

The cumulative three-dimensional velocity distribution for the maximum likelihood parameters for each model considered here is shown in Figure \ref{cum v plot}.
The distributions are consistent with the observed retention rates of neutron stars in globular clusters. 
In fact, the results of \cite{IVA04} on the dynamical formation and evolution of binaries containing neutron stars in dense clusters suggest that concerns regarding the fraction of neutron stars retained in globular clusters are not justified \citep[the so-called ``retention problem"; for a discussion, see for example][]{2002ApJ...573..283P}.
This is thanks to neutron stars in clusters being very effectively recycled and the total number of neutron stars in clusters being much smaller than previously thought (F.~Rasio 2005, private communication).
In particular, in light of a deep \emph{Chandra X-Ray Observatory} observation, the globular cluster \mbox{47 Tucanae} is now estimated to contain $\sim$25 millisecond pulsars \citep{2005ApJ...625..796H}, compared to a previous estimate based on radio observations of $\gtrsim$200 \citep{2000ApJ...535..975C}.

In most of the remainder of this paper, we adopt the single-parameter model with the smallest fractional uncertainty (15\%): the exponential one.
The corresponding three-dimensional mean velocity is \mbox{$380^{+40}_{-60}$ km s$^{-1}$}.
We consider how our subsequent results would be affected if we had chosen one of the other models with similar evidence in section \ref{birth velocity distribution}.

\section{POPULATION SYNTHESIS}
\label{pop synth}
We now proceed with a more general pulsar population synthesis.
We consider only isolated, non-recycled radio pulsars in the Milky Way field.

\subsection{Galactic Model}
\label{galactic model}
As the first step, we must define a model for the galaxy in which the synthetic pulsars will be placed.
We model the spiral arm structure of the Milky Way, its large-scale gravitational potential, and its distribution of free electrons.
We introduce a Galactocentric system of coordinates. 
The origin is defined to be the Galactic Center (GC). 
The $x$, $y$, and $z$ axes, respectively parallel to $(l, b)=(90^{\circ}, 0^{\circ}), (180^{\circ}, 0^{\circ}), (0^{\circ}, 90^{\circ})$, form a right-handed Cartesian frame. 
As usual, \mbox{$r=\sqrt{x^{2}+y^{2}}$}, \mbox{$\theta=\arctan(y/x)$}, and \mbox{\boldmath$x$}~$=(x, y, z)$.

\subsubsection{Spiral Arm Structure}
Isolated neutron stars are thought to be formed in core-collapse supernova events.
These mark the deaths of massive, short-lived Population I stars associated with the arms of spiral galaxies.
It is thus expected that the birth sites of isolated pulsars will be highly correlated with Galactic spiral arms.
In fact, \citet[][see especially their Figure 5]{2003MNRAS.342.1299K} note that following the PM survey, the Galactic spiral structure is now clearly visible in the distribution in Galactic longitudes of young pulsars, making it necessary for any realistic pulsar population synthesis study to model this spiral structure.
This has not previously been done in any of the major published works.

We model the spiral structure of the spatial distribution of the pulsar progenitors by four major arm centroids whose loci are described analytically by equations of the form
\begin{equation}
\theta(r)=k~\ln(r/r_{0})+\theta_{0}.
\end{equation}
The values for the parameters for each arm, given in Table \ref{spiral params}, are taken from Wainscoat et al. (1992) and are consistent with those of the NE2001 model.  
We do not, however, model the ``local arm" nor the perturbations to the spirals included NE2001.
We adopt a Sun-GC distance of \mbox{$R_{\odot}=8.5$ kpc}.
This distance is consistent with NE2001, \emph{Hipparcos} proper motions of Cepheids \citep{1997MNRAS.291..683F}, and a single-step measurement using red clump stars \citep{1998ApJ...494L.219P}.

\begin{deluxetable}{ccccc}
\tablecaption{Spiral Arm Parameters \label{spiral params}}
\tablewidth{0pt}

\tablehead{
\colhead{Arm Number} & 
\colhead{Name} &
\colhead{$k$} &
\colhead{$r_{0}$} & 
\colhead{$\theta_{0}$}
\\
\colhead{} & 
\colhead{} &
\colhead{(rad)} &
\colhead{(kpc)} & 
\colhead{(rad)}
}

\startdata

1 & Norma              & 4.25 & 3.48 & 1.57 \\
2 & Carina-Sagittarius & 4.25 & 3.48 & 4.71 \\
3 & Perseus            & 4.89 & 4.90 & 4.09 \\
4 & Crux-Scutum        & 4.89 & 4.90 & 0.95 \\

\enddata

\tablecomments{The numerical values differ from those given by \cite{1992ApJS...83..111W} due to differences between the coordinate systems used in that paper and in this one.}

\end{deluxetable}

\begin{figure}[ht]
\begin{center}
\includegraphics[width=0.45\textwidth]{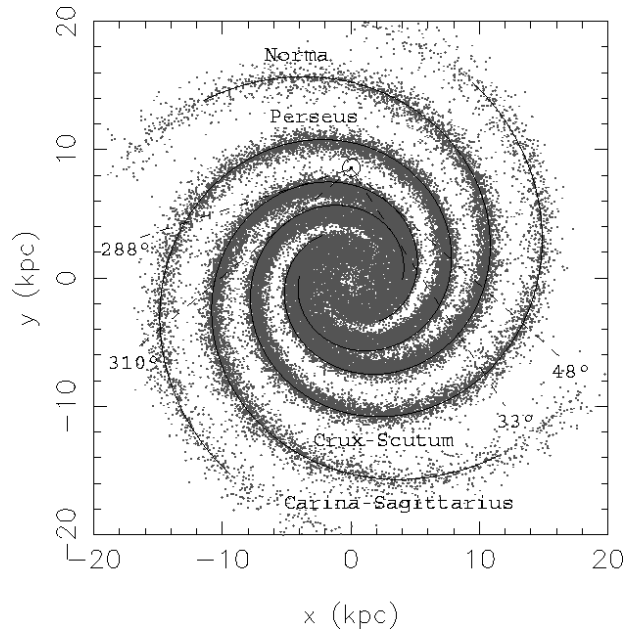}
\caption{Example of a simulated initial distribution of pulsars in the $xy$ plane.
Each point represents the birth position of a pulsar projected onto the Galactic plane.
The location of the Sun is indicated by the symbol $\odot$.
Solid lines trace the spiral arm centroids.
Dashed lines are tangent to arms in the vicinity of the Sun.
}
\label{xyi}
\end{center}
\end{figure}

\subsubsection{Gravitational Potential}
We adopt the modification by \cite{1989MNRAS.239..651K} of the \cite{1987AJ.....94..666C} fit of the Galactic gravitational potential. 
This model consists of a disc-halo component, a bulge component, and a nucleus component:
\begin{equation}
\label{gravitational potential}
\phi_{G}(r, z)=\phi_{dh}(r, z)+\phi_{b}(r)+\phi_{n}(r),
\end{equation}
where 
\begin{equation}
\phi_{dh}(r, z)=
\frac{-GM_{dh}}
{\sqrt{(a_{G}+\sum_{i=1}^3\beta_{i}\sqrt{z^{2}+h_{i}^{2}})^{2}+b_{dh}^{2}+r^{2}}}
\end{equation}
and
\begin{equation}
\phi_{b,n}(r)=
\frac{-GM_{b,n}}
{\sqrt{b_{b,n}^{2}+r^{2}}}
\textrm{.}
\end{equation}
The numerical values of the constants are given in Table \ref{gp params}.

\begin{deluxetable}{cccc}
\tablecaption{Parameters for the Galactic Gravitational Potential Model \label{gp params}}
\tablewidth{0pt}
\tablehead{
\colhead{Constant} & 
\colhead{Disc-Halo (dh)} &
\colhead{Bulge (b)} & 
\colhead{Nucleus (n)}
}
\startdata

$M$ & $1.45\times10^{11}~\textrm{M}_{\odot}$ & $9.3\times10^{9}~\textrm{M}_{\odot}$ & $1.0\times10^{10}~\textrm{M}_{\odot}$ \\
$\beta_{1}$ & $0.4$ & & \\
$\beta_{2}$ & $0.5$ & & \\
$\beta_{3}$ & $0.1$ & & \\
$h_{1}$ & $0.325$ kpc & & \\
$h_{2}$ & $0.090$ kpc & & \\
$h_{3}$ & $0.125$ kpc & & \\
$a_{G}$ & $2.4$ kpc & & \\
$b$ & $5.5$ kpc & $0.25$ kpc & $1.5$ kpc \\
\enddata

\tablecomments{Values taken from \cite{1989MNRAS.239..651K}.}

\end{deluxetable}

\subsubsection{Free Electron Density}
The propagation of radio pulses from the pulsars to us is affected by the intervening interstellar medium.
First, it introduces a frequency dependence of the group velocity of the radio waves which causes the observed pulses to be dispersed by an amount proportional to the integrated column density of free electrons (a quantity known as the dispersion measure, DM).
Second, the density fluctuations of the free electrons scatter the radio waves and the resulting multipath propagation broadens the pulses.
These two effects hamper the detectability of the pulsars and must be modelled.
To do so, we use the software implementation of the NE2001 model of Galactic free electron density \citep{NE2001}, which, given the position of a synthetic pulsar in the Galaxy and the observing frequency, computes its modelled DM and pulse broadening scattering time ($\tau_{scatt}$).
To reduce the computational cost, we have changed the default integration distance step of \mbox{10 pc} to \mbox{50 pc}.

\subsection{Pulsar Birth Properties}
The modelled Galaxy is populated by synthetic pulsars.
We describe how birth properties are assigned to them.

\label{birth properties}
\subsubsection{Spatial and Kinematic Distributions}
The birth locations of the Monte Carlo pulsars are specified by a radial ($r$) distribution and a vertical ($z$) distribution.

The spiral structure is realized by choosing the birth locations so that their projections lie on arms (assuming an equal pulsar birthrate in each arm) and subsequently altering them to simulate a spread about the arm centroids.
More precisely, an arm number, \mbox{$1\leq i \leq 4$}, is first randomly\footnote{In this paper, the unqualified term ``random" implies an uniform distribution.}
chosen.
A distance $r_{raw}$ from the GC is then chosen according to the specified radial distribution (see below) and the corresponding polar angle $\theta_{raw}$ is calculated so that the pulsar lies on the centroid of arm $i$. 
To avoid artificial features near the GC, the distribution is blurred by applying a correction of magnitude \mbox{$\theta_{corr}\exp(-0.35r_{raw}/\textrm{kpc})$}, where $\theta_{corr}$ is randomly chosen in the interval \mbox{$[0, 2\pi)$ rad}, to the polar angle of each pulsar.
Finally, to spread the pulsars about the spiral centroids, the $(x, y)$ coordinates of each pulsar on the Galactic plane are further altered by translating it by a distance $r_{corr}$ drawn from a normal distribution centered at zero with standard deviation $0.07r_{raw}$, without preference with respect to direction.
These corrections are somewhat arbitrary, chosen so that they produce a reasonably natural distribution.
Figure \ref{xyi} illustrates a resulting simulated birth distribution in the $xy$ plane.

For the radial distribution of the pulsar progenitors, we adopt the functional form suggested by \cite{2004A&A...422..545Y} following an analysis based on the NE2001 model and including data from the PM and SM surveys:
\begin{equation}
\label{radial distr eq}
\rho(r)=A \left( \frac{r+R_{1}}{R_{\odot}+R_{1}} \right)^{a}
\exp{\left[ -b \left( \frac{r-R_{\odot}}{R_{\odot}+R_{1}} \right) \right]}
,
\end{equation}
where $\rho(r)$ is the surface density at Galactocentric radius $r$; $R_{1}$, $a$ and $b$ are model parameters; and $A$ is a normalization constant. 
\cite{2004A&A...422..545Y} found the best-fit values $a=1.64 \pm 0.11$, $b=4.01 \pm 0.24$, and \mbox{$R_{1}=0.55 \pm 0.10$ kpc}. 
Within the limits of the estimated errors, this radial distribution is nearly the same as the one derived by \cite{lor03} using a different method. 
\cite{2004A&A...422..545Y} and \cite{lor03} actually investigated the distribution of evolved pulsars, rather than that of their progenitors, and it is \textit{a priori} unclear whether it is justified to adopt this model for the birth distribution.
We address this question \textit{a posteriori} in section \ref{radial distribution}.

The distance from the Galactic plane is simply chosen from a exponential distribution with specified mean, $\langle z_{0} \rangle$, and is assigned a random sign.

To model the birth space velocities of the pulsars, we adopt the exponential velocity distribution determined in section \ref{kick vel}.
We examine the importance of this assumption in section \ref{birth velocity distribution}.

\subsubsection{Spin Period and Magnetic Field}
The birth spin period $P_{0}$ of each pulsar is chosen from a normal distribution with mean $\langle P_{0} \rangle$ and standard deviation $\sigma_{P_{0}}$.
Negative spin periods are rejected and redrawn.
The equatorial surface magnetic field $B$ is chosen from a normal distribution with mean $\langle \log{B} \rangle$ and standard deviation $\sigma_{\log{B}}$ in the logarithm to the base 10.

\subsubsection{Radio Luminosity}
Despite nearly four decades of study, the radio emission mechanism and the geometry of pulsar emission beams remain poorly understood.
Nonetheless, we must model the pulsar luminosity in our simulations in order to determine which objects are detectable.

To avoid modelling potential apparent luminosity variations due to complex emission beam and viewing geometries, each pulsar is assigned a (pseudo-)luminosity $L$ defined such that \mbox{$L=SD^{2}$}, where $S$ is the pulsar's flux density at the Earth, instead of a true physical luminosity.

Given the lack of well defined correlation of pulsar radio luminosities with other properties in the observed pulsar sample, a model that must be tested is one in which the luminosities are assumed independent of all other properties. 
This model will be referred to as the ``random" model.
\cite{1998MNRAS.295..743L} investigated the luminosity distribution of pulsars at 400 MHz. 
For \mbox{$L_{400} \geq 10$ mJy kpc$^{2}$}, they found that \mbox{$N(L_{400})\textrm{d}L_{400} \propto L_{400}^{-2}\textrm{d}L_{400}$}. 
As $L_{400} \rightarrow 0$, this law must break down, for otherwise the integral giving the total number of pulsars in the Galaxy is divergent.
We model a flattening of the luminosity distribution by a second power-law component for luminosities below a certain turn-over point $L_{to}$, with continuity enforced at $L_{to}$, and a low-luminosity cut-off $L_{low}$ ensuring convergence:
\begin{equation}
p(L)\propto
\left\{
\begin{array}{cl}
L^{\alpha_{1}} & \textrm{for } L \in [L_{low}, L_{to}) \\
L^{\alpha_{2}} & \textrm{for } L \in [L_{to}, \infty) \\
0 & \textrm{otherwise}
\end{array}
\right. .
\end{equation}
Assuming that the spectral index is independent of the luminosity at a particular frequency, the distribution of luminosities at \mbox{1.4 GHz} should have the same shape as at \mbox{400 MHz}, except for being shifted toward lower luminosities. 
From Figure 8 of \cite{1998MNRAS.295..743L}, we estimate that for spectral indexes $\sim-1.8$ \citep{2000A&AS..147..195M}, \mbox{$L_{to}=2$ mJy kpc$^{2}$}, \mbox{$\alpha_{1}=-19/15$}, and \mbox{$\alpha_{1}=-2$} at \mbox{1.4 GHz}.
As discussed in section \ref{inadequacy of random luminosities}, our conclusions are unaffected by this crude approximation.
We set \mbox{$L_{low}$=0.1 mJy kpc$^{2}$}, roughly corresponding to the faintest observed pulsars.

Since the radiated energy is thought to be derived from the loss of rotational kinetic energy due to magnetic braking, another natural model to test is one in which the radio luminosity is related to pulsar's period and period derivative.
\cite{1975MNRAS.171..579L} argued that the pulsar luminosity has a power-law dependence on $P$ and $\dot{P}$.
\cite{1981JApA....2..315V} and \cite{1984bens.work..151P} reached similar conclusions, and this has often been assumed in later works.
The correct power-law exponents, though, are still undetermined.
Following the common practice, we also investigate a model in which $L$ is proportial to a power law in $P$ and $\dot{P}$.
We dither the standard candle luminosity by adding to it a correction in the logarithm.
This presumably accounts for both physical variations about the modelled luminosity and observed variations due to differing viewing geometries:
\begin{equation}
\log{L} = 
\log{
\left(
L_{0} P^{\epsilon_{P}}
\dot{P_{15}}^{\epsilon_{\dot{P}}} 
\right)
} 
+ L_{corr}
,
\end{equation}
where $P$ is in s, $\dot{P_{15}}$ is in \mbox{$10^{-15}$ s s$^{-1}$}, and $L_{corr}$ is chosen from a zero-centered normal distribution with standard deviation $\sigma_{L_{corr}}$.
The required dither about the luminosity law is also likely to be partially due to errors in the distance measurements to actual pulsars (see section \ref{radial distribution}), which we do not treat explicitly here.
In this ``$P-\dot{P}$ power law" model, luminosities are time-dependent.

Since we simulate only radio surveys at \mbox{1.4 GHz} (see section \ref{radio surveys}), we consider only luminosities at that frequency.

\subsection{Pulsar Evolution}
\label{evolution}
After birth properties are chosen, the synthetic pulsars are evolved in time.

\subsubsection{Spatial Evolution}
\label{evolution spatial}
The empty space between stars dominates the volume of the Galactic field, so that encounters are extremely rare \citep{1987gady.book.....B}.
As a consequence, the orbits of pulsars are determined by the large-scale, smooth Galactic gravitational potential according to the equation
\begin{equation}
\mbox{\boldmath$\ddot{x} = - \nabla$}
\phi_{G}
,
\end{equation}
where $\phi_{G}$ is given by equation \ref{gravitational potential}.
The orbits are computed numerically using the adaptive time-stepping \verb+LSODA+ 
solver of the Lawrence Livermore National Laboratory \verb+ODEPACK+\footnote{http://www.llnl.gov/CASC/odepack/} collection of ordinary differential equation solvers.

\begin{deluxetable}{cccc}
\tablecaption{Measured Pulsar Braking Indexes\label{measured n}}
\tablewidth{0pt}
\tablehead{
\colhead{Pulsar Name} & 
\colhead{Braking Index} &
\colhead{References} 
}
\startdata

B0531+21 (Crab) & 2.51$\pm$0.01 & 1, 2 \\
B0540$-$69 & 2.140$\pm$0.009 & 3, 4, 5 \\
B0833$-$45 (Vela) & 1.4$\pm$0.2 & 6 \\
J1119$-$6127 & 2.91$\pm$0.05 & 7 \\
B1509$-$58 & 2.839$\pm$0.003 & 8, 9

\enddata

\tablerefs{ 
(1) \cite{1988MNRAS.233..667L};
(2) \cite{1993MNRAS.265.1003L};
(3) \cite{2005ApJ...633.1095L};
(4) \cite{1992ApJ...394..581G};
(5) \cite{1999ApJ...512..300D};
(6) \cite{1996Natur.381..497L};
(7) \cite{2000ApJ...541..367C};
(8) \cite{1994ApJ...422L..83K};
(9) \cite{2005ApJ...619.1046L}
}

\end{deluxetable} 

\begin{figure}[ht]
\begin{center}
\includegraphics[angle=-90, width=0.45\textwidth]{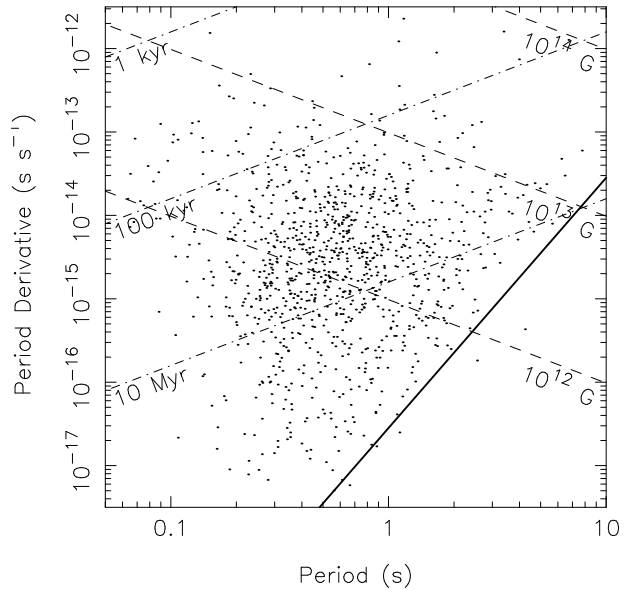}
\caption{$P-\dot{P}$ diagram for the real pulsars retained in our analysis.
Each point indicates the period and period derivative of a pulsar.
Contours of constant surface magnetic field and of constant characteristic age are indicated by dashed and dotted-dashed lines, respectively.
The thick solid line marks the modelled death line.
}
\label{ppdot real}
\end{center}
\end{figure}

\subsubsection{Rotational Evolution}
\label{evolution rotational}
The losses of rotational kinetic energy due to radiation by pulsars are inevitably associated with losses of angular momentum and the lengthening of the spin period.
These losses are generally assumed to be dominated by magnetic dipole braking.
In the case of a perfect dipole rotating in a vacuum,
\begin{equation}
\label{dipolar spindown law}
P \dot{P} =
\left(
\frac{8 \pi^{2} R^{6}}{3 I c^{3}}
\right)
B^{2} \sin^{2}{\chi}
,
\end{equation}
where $R$ is the radius of the star, $I$ is its moment of inertia, $c$ is the speed of light, and $\chi$ is the angle determined by the magnetic and spin axes \citep{1969ApJ...157.1395O}.
Despite the extreme gravitational binding energies of electrons and protons at the surface of a neutron star, the magnetic field of a pulsar is so strong (with a typical value \mbox{$\sim10^{12}$ G}) that a surface charge layer cannot be in dynamical equilibrium, causing the magnetosphere to be filled with plasma.
A torque such that $P \dot{P} \propto R^{6} B^{2} / I$ is then exerted on the rotating pulsar even if the spin and magnetic axes are aligned \citep{1969ApJ...157..869G}, so that equation \ref{dipolar spindown law} with the factor $\sin^{2}{\chi}$ is invalid in the limit $\chi \to 0$.
This suggests that even orthogonal rotators are fairly well approximated by setting $\sin^{2}{\chi}\sim1$ in equation \ref{dipolar spindown law}.
Realistic pulsar spin down in a plasma is still poorly understood and is being actively researched \citep[see, e.g.,][]{2004IAUS..218..357S}.
For simplicity, we assume that the pulsar spin down is determined by equation \ref{dipolar spindown law} with $\sin^2{\chi} \equiv 1$ and we adopt the canonical values \mbox{$R=10^{6}$ cm} and \mbox{$I=10^{45}$ g cm$^ {2}$} \citep{1977ApJ...215..885T}.
Any torque scatter due to the distribution of the magnetic inclination angles, if in fact significant, can be thought of being absorbed in the distribution of the $B^2$ factor, although we will continue to refer to $B$ simply as the magnetic field to alleviate the text.

These assumptions have the benefit that the magnetic field $B$ of a synthetic pulsar is equal at all times to the magnetic field of a real pulsar with the same $P$ and $\dot{P}$ inferred from pulse timing using the conventional formula \mbox{$B_{timing}=3.2 \times 10^{19} \sqrt{P \dot{P} / \textrm{s}}$ G}.
They also imply a braking index $n=3$, where $\dot{P} \propto P^{2-n}$.
The braking indexes of several pulsars have been measured and are in fact in the range \mbox{1.4$-$2.91} (see Table \ref{measured n}).
We experimented with normal distributions of braking indexes in our simulations.
The results were weakly affected and it was not possible to constrain the correct mean and standard deviation of the underlying distribution.
Again for simplicity, we chose to keep $n\equiv3$ for all synthetic pulsars.
We discuss the impact of this choice on our results in section \ref{braking index distribution}.

Integrating equation \ref{dipolar spindown law} shows that, assuming a constant magnetic field, the spin period of a pulsar of age $t$ given by
\begin{equation}
\label{evolved period}
P = 
\sqrt{
P_{0}^{2} 
+ 
\frac{3 I c^{3}}
{8 \pi ^{2} R^{6}}
t
}
.
\end{equation}
In the case of the $P-\dot{P}$ power-law model, the evolved luminosity is calculated according to the evolved period and period derivative.

\subsubsection{Radio Emission Shut-Off}
\label{death line section}
The production of electron-positron pairs, as a source of particles to accelerate, has been assumed essential for radio pulsar emission.
The polar caps, located just above the star's surface at the magnetic poles, are the presumed sites from which radio waves originate.
In most polar-cap models, radio emission ceases when the potential drop required for pair production exceeds the maximum which can be achieved in the rotating pulsar's magnetosphere \citep{1971ApJ...164..529S, 1975ApJ...196...51R, 1993ApJ...402..264C}.
The precise time at which the turn-off occurs depends on both the structure and magnitude of the star's magnetic field.
This has led to the calculation of several theoretical ``death lines" in the $P-\dot{P}$ diagram, after crossing which pulsars become radio-quiet\footnote{We define the radio-loud pulsars as those that are active radio emitters, but that are not necessarily beamed toward us (``potentially observable"). The pulsars that are not radio-loud are said to be radio-quiet.}.
These are generally well approximated by the equation
\begin{equation}
\label{death line}
\frac{B}
{P^2}
= 0.17 \times 10^{12} 
~\textrm{G s}^{-2}
\end{equation}
\citep{1992A&A...254..198B}.
This theoretical prediction is well observed empirically, as there is in fact a well defined cut in the distribution of observed pulsars in the $P-\dot{P}$ diagram (see Figure \ref{ppdot real}).
In our simulations, equation \ref{death line} will be used to classify synthetic pulsars as either active or extinguished radio sources.
Actual radio pulsar surveys, though, generally have reduced sensitivity to long-period pulsars, due to radio interference and hardware and software high-pass filters. 
These effects, which we do not model, may play a role in the paucity of pulsars observed in the long-period part of the $P-\dot{P}$ diagram.
We consider in section \ref{p pdot power law} whether a sudden radio emission shut-off is really required by the data.

\subsection{Radiation Beaming}
\label{radiation beaming}
The pulsed flux observed from the pulsars is due to their highly beamed radio emission, which sweeps our line of sight as the star rotates.
A consequence of this beaming is that only a small fraction of the pulsars emit radio waves in our direction and are observable.
\cite{1998MNRAS.298..625T} have analysed polarization data for a large number of isolated radio pulsars and developed a model (TM98) for the fraction $f$ of pulsars that are beamed toward us as a function of the spin period:
\begin{equation}
\label{f}
f(P)=0.09[\log{(P/s)}-1]^{2} + 0.03
.
\end{equation}
In this model, the average beaming fraction is \mbox{$\sim10$\%} (both for the pulsar sample analysed by Tauris \& Manchester 1998\nocite{1998MNRAS.298..625T} and for the optimal population model presented in this work; see section \ref{population synthesis results}).
We implement it in our simulations by assigning to each Monte Carlo pulsar a probability $f(P)$ that its radio beam crosses our line of sight.

\begin{deluxetable}{ccc}
\tablecaption{Parkes and Swinburne Multibeam Survey Parameters \label{survey params}}
\tablewidth{0pt}

\tablehead{
\colhead{Parameter} & 
\colhead{Parkes\tablenotemark{a}} &  
\colhead{Swinburne\tablenotemark{b}} 
}

\startdata

Longitude Coverage                  & \multicolumn{2}{c}{$260^{\circ} \leq l \leq 50^{\circ}$} \\
Latitude Coverage                   & $|b| \leq 5^{\circ}$ & $5^{\circ} \leq |b| \leq 15^{\circ}$ \\
Aperture Diameter                   & \multicolumn{2}{c}{64 m} \\
Central Observing Frequency ($\nu$) & \multicolumn{2}{c}{1.4 GHz} \\
Bandwidth ($\Delta \nu$)            & \multicolumn{2}{c}{288 MHz} \\
Number of Channels ($N_{ch}$)       & \multicolumn{2}{c}{96} \\
Antenna Gain ($G$)                  & \multicolumn{2}{c}{0.64 K Jy$^{-1}$ \tablenotemark{c}} \\
Number of Polarizations ($N_{p}$)   & \multicolumn{2}{c}{2} \\
Integration Time ($t_{int}$)        & 35 min & 265 s \\
Sampling Interval ($t_{samp}$)      & 250 $\mu$s & 125 $\mu$s \\
S/N Detection Threshold ($\sigma$)  & \multicolumn{2}{c}{8} \\
System Losses Factor ($\beta$)      & \multicolumn{2}{c}{1.5} \\
Receiver Temperature ($T_{rec}$)    & \multicolumn{2}{c}{21 K} \\

\enddata

\tablenotetext{a}{Parameters taken from \cite{2001MNRAS.328...17M}.}
\tablenotetext{b}{Parameters taken from \cite{2001MNRAS.326..358E}.}
\tablenotetext{c}{Average over the 13 beams of the Parkes Multibeam receiver.}

\end{deluxetable}

\begin{figure*}[ht]
\begin{center}
\includegraphics[angle=-90, width=0.90\textwidth]{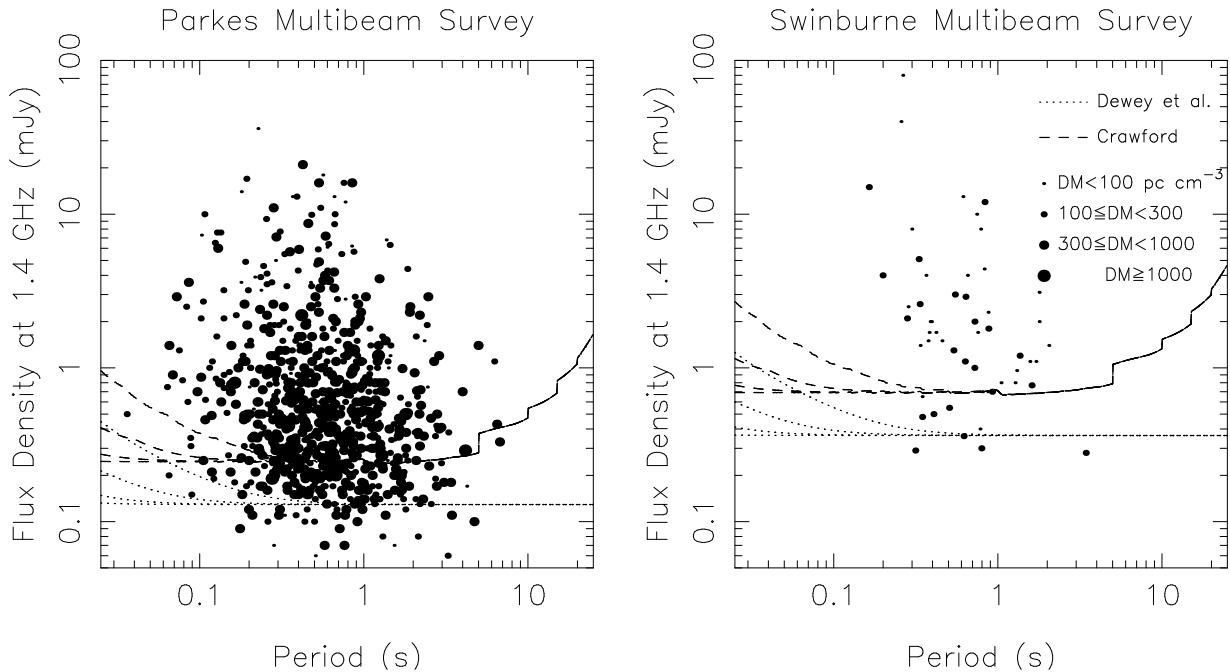}
\caption{\cite{dss+84} (dotted) and Crawford's (dashed) $S_{min}$ curves for the Parkes (left) and Swinburne (right) Multibeam surveys.
In each case, the curves are drawn for $DM=0, 100, 300, 1000$ pc cm$^{-3}$ (in increasing order of $S_{min}$).
The dots represent pulsars detected in the actual surveys that were retained in our comparison sample.
The size of the dot indicates whether $0 \leq DM < 100$, $100 \leq DM < 300$, $300 \leq DM < 1000$, or $DM \geq 1000$ \mbox{pc cm$^{-3}$}, with the larger dots corresponding to the larger $DM$.
In this figure, $S_{min}$ has been averaged over the 13 beams of the Parkes Multibeam receiver, accounting for a Gaussian beam power pattern, sky background brightness temperature, and interstellar scattering as in \cite{2001MNRAS.328...17M}.}
\label{s_min plots}
\end{center}
\end{figure*}

\subsection{Radio Surveys}
\label{radio surveys}
The goal of our population synthesis is to reproduce the actual Galactic pulsar population.
To assess its success, it is necessary to compare the results of our simulations with observational data.
Because the observed pulsar sample is heavily biased with respect to the underlying population \citep[for a discussion, see e.g.][]{1993MNRAS.263..403L}, care must be taken to ensure that equivalent real and simulated subsets are compared.
We do so by restricting the observed real sample to pulsars detected in surveys with well defined sensitivity and performing simulations of the same surveys on our synthetic Galaxy.
We first motivate and precisely define our chosen comparison sample, then describe how the radio surveys are modelled.

\subsubsection{Comparison Sample}
\label{comparison sample}
The ATNF Pulsar Database is currently the most comprehensive pulsar catalog available.
It contains over 1500 objects detected in more than a dozen surveys.
About $2/3$ of the pulsars were detected in the PM or the SM surveys at \mbox{1.4 GHz}.
These two surveys together covered the well defined sky rectangle \mbox{$260^{\circ} \leq l \leq 50^{\circ}$}, \mbox{$|b| \leq 15^{\circ}$} and used the same observing system \citep{2001MNRAS.328...17M, 2001MNRAS.326..358E}.
Apart from complementary sky coverages (\mbox{$|b|\leq 5^{\circ}$} and \mbox{$5^{\circ} \leq |b| \leq 15^{\circ}$}) and different integration times (\mbox{35 min} vs. \mbox{265 s}) and sampling intervals (\mbox{250 $\mu$s} vs. \mbox{125 $\mu$s}), the observing setups are in fact identical, providing a remarkably homogeneous pair of surveys dominating the available statistics.
The parameters of the PM and SM surveys are given in Table \ref{survey params}.
Most of the earlier surveys \citep[e.g.,][]{1971MNRAS.151..277L, 1975ApJ...201L..55H, 1977MNRAS.179..635D, 1978MNRAS.185..409M, 1978ApJ...225L..31D, 1985ApJ...294L..25D, 1986ApJ...311..694S, 1995ApJ...449..156N, 1997ApJ...474..426S, 1996MNRAS.279.1235M}
had a central observing frequency near \mbox{400 MHz}.
\cite{1995MNRAS.273..411L} have suggested that pulsar spectral indexes may be correlated with other properties, in particular the spin period.
\cite{2000A&AS..147..195M}, basing their analysis on an extension of the data set of \cite{1995MNRAS.273..411L}, on the other hand, have not found such a correlation.
In the early stages of this project, we experimented with simulations of both \mbox{400 MHz} and \mbox{1.4 GHz} surveys, assuming spectral indexes independent of other pulsar characteristics, and with a normal distribution with mean $-1.8$ and standard deviation 0.2 \citep{2000A&AS..147..195M}.
Discrepant results were obtained depending on whether the pulsar birthrate was calibrated based on the \mbox{400 MHz} or \mbox{1.4 GHz} detections,
 suggesting an inadequate modelling of the spectral dependence.
The correct way of simulating spectral variations is unclear.
Another difficulty in modelling a heterogeneous collection of surveys is ensuring that the relative sensitivities of the surveys are accurately represented.
An underestimated flux detection threshold for a survey covering a given portion of the sky, for example, would result in an artificially high number of simulated detections in that area.
Such spurious features would be difficult to distinguish from genuine ones.
For these reasons, we limit ourselves to the pulsars detected in the PM and SM surveys.
Of this reduced sample, we further ignore the pulsars that lie outside the documented survey boundaries, those with \mbox{$P<30$ ms} or \mbox{$\dot{P}<0$}, and those in binary systems.
The latter restrictions limit the contamination of our sample by recycled objects.
The resulting sample contains 1065 pulsars, the PM and SM surveys contributing 914\footnote{Data for 117 PM detections were provided to us prior to public release by D.~R. Lorimer.} and 151, respectively.
A significant fraction of the retained pulsars (126 and 97 for the PM and SM surveys, respectively) have missing flux or period derivative measurements.
The pulsars with missing data must nonetheless be taken into account to ensure that the PM and SM detections are represented in accurate proportion and in order to determine the correct number of Monte Carlo pulsars necessary to reproduce the number of actual survey detections.
When comparing the simulated and real samples (see section \ref{model assessment}), we thus consider all the real pulsars, except for the distributions that require knowledge of missing data (flux densities, magnetic fields, and $P-\dot{P}$ diagram), for which we assume that the subsample with data is unbiased with respect to the detections.
In the case of the PM survey, there is in fact no expected bias (R.~N. Manchester and D.~R. Lorimer 2005, private communication).
Unless the bias is very important in the case of the SM survey, our results are most likely unaffected, as the total subsample with missing data represents only 9\% of the comparison sample.

\subsubsection{Sensitivity Modelling}
\label{sensitivity modelling}
The detectability of a pulsar depends on its intrinsic properties (brightness, pulse period, and duty cycle), on its location (distance, $DM$, interstellar scattering, and brightness temperature of the background sky), and on the details of the observing system.
The minimum flux theoretically detectable from a radio pulsar is usually estimated by the formula
\begin{equation}
S_{min}=\delta_{beam} \frac{\beta \sigma (T_{rec}+T_{sky})}{G\sqrt{N_{p}\Delta \nu t_{int}}}\sqrt{\frac{W_{e}}{P-W_{e}}},
\end{equation}
where $\delta_{beam}$ is a factor accounting for the reduction in sensitivity to pulsars located away from the center of the telescope beam, $T_{rec}$ is the receiver temperature on cold sky, $T_{sky}$ is the sky background temperature, $G$ is the antenna gain, $N_{p}$ is the number of polarizations, $\Delta \nu$ is the receiver bandwidth, $t_{int}$ is the integration time, $P$ is the pulse period, $W_{e}$ is the effective pulse width, $\sigma$ is the signal-to-noise detection threshold, and $\beta$ is a constant accounting for various system losses \citep[][]{dss+84}. 
Further details of the $S_{min}$ calculation are provided in Appendix 1.
\cite{2000PhDT.........1C} performed a detailed analysis of the observing system used in the PM and SM surveys, providing an alternative, ostensibly more accurate, $S_{min}$ for these surveys.
Figure \ref{s_min plots} shows sensitivity curves calculated using both the \cite{dss+84} and the \cite{2000PhDT.........1C} $S_{min}$ formulae, for the PM and SM surveys.
148 pulsars detected in the actual PM survey (out of the 806 with flux measurements retained in our analysis) have measured 1.4 GHz flux \mbox{$<0.22$ mJy}, the average nominal survey sensitivity for non-millisecond pulsars according to Crawford's formalism \citep{2001MNRAS.328...17M}.
Similarly, 8 pulsars detected in the SM survey (out of the 54 with flux measurements retained) have measured flux below the average survey sensitivity given by Crawford's $S_{min}$, \mbox{$0.63$ mJy}.
While interstellar scintillation is likely responsible for some of these detections, it appears (barring a systematic error in the catalogued pulsar fluxes) that the flux threshold is overestimated by Crawford's analysis.
The threshold predicted by the \cite{dss+84} formula is lower and empirically seems more accurate.
It is thus the one we adopt.
\cite{2004ApJ...604..775G} used the Crawford formalism in their analysis of the PM data.
They remarked that the modelled sensitivity differs from that predicted by the \cite{dss+84} formula, but apparently did not consider which best empirically accounts for the actual detections.

A synthetic pulsar is detected if and only if it lies in the sky area covered by either the PM or SM survey and the radio flux density of the pulsar at \mbox{1.4 GHz} at the Earth ($S_{1400}$) exceeds the survey threshold, $S_{min}$.

\subsection{Simulation Procedure}
\label{simulation procedure}
After describing the components of our population synthesis, we now explain how we proceed with the simulations.
We first choose values for the model parameters.
Synthetic pulsars are then created one at a time.
The age ($t$) of the pulsar is chosen randomly in $[0, t_{max}]$, where $t_{max}$ is such that practically all pulsars allowed by the model parameters would cross the death line before reaching that age.
More precisely, the spin-down equation \ref{dipolar spindown law} and the death line equation \ref{death line} determine the age of radio emission cessation $t_{death}(P_{0}, B)$ for a pulsar with initial spin period $P_{0}$ and magnetic field $B$.
We maximize $t_{death}$ over the parameter space rectangle $[\langle P_{0} \rangle - 3\sigma_{P_{0}}, \langle P_{0} \rangle + 3\sigma_{P_{0}}] \times [\langle \log{B} \rangle - 3\sigma_{\log{B}}, \langle \log{B} \rangle + 3\sigma_{\log{B}}]$ and set $t_{max}$ equal to the maximum.
Because $P_{0}$ and $B$ are chosen independently, only a negligible fraction \mbox{$\sim0.006$} of the synthetic pulsars have a lifetime exceeding $t_{max}$.
Typically, \mbox{$t_{max}<1$ Gyr}.
Astrophysically, this is equivalent to assuming a constant birthrate in the Galaxy over the life span of the longest lived objects allowed by the model, i.e. that the population is in a steady state.
This is a well motivated assumption, given the Galaxy's age of \mbox{$\sim10$ Gyr}.
Birth characteristics are assigned to the pulsar following section \ref{birth properties} and are evolved as in section \ref{evolution}.
If the resulting evolved pulsar lies beyond the death line or is not beamed toward us, we subsequently ignore it (except for keeping count of the number of generated MC pulsars, $N_{MC}$).
Otherwise, we test it for detection in the PM and SM surveys as in section \ref{sensitivity modelling}.
We repeat the procedure until the number of detections in the simulation equals the number of detections in the actual surveys.
The estimated pulsar birthrate is then \mbox{$\dot{N}=N_{MC}/t_{max}$}.
The obtained sample of observed synthetic pulsars is compared with the real observed sample to assess the realism of the model population.

\subsection{Model Assessment and Search for ``Optimal" Parameters}
\label{model assessment}
Before adopting a set of model parameters and drawing conclusions from it, we must verify that it reproduces the actual observations to a satisfactory level.
We do so as follows.

For each simulation, the observed marginal distributions of $l$, $b$, $DM$, $S_{1400}$, $P$, and $B$ in the simulation are compared with the equivalent distributions for the real sample.
We also consider the simulated $P-\dot{P}$ diagram.
Except for the surface magnetic field, we have selected directly observable quantities that characterize the spatial, rotational, and brightness properties of the pulsars.

Unfortunately, the comparison does not lend itself well to a rigorous, fully quantitative analysis.
We have considered two statistical methods for quantifying our analysis.

The Kolmogorov-Smirnov goodness-of-fit test, that we have already used in section \ref{kick vel}, is one of the most popular tests for comparing statistical distributions.
Given two empirical cumulative distribution functions $F_{1}$ and $F_{2}$ for real-valued random variables $X_{1}$ and $X_{2}$, the \mbox{``$D$-statistic"} is defined as \mbox{$D=\textrm{sup}_{x \in \mathbb{R}} |F_{1}(x)-F_{2}(x)|$}.
Remarkably, in the case of the null hypothesis that the samples 1 and 2 are identically distributed, the \mbox{$D$-statistic} has a distribution which is independent of that of the compared random variables. 
We may thus test the hypothesis that $X_{1}$ and $X_{2}$ are identically distributed by computing the probability of observing a deviation $D$ greater than or equal to the one observed ($D_{obs}$), i.e. the $P$-value \mbox{$P_{\textrm{K-S}}=P(D \geqslant D_{obs})$}.
According to the K-S test, we reject the null hypothesis with significance level $\alpha$ if $P_{\textrm{K-S}}<\alpha$.
Two problems prevent us from relying exclusively on the \mbox{K-S} test.
First, a pulsar is not adequately characterized by a single real number as its location, spin characteristics, and brightness are all of interest.
It is a mathematical fact that the K-S $D$-statistic loses its distribution-free virtue for multidimensional random variables and no practical generalization to higher dimensions is available \citep{JUS97}.
Alternatively, we may perform a K-S test on each of the marginal distributions of interest.
Because they are not independent, we may not however meaningfully combine the results to obtain a global goodness-of-fit figure for the model.
The second problem with the K-S test, in our context, is that it tests whether the two samples are drawn from \emph{exactly} the same distribution.
In the case of large samples, the test is sensitive to small deviations and, if present, they often lead to a formal rejection of the null hypothesis to a very low significance level.
Such deviations may turn out to be of little astrophysical importance and are expected given the complexity of the system that we are attempting to model in a relatively simple way. 
Some human judgement is thus required when interpreting the quantitative results.

Bayesian maximum likelihood analyses, such as the one performed by \cite{2002ApJ...568..289A} or our treatment of the birth velocity distribution in section \ref{kick vel}, do offer a procedure for optimizing a given parameteric model and are not limited to unidimensional distributions.
A major disadvantage, though, is that they do not provide any absolute measure of the correctness of the assumed functional form.
Additional checks, e.g. visual and using K-S tests as we did in section \ref{kick vel}, are required for that purpose.
Because of this, we choose not to tackle the computational challenges associated with the implementation of maximum likelihood formalism for multidimensional distributions.

In light of these remarks, we adopt a semi-quantitative approach.
We start with artibrary, but reasonable, model parameters and iterate.
At each step, we perform independent K-S tests on the marginal distributions (of $l$, $b$, $DM$, $S_{1400}$, $P$, and $B$) and use the K-S $P$-values to guide our search for ``optimal" model parameters.
The choice of subsequent model parameters, though, is based on physical intuition and involves qualitative considerations.
We stop when no significant further improvement appears possible.
We do not have a well defined metric for the ``best" model parameters and there is no guarantee that we have in fact found them.
We do however believe that substantial improvement would require revision of the assumed functional forms in the population model.

Hereafter, we commit the abuse of language of referring to the final model as ``optimal" or ``best", cautioning that we do so very informally.

\begin{figure*}[ht]
\begin{center}
\includegraphics[angle=-90.0, width=0.9\textwidth]{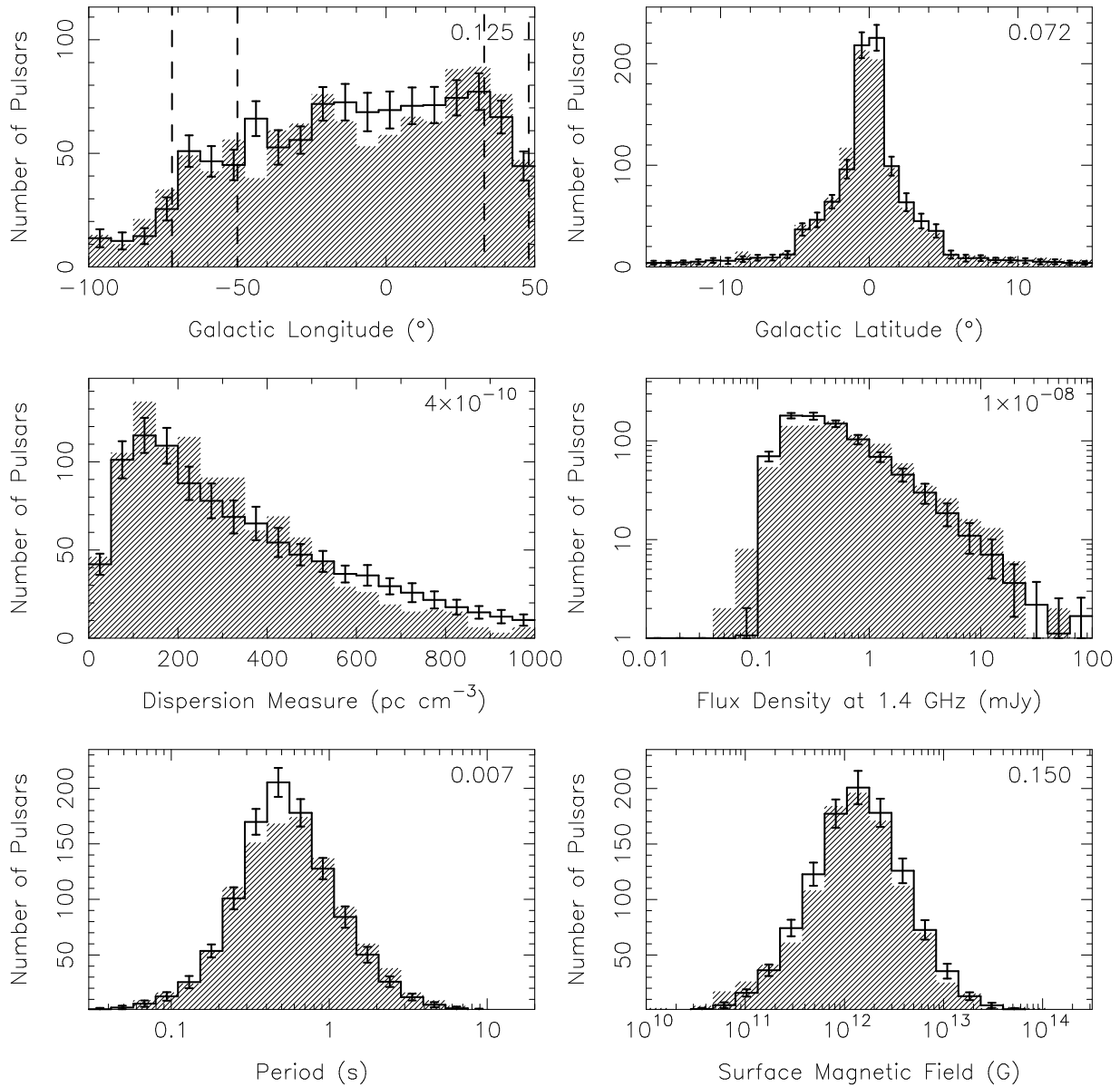}
\caption{Distributions of observed pulsar Galactic longitudes and latitudes, dispersion measures, flux densities at 1.4 GHz, pulse periods, and surface magnetic fields for our optimal model (solid lines) compared to the real distributions (hatched histograms). 
 For the simulation histograms, the number of pulsars in each bin is the average over 50 Monte Carlo realizations of the model.
 The error bars indicate the corresponding standard deviations.
 On each histogram, the associated K-S $P$-value is displayed in the upper right corner.
 For its computation, all $50\times1065$ MC pulsars are used to construct the empirical distribution function for the model.
}
\label{marginals all best}
\end{center}
\end{figure*}

\begin{figure}[ht]
\begin{center}
\includegraphics[angle=-90.0, width=0.45\textwidth]{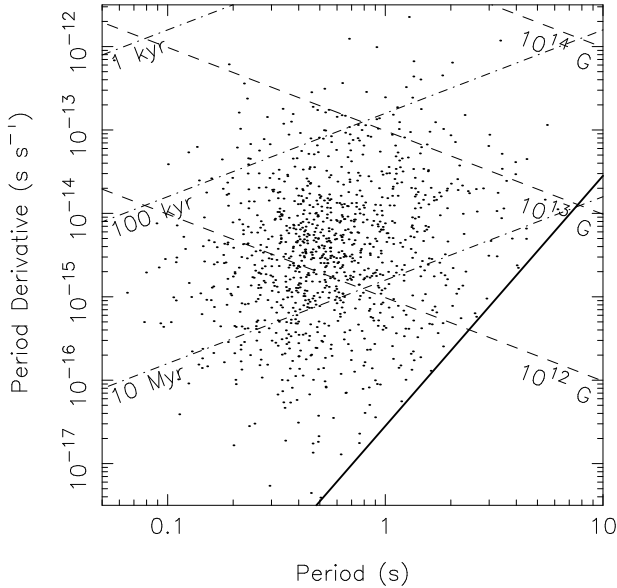}
\caption{$P-\dot{P}$ diagram for a typical Monte Carlo realization of our optimal model.
Each point represents a simulated pulsar.
Contours of constant surface magnetic field and of constant characteristic age are indicated by dashed and dotted-dashed lines, respectively.
The thick solid line marks the modelled death line.
}
\label{ppdot best}
\end{center}
\end{figure}

\begin{figure}[ht]
\begin{center}
\includegraphics[angle=-90.0, width=0.4\textwidth]{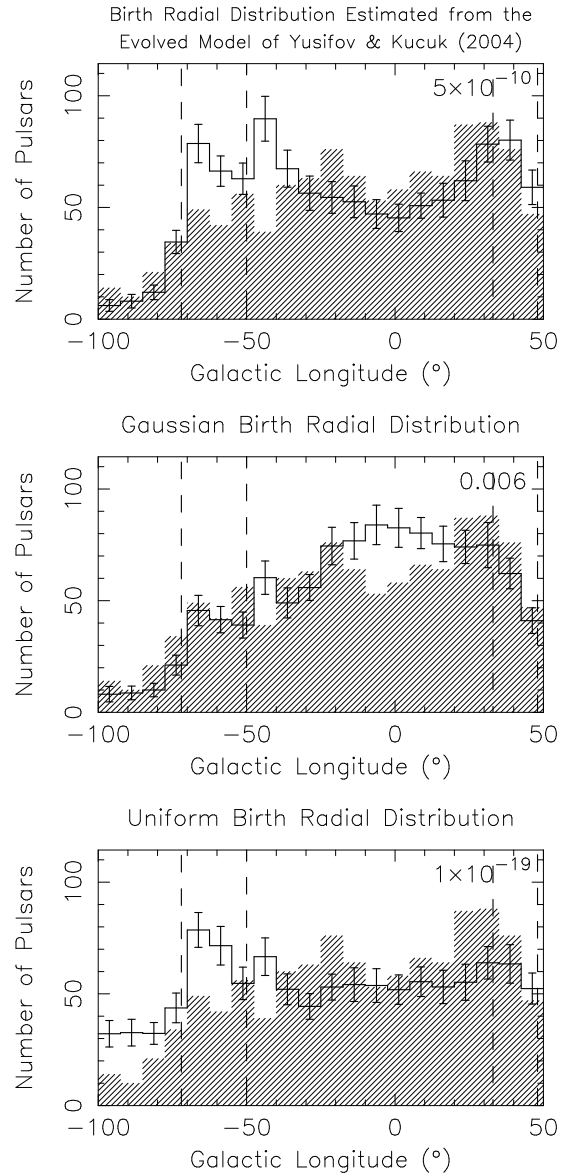}
\caption{Distributions of observed pulsar Galactic longitudes (solid lines) compared to the real distributions (hatched histograms) for alternative models of the birth radial distribution of pulsars.
The top panel shows the distribution obtained when the \cite{2004A&A...422..545Y} model is taken as a prescription for the evolved distribution and the corresponding birth distribution is estimated as in Appendix 2.
Clear peaks in the detected simulated pulsar density are seen near $\sim295^{\circ}$, $\sim215^{\circ}$, and $\sim30^{\circ}$ and the general ``U" shape of the distribution is suggestive that pulsars are more preferentially detected in a thin ring around the Galactic Center in the simulations than in reality.
In the center panel, a Gaussian distribution peaking at the GC and with scale radius 5 kpc was assumed.
An excess of simulated detections is seen near the GC ($|l| \lesssim 20^{\circ}$).
In the bottom panel, an uniform distribution on a disc of radius 15 kpc centered on the GC was assumed.
An excess of simulated detections is seen for $|l| \gtrsim 40^{\circ}$ and there is a paucity elsewhere.
In each case, all other parameters are as in the optimal model.
For the simulation histograms, the number of pulsars in each bin is the average over 50 Monte Carlo realizations of the model.
The error bars indicate the corresponding standard deviations.
On each histogram, the associated K-S $P$-value is displayed in the upper right corner.
For its computation, all $50\times1065$ MC pulsars are used to construct the empirical distribution function for the model.
}
\label{marginal Gl comparison}
\end{center}
\end{figure}

\begin{figure}[ht]
\begin{center}
\includegraphics[angle=-90.0, width=0.45\textwidth]{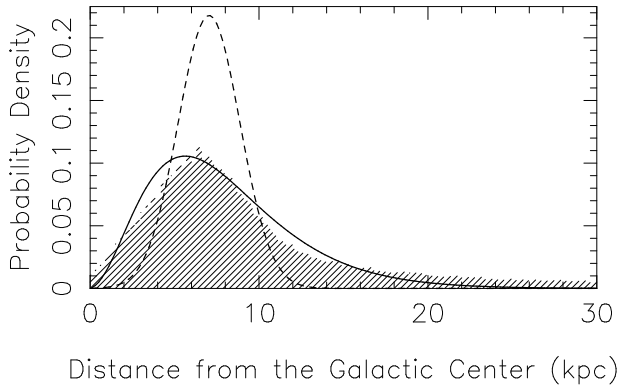}
\caption{Probability density function for the evolved Galactocentric radial coordinate of a pulsar for the \cite{2004A&A...422..545Y} model (solid line) and estimate of the corresponding birth PDF as computed in Appendix 2 (dashed line).
The shaded histogram shows the empirical evolved distribution for simulated pulsars with birth radial distribution given by the dashed curve.
All the other simulation parameters are as in the optimal model.
}
\label{f birth ev}
\end{center}
\end{figure}

\begin{figure}[ht]
\begin{center}
\includegraphics[angle=-90.0, width=0.45\textwidth]{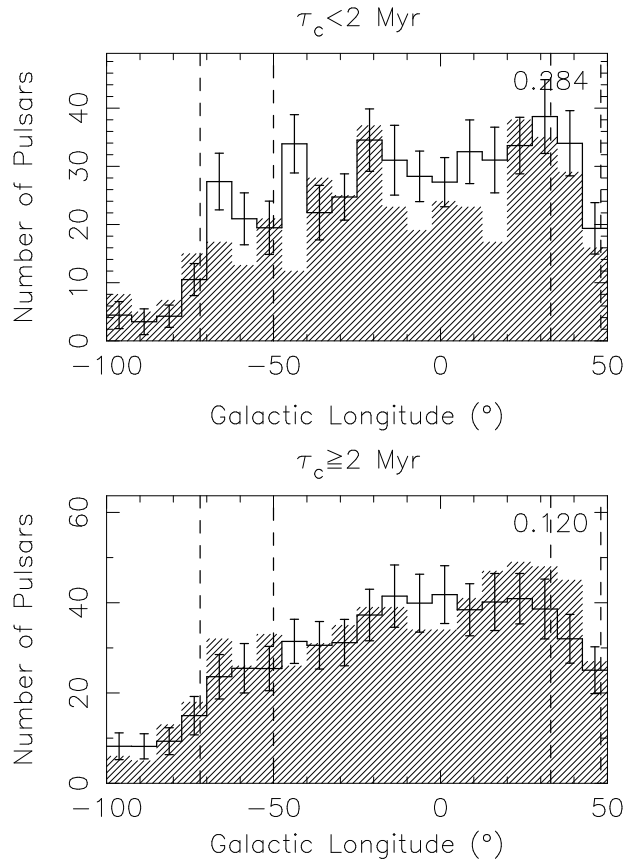}
\caption{Distributions of observed pulsar Galactic longitudes for our optimal model (solid lines) compared to the real distributions (hatched histograms), for the pulsars with characteristic age \mbox{$\tau_{c}<2$ Myr} (top) and for those with \mbox{$\tau_{c}\geq2$ Myr} (bottom).
For the simulation histograms, the number of pulsars in each bin is the average over 50 Monte Carlo realizations of the model.
The error bars indicate the corresponding standard deviations.
On each histogram, the associated K-S $P$-value is displayed in the upper right corner.
For its computation, all $50\times1065$ MC pulsars are used to construct the empirical distribution function for the model.
Near \mbox{$\pm30^{\circ}$} (corresponding to the Norma and Perseus, and to the Crux-Scutum and Carina-Sagittarius arms, respectively), there appears to be an enhanced pulsar detection density in both the real and simulated samples in the sample of young pulsars, although the simulated distribution lacks a pronounced dip near the Galactic Center.
For the pulsars older than \mbox{2 Myr}, the correlation between the spiral arms and the Galactic longitudes is greatly diminished.
}
\label{marginal Gl tau_c}
\end{center}
\end{figure}

\subsection{Results}
\label{population synthesis results}
Table \ref{pop model params} lists the parameters of the optimal model.
The marginal distributions of observed $l$, $b$, $DM$, $S_{1400}$, $P$, and $B$ for the model are shown in Figure \ref{marginals all best}.
For the simulation histograms, the number of pulsars in each bin is the average over 50 MC realizations of the model.
The error bars indicate the corresponding standard deviations.
On each histogram, the associated K-S $P$-value is displayed in the upper right corner.
For its computation, all $50\times1065$ MC pulsars are used to construct the empirical distribution function for the model.
A MC realization of the $P-\dot{P}$ diagram for the model is shown in Figure \ref{ppdot best}.
The agreement with the actual distributions and $P-\dot{P}$ diagram (c.f. Figure \ref{ppdot real}) is generally qualitatively good.
We discuss the results of our population synthesis and their implications in the next section.

\begin{deluxetable}{cc}
\tablecaption{Optimal Population Model Parameters \label{pop model params}}
\tablewidth{0pt}
\tablehead{
\colhead{Model Parameter} & 
\colhead{Value} 
}
\startdata

Radial Distribution Model & Yusifov \& K{\" u}{\c c}{\" u}k \\
$R_{1}$ & 0.55 kpc \\
$a$ & 1.64 \\
$b$ & 4.01 \\
\hline
Birth Height Distribution & Exponential \\
$\langle z_{0} \rangle$ & 50 pc \\
\hline
Birth Velocity Distribution & Exponential \\
$\langle v_{3D} \rangle$ & 380 km s$^{-1}$\\
\hline
Birth Spin Period Distribution & Normal\\
$\langle P_{0} \rangle$ & 300 ms \\
$\sigma _{P_{0}}$ & 150 ms \\
\hline
Magnetic Field Distribution  & Log-Normal \\
$\langle \log{(B/\textrm{G})} \rangle$ & 12.65 \\
$\sigma _{\log B}$ & 0.55 \\
\hline
Luminosity Model & $P-\dot{P}$ Power Paw \\
$L_{0}$ & 0.18 mJy kpc$^{2}$ \\
$\epsilon_{P}$ & $-$1.5 \\
$\epsilon_{\dot{P}}$ & 0.5 \\
$\sigma_{L_{corr}}$ & 0.8 \\

\enddata
\end{deluxetable}

\section{DISCUSSION}
\label{discussion}
\subsection{Spatial Distribution of the Pulsars}
We have taken the first step in modelling the rich Galactic structure by simulating the four major spiral arms and a non-trivial Galactocentric radial distribution of pulsar birth sites.

\subsubsection{Galactocentric Radial Distribution}
\label{radial distribution}
Following the completion of the first high-frequency (1.4 GHz) radio pulsar surveys capable of probing the inner Galaxy \citep{1992MNRAS.254..177C, 1992MNRAS.255..401J}, \cite{1994MNRAS.268..595J} concluded that the Gaussian radial distribution models peaking at the GC often assumed in pulsar population studies were incompatible with the observations.
Instead, he proposed a model in which the surface density of pulsars peaks at a distance of \mbox{4 kpc} from the GC.
In spite of this, the practice of assuming simplistic radial distributions in pulsar population studies has generally persisted.

Analysing recent data from the PM and SM surveys, \cite{lor03} and \cite{2004A&A...422..545Y} reiterated the pulsar deficit near the GC.
In our population synthesis, we have adopted the Galactocentric radial distribution derived by \cite{2004A&A...422..545Y} for the true, evolved Galactic population as the distribution for the birth locations of our synthetic pulsars.
We first examine whether it is indeed consistent with the observations.

The observed distributions of pulsars in Galactic longitudes (as angular position indicators) and in dispersion measures (as distance indicators) provide a test for the assumed radial distribution of the synthetic pulsars.
The agreement between the simulated and real distributions is reasonably good in both cases (see Figure \ref{marginals all best}), the major features (fall-off in detection density for $|l| \gtrsim 40^{\circ}$ and approximate plateau in-between; rapid rise of the $DM$ distribution for \mbox{$DM\leq200$ pc cm$^{-3}$} and slow decline afterward) being well reproduced.
In the case of the Galactic longitudes, a noticeable discrepancy is the lack of a significant decrease in simulated detection density for $|l|\lesssim20^{\circ}$ that is observed in reality and which suggests spiral arm tangents near $l=\pm30^{\circ}$.
The source of the discrepancy is unclear.
Three possibilities are an inadequate modelling of the spiral arm structure (see section \ref{discussion spiral arm structure}), an incorrectly modelled Galactocentric radial distribution, or underestimated selection effects against pulsars near the GC (e.g., underestimated scattering).
Unfortunately, we cannot isolate or exclude any of them on the basis of our analysis, for the parameters of our model are too intertwined with each other to single out the effect of each one.
In the case of the dispersion measures, the simulated distribution appears somewhat skewed toward higher values with respect to the real one.
This is in qualitative agreement with a critique by \cite{2003MNRAS.342.1299K} of the NE2001 model.
They considered the distribution of derived distances from the Galactic plane ($z$) of known pulsars with $|b|\leq20^{\circ}$ as a function of distance from the Sun using NE2001.
They found a slight decreasing trend, which is opposite to what is expected, as the maximum $z$-distance probed increases with distance from the Sun.
This suggests that NE2001 has a tendency to underestimate the distances to pulsars, i.e. to overestimate the free electron density.
The minor discrepancies between the real and simulated distributions appear equally possible as being due to remaining imperfections in the NE2001 model rather than the radial distribution model.
Thus, within the limitations imposed by our method, the Galactocentric radial distribution of pulsar birth sites is consistent with the model of \cite{2004A&A...422..545Y} and has a deficit near the GC.

We now address the question of whether it was justified to use the \cite{2004A&A...422..545Y} model for the evolved pulsars as the birth distribution in the first place.
In Appendix 2, we compute the birth radial distribution, taking the \cite{2004A&A...422..545Y} model for the evolved distribution as a prescription, under the assumption that the kinematic and rotational evolution of the pulsars is accurately described by our optimal model parameters.
The top panel of Figure \ref{marginal Gl comparison} shows the observed distribution in Galactic longitudes obtained in simulations when this birth distribution is assumed and all the other parameters are as in the optimal model.
There are clear peaks in the detected simulated pulsar density at Galactic longitudes $\sim295^{\circ}$, $\sim215^{\circ}$, and $\sim30^{\circ}$.
Moreover, the general ``U" shape of the distribution is strongly suggestive that pulsars are more preferentially detected in a thin ring around the GC in the simulations than in reality.
The most plausible explanation is the that birth radial distribution assumed (see Figure \ref{f birth ev}) is too strongly concentrated in such a ring.
While this suggests an inconsistency between our optimal model and the studies of \cite{lor03} and \cite{2004A&A...422..545Y}, we argue that this discrepancy may easily be understood if the radio luminosities of pulsars are a simple function of their period and period derivative such that younger pulsars have a tendency to be more luminous and therefore preferentially detected.
In fact, if this is the case, then \cite{lor03} and \cite{2004A&A...422..545Y}, who have not treated this bias in their studies, may have derived radial distributions that are biased toward younger pulsars.
If this bias is important, then their distributions may actually be better approximations to the birth radial distribution than to the unbiased evolved one.
In any case, the survey results are more closely reproduced in the simulations presented here when taking the \cite{2004A&A...422..545Y} distribution as the birth radial distribution than with the more concentrated distribution inferred from prescribing the \cite{2004A&A...422..545Y} model for the evolved pulsars, and this motivates us to adopt the former case.
We shall see in section \ref{pulsar luminosity} that the data strongly suggest that the radio luminosity of pulsars must indeed be somehow correlated with their age.
If this is correct, it may also explain discrepancies regarding the torque decay in isolated pulsars between various studies in the literature (see section \ref{spin-down torque magnetic field}).

To further verify that the consistency of the \cite{2004A&A...422..545Y} model used as the birth radial distribution of pulsars is not an artefact of the potential insensitivity of our simulation results to the details of the birth spatial distribution of the pulsars and that the complexity of the model is justified, we have considered two simpler models.
Namely, we have tried a birth radial distribution in which the pulsar surface density decreases like a Gaussian with scale radius \mbox{5 kpc} with distance from the GC and one in which the pulsar density is uniform on a disc of radius \mbox{15 kpc} centered at the GC, keeping the rest of our optimal simulation parameters.
The resulting distributions in Galactic longitudes are shown in the center and bottom panels of Figure \ref{marginal Gl comparison}.
A clear excess of simulated detections near the GC (\mbox{$|b| \lesssim 20^{\circ}$}) is seen in the Gaussian case.
In the uniform case, there is an excess away from the GC (\mbox{$b \gtrsim 40^{\circ}$}) and a paucity elsewhere.
In both cases, these discrepancies are qualitatively expected from a comparison of the assumed models with the \cite{2004A&A...422..545Y} distribution and so the complexity of the latter model well motivated.

As discussed by \cite{2004A&A...422..545Y}, the modelled radius of maximum pulsar density, \mbox{3.15 kpc}, is smaller by nearly \mbox{1.5 kpc} than the peak radii of several Population I objects.
\cite{2000A&A...358..521B} found \mbox{4.7 kpc} and \mbox{4.3 kpc} for the radii of maximum far infrared luminosity produced by massive stars and of the \mbox{H$_{2}$} surface density, respectively.
The shell supernova remnant distribution inferred by \cite{1998ApJ...504..761C} peaks at a distance of about \mbox{4.8 kpc} from the GC.
Nevertheless, the qualitative concentration in a ring midway between the Sun and the GC is common to both the pulsars and Population I objects.
The quantitative discrepancy may again be attributable to remaining imperfections in the modelling of the free electron density in NE2001.
However, if NE2001 indeed has a tendency to underestimate distances, then the distance to the pulsar ring most likely has been underestimated, exacerbating the discrepancy with the location of massive stars.
While the origin the quantitative discrepancy is unclear, the uncertainties remaining in the determination of the exact pulsar radial distribution prevent us from concluding that the disagreement is genuine.
As the birth sites of pulsars are also correlated with the spiral arms (section \ref{discussion spiral arm structure}), active star-forming regions, and their scale height is also consistent with that of massive stars (see \ref{discussion scale height}), it is safe to assume that, as theoretically expected, pulsar birth locations do coincide with those of massive star deaths.  

\subsubsection{Spiral Arm Structure}
\label{discussion spiral arm structure}
From the Galactic longitude distribution of our optimal model in Figure \ref{marginals all best}, on which we have indicated the tangent points to the modelled spiral arms (c.f. Figure \ref{xyi}), we see that near \mbox{$\pm30^{\circ}$} (corresponding to the Norma and Perseus, and to the Crux-Scutum and Carina-Sagittarius arms, respectively), there is a hint of enhanced pulsar detection density in both the real and simulated samples.
This correlation is more apparent when only young pulsars (\mbox{$\tau_{c}<2$ Myr}) are considered (see Figure \ref{marginal Gl tau_c}), although the simulated distribution lacks a pronounced dip near the GC, as in section \ref{radial distribution}.
For the pulsars older than \mbox{2 Myr}, the correlation between the spiral arms and the Galactic longitudes is greatly diminished.

Interestingly, the correlation between the young simulated pulsars and the tangent points near $-50^{\circ}$ and $-72^{\circ}$ seems stronger than between the real young pulsars and the tangents.
This may be due to an important assumption that we have made in modelling the spiral arms.
We have assumed in our simulations that the spiral pattern is fixed in time in the coordinate system corotating with the Sun.
In reality, the matter in the Galactic disc does not in general corotate with the spiral pattern \citep[see, e.g.,][]{1987gady.book.....B}.
In our Galaxy, the corotation radius is \mbox{$\sim14$ kpc} \citep{1971A&A....10...76B}, so that from the vantage point of the Sun, the spiral pattern and hence the presumed birth sites of pulsars move with time.
This effect is significant, as was demonstrated by \cite{1994JApA...15...69R} who found a correlation between the current pulsar distribution and the expected locations of the spiral arms \mbox{$\sim$60 Myr} ago, confirming that the relative angular velocity between the spiral pattern and the Sun motion spreads the pulsars over the disc.
That this effect was not included in the simulations may explain the stronger correlation between the current spiral arm tangents and the young simulated pulsars.
That the correlation between the real young pulsars and the spiral arms near $l=\pm30^{\circ}$ is preserved may due to the fact that the Norma and Perseus, and the Crux-Scutum and Carina-Sagittarium arms are tightly wound in those regions, acting as larger concentrations that take longer to move a distance exceeding their size.

Thus, while spiral arm structure is certainly required to reproduce the spatial distribution of pulsars, our simple modelling using fixed spirals appears deficient.
A treatment of the kinetics of the spiral pattern may be needed for more accurate results. 

\subsubsection{Scale Height}
\label{discussion scale height}
The distribution of distances of the pulsars from the Galactic plane is most directly reflected in the observed distribution in Galactic latitudes (Figure \ref{marginals all best}).
We find that \mbox{$\langle z_{0} \rangle=50$ pc}, corresponding to a $e^{-1}$ distribution thickness of \mbox{100 pc}, accomodates the observations well.
However, due to the high birth space velocities of the pulsars, our simulations are not sensitive to the mean birth distance from the Galactic plane to a precision better than a few tens of parsecs.
Massive star formation is distributed on a disc layer of thickness \mbox{$\sim$70 pc} \citep[full width at half maximum;][]{2000A&A...358..521B}.
Using a vertical velocity \mbox{10 km s$^{-1}$} and a main sequence lifetime \mbox{10 Myr} characteristic of massive stars, these move \mbox{$\sim100$ pc} away from their formation site perpendicular to the Galactic plane before exploding in a supernova, somewhat thickening the expected spatial distribution of these events, but not enough to create any serious discrepancy with the birth scale height of pulsars in our simulations.
Thus, up to the precision provided by our simulations, the birth scale height of pulsars coincides with that of massive stars ending their lives in supernovae.

\subsection{Birth Spin Period Distribution}
The birth spin period of the Crab pulsar, whose age is known from its association with the supernova of \mbox{A.D.~1054} and whose braking index has been measured, was the first to be estimated, with a value \mbox{$P_{0}\sim$19 ms} \citep{1977QB843.P8M36....}.
This has led to the generally made assumption that the initial spin periods of pulsars are much smaller than observed ones.
The existence of a young pulsar with period \mbox{16 ms} has provided additional support to that assumption \citep{1998ApJ...499L.179M, 2004ApJ...603..682M}.
However, a number of recent measurements suggest that initial periods in the range \mbox{$\sim50-150$ ms} are not uncommon (see Table \ref{known birth spin periods}, which lists the pulsars with estimated birth spin periods).
In addition, it has been argued that the \mbox{105 ms} pulsar \mbox{PSR J1852$-$0040} may have had a birth period close to the present value \citep{2005ApJ...627..390G}.
If, as our results suggest (see section \ref{pulsar luminosity}), short-period pulsars tend to be brighter, it is possible that only the lower part of the birth spin period distribution has been sampled and it may extend to considerably larger values.
Several population studies have supported this idea, either assuming or requiring a large fraction of pulsars to be born with periods \mbox{$\gtrsim 100$ ms} \citep[e.g.,][]{1981JApA....2..315V, 1987ApJ...319..162N, 1989ApJ...345..931E, 1990ApJ...352..222N, 2001A&A...374..182R, 2004ApJ...604..775G}.
From a pulsar current analysis of the PM data, \cite{2004ApJ...617L.139V} estimated that perhaps as many as 40\% of the pulsars may be born with periods in the range \mbox{$0.1-0.5$ s}.
In our optimal model, the birth spin period distribution is normal, centered at \mbox{300 ms} and with standard deviation \mbox{150 ms}.
Unfortunately, this distribution is not precisely constrained by our method, as the spin period of a pulsar loses the imprint of its initial value as the pulsar ages, being asymptotically determined only by the star's age and magnetic field (Eq. \ref{evolved period}).
Moreover, the derived distribution is clearly dependent on the details of the assumed luminosity law, as we favor one in which the luminosity is dependent the spin period of the pulsar (see section \ref{pulsar luminosity}).
For example, our simulations are incompatible with a large fraction of pulsars born with periods \mbox{$\lesssim 100$ ms}, since these are generally more luminous and easily detected in our $P-\dot{P}$ luminosity model, resulting in an excess of detections at short periods in the simulations with respect to the actual observations.
Nonetheless, as we will argue, the data do put constraints on the possible luminosity laws which suggest that our results are probably at least qualitatively robust.
We do not find evidence for a multimodal spin period distribution (injection), although we cannot rule it out.

\begin{deluxetable}{ccc}
\tablecaption{Estimated Birth Spin Periods of Young Pulsars \label{known birth spin periods}}
\tablewidth{0pt}

\tablehead{
\colhead{Pulsar Name} & 
\colhead{Estimated Birth Spin Period} &  
\colhead{References}
\\
\colhead{} & 
\colhead{(ms)} &  
\colhead{}
}

\startdata

J0205+6449                    & 60             & 1 \\
B0531+21 (Crab)               & 19             & 2 \\
J0537$-$6910\tablenotemark{a} & $\sim$11       & 3, 4 \\
J0538+2817                    & 139            & 5, 6 \\
B0540$-$69                    & 30 $\pm$ 8     & 7 \\
J1124$-$5916                  & $\gtrsim$90    & 8 \\
J1811$-$1925\tablenotemark{a} & 62             & 9, 10 \\
J1833$-$1034                  & $\gtrsim$55    & 11 \\
B1951+32                      & 27 $\pm$ 6     & 12 \\

\enddata

\tablenotetext{a}{Not observed as a radio pulsar \citep{1998MmSAI..69..951C}.}
\tablerefs{ 
(1) \cite{2002ApJ...568..226M};
(2) \cite{1977QB843.P8M36....};
(3) \cite{1998ApJ...499L.179M}; 
(4) \cite{2004ApJ...603..682M};
(5) \cite{2003ApJ...593L..31K};
(6) \cite{2003ApJ...585L..41R}
(7) \cite{1985ApJ...291..152R};
(8) \cite{2002ApJ...567L..71C};
(9) \cite{1999ApJ...523L..69T};
(10) \cite{2001ApJ...560..371K};
(11) \cite{CRG+05};
(12) \cite{2002ApJ...567L.141M}
}

\end{deluxetable}

\begin{figure}[ht]
\begin{center}
\includegraphics[angle=-90.0, width=0.45\textwidth]{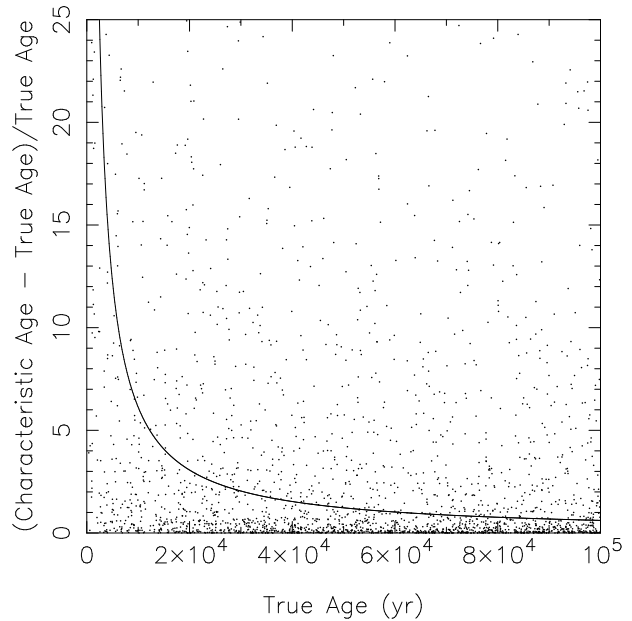}
\caption{Fractional difference between the characteristic age and the true age of synthetic pulsars younger than \mbox{100,000} yr in a Monte Carlo realization of the optimal model as a function of true age.
The solid curve (60,000 yr/true age) represents the model median value.
}
\label{delta t fig}
\end{center}
\end{figure}

\cite{1999ApJ...523L..69T} and \cite{2001ApJ...560..371K} have noted that \mbox{PSR J1811$-$1925} inside the supernova remnant \mbox{G11.2$-$0.3} has an actual age (inferred from a probable association with a supernova witnessed by Chinese astronomers in \mbox{A.D.~386})  smaller by a factor \mbox{$\sim$12} than its characteristic age obtained from timing (24,000 yr).
An important assumption in the determination of the age of a pulsar from timing is that its initial spin period is negligible with respect to the observed one \citep[$P_{0}/P \ll 1$; see, e.g.,][]{1977ApJ...215..885T}.
A discrepancy between the actual and characteristic ages suggests that this assumption is incorrect and that the characteristic age is a poor indicator of the pulsar's true age.
As the distribution of pulsar birth spin periods appears to extend much above \mbox{100 ms}, that characteristic ages are poor age indicators for young pulsars may turn out to be the rule rather than the exception.
In Figure \ref{delta t fig}, we have plotted the fractional age difference $\Delta t / t_{real} \equiv (\tau_{c}-t_{real})/t_{real}$, where $t_{real}$ is the true age of the pulsar, for pulsars with $t_{real}\leq100,000$ yr in a MC realization of our optimal model.
For given birth spin period $P_{0}$ and magnetic field $B$, the spin-down equation \ref{dipolar spindown law} implies
\begin{equation}
\frac{\Delta t}
{t_{real}}
=
\frac{\eta(P_{0}, B)}
{t_{real}},
\end{equation}
where
\begin{equation}
\eta(P_{0}, B)= 
\frac{3 I c^{3} P_{0}^{2}}
{16 \pi^{2} R^{6} B^{2} \sin^{2}{\chi}}.
\end{equation}
In particular, the characteristic age with this spin-down law is a systematic overestimate of the true age.
The median value of $\eta(P_{0}, B)$ for our optimal model is \mbox{$\sim60,000$ yr}.
As the Figure shows, pulsars with true age \mbox{$\lesssim30,000$ yr} have median characteristic age differing by a factor significantly larger than unity from the true age, if the optimal model presented here accurately represents the true pulsar population.

We have assumed that the spin down of pulsars is dominated by magnetic braking.
Some work has suggested that a significant amount of rotational kinetic energy in young neutron stars may be carried away by gravitational waves generated by unstable $r$-mode oscillations \citep{1998ApJ...502..708A, 1998PhRvL..80.4843L, 1998PhRvD..58h4020O, 1999ApJ...510..846A, 2000ApJ...543..386H}.
The spin down due to gravitational radiation by a neutron star is, however, expected to be important only during the first year after its formation \citep{1998PhRvL..80.4843L} and more recent investigations suggest that $r$-modes may saturate at low amplitudes and therefore not be as important as initially thought \citep{2002PhRvD..66d1303G, 2003ApJ...591.1129A, 2004PhRvD..70l1501B}.
Thus, for the purpose of age determination, it appears sufficient to a very good approximation to consider only the magnetic spin-down history.
What we have been calling the birth spin period, though, is more accurately defined as the period at which magnetic braking becomes dominant.

\begin{figure}[ht]
\begin{center}
\includegraphics[angle=-90.0, width=0.45\textwidth]{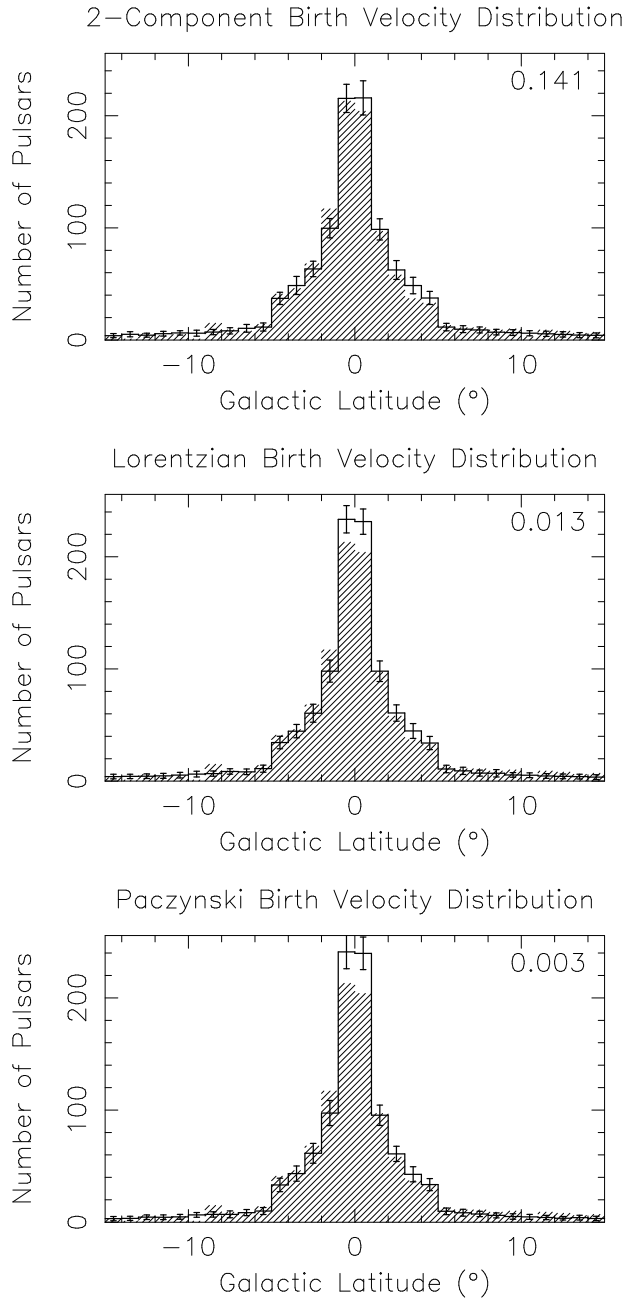}
\caption{Distributions of observed pulsar Galactic latitudes (solid lines) compared to the real distributions (hatched histograms), for Gaussian two-component (top), Lorentzian (center), and \cite{1990ApJ...348..485P} (bottom) birth velocity distributions with parameters derived in section \ref{kick vel}. 
All other parameters are as in the optimal model.
For the simulation histograms, the number of pulsars in each bin is the average over 50 Monte Carlo realizations of the model.
The error bars indicate the corresponding standard deviations.
On each histogram, the associated K-S $P$-value is displayed in the upper right corner.
For its computation, all $50\times1065$ MC pulsars are used to construct the empirical distribution function for the model.
Each birth velocity distribution is seen to agree with the observational data similarly well as the exponential one assumed in the optimal model (c.f. Figure \ref{marginals all best}).
}
\label{marginal Gb different v_0}
\end{center}
\end{figure}

\begin{figure}[ht]
\begin{center}
\includegraphics[angle=-90.0, width=0.45\textwidth]{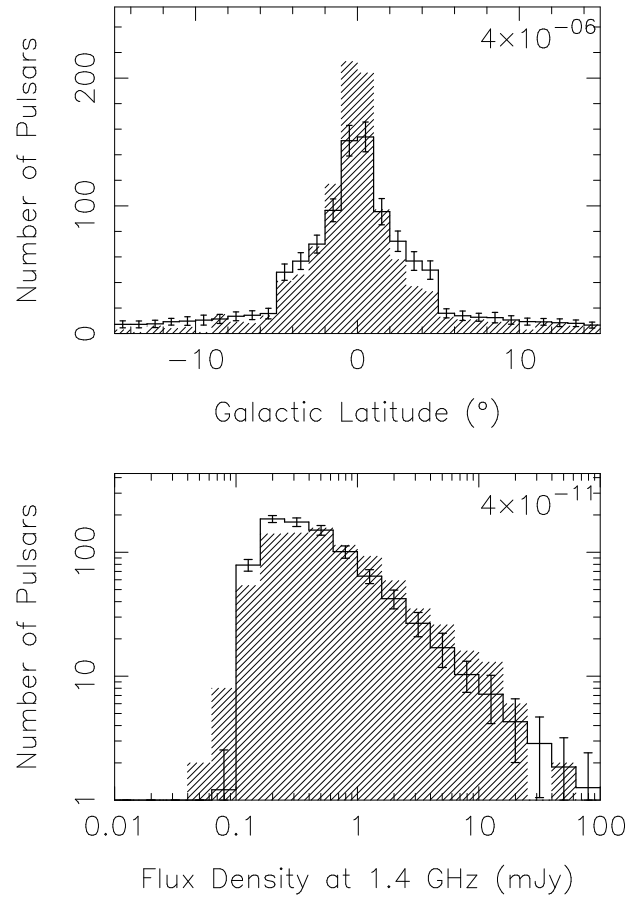}
\caption{Distributions of observed pulsar Galactic latitudes and flux densities at \mbox{1.4 GHz} for our optimal model with the $P-\dot{P}$ power-law luminosity model replaced by the random model (solid lines) compared to the real distributions (hatched histograms).
For the simulation histograms, the number of pulsars in each bin is the average over 50 Monte Carlo realizations of the model.
The error bars indicate the corresponding standard deviations.
On each histogram, the associated K-S $P$-value is displayed in the upper right corner.
For its computation, all $50\times1065$ MC pulsars are used to construct the empirical distribution function for the model.
The observed Galactic latitude distribution in the simulation is clearly thicker than the real one.
The sharp drop near $|b|=5^{\circ}$ is due to the reduced sensitivity of the Swinburne Multibeam survey relative to the Parkes Multibeam survey.
The agreement between the flux density distributions is comparable as in the optimal case with the $P-\dot{P}$ power-law model (c.f. Figure \ref{marginals all best}).
}
\label{marginals Gb S_1400 L rand}
\end{center}
\end{figure}

\subsection{Birth Velocity Distribution}
\label{birth velocity distribution}
In our population synthesis, we have assumed the exponential birth velocity distribution favored in section \ref{kick vel}, although it was shown that a two-component Gaussian and single-component Lorentzian and \cite{1990ApJ...348..485P} models were equally consistent with the proper-motion data considered.
To estimate the sensitivity of our results to this choice, we repeated our ``optimal" simulations, but replacing the exponential model with each of the other models in turn.
The results were by and large unaffected.
Figure \ref{marginal Gb different v_0} shows the observed distribution of Galactic latitudes, which is the most sensitive to the assumed birth velocity model, in each case.

\begin{figure}[ht]
\begin{center}
\includegraphics[angle=-90.0, width=0.45\textwidth]{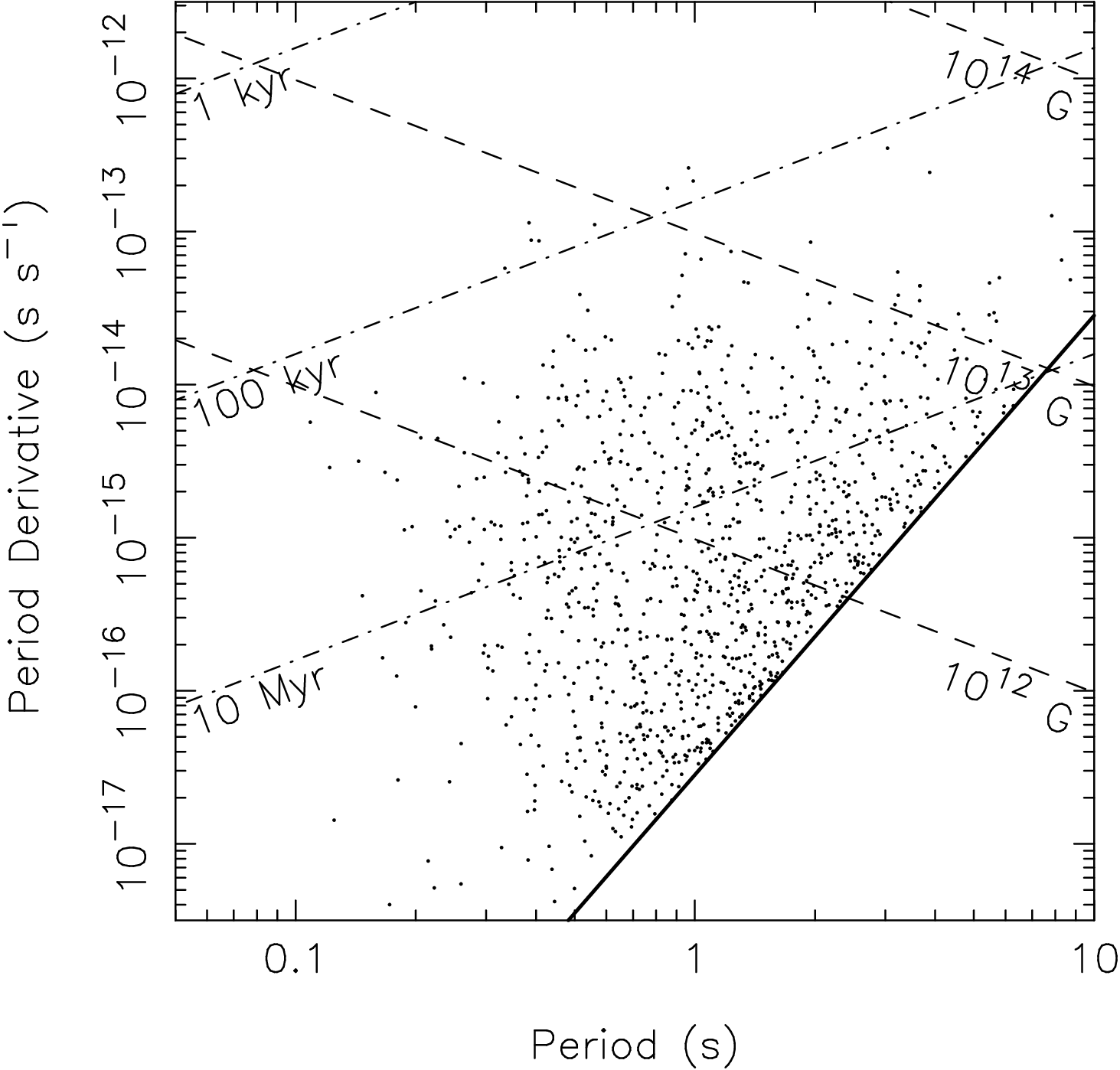}
\caption{
$P-\dot{P}$ diagram for a typical Monte Carlo realization of our optimal model with the $P-\dot{P}$ power-law luminosity model replaced by the random model.
Each point represents a simulated pulsar.
Contours of constant surface magnetic field and of constant characteristic age are indicated by dashed and dotted-dashed lines, respectively.
The thick solid line marks the modelled death line.
There is a clear pile-up of pulsars near the death line that is not seen in the real diagram (c.f. Figure \ref{ppdot real}).
}
\label{ppdot L rand}
\end{center}
\end{figure}

\subsection{Radio Luminosity}
\label{pulsar luminosity}

\subsubsection{Inadequacy of Random Luminosities}
\label{inadequacy of random luminosities}
The random luminosity model is clearly inadequate.
While it reproduces the observed flux distribution reasonably well in comparison with our optimal model (c.f. Figure \ref{marginals all best}), it leads to an observed scale height that is too large (Figure \ref{marginals Gb S_1400 L rand}).
Moreover, it produces a clear pile-up of observed objects near the death line in the $P-\dot{P}$ diagram (Figure \ref{ppdot L rand}) that is not seen in reality (c.f. Figure \ref{ppdot real}).
These discrepancies are largely unaffected by the choice of parameter values.
These two major discrepancies suggest that, more generally, random luminosities lead to the predicted detection of too many old pulsars and that the correct model must favor the detection of young objects.

\subsubsection{Period-Period Derivative Power Law}
\label{p pdot power law}
\begin{figure}[ht]
\begin{center}
\includegraphics[angle=-90.0, width=0.45\textwidth]{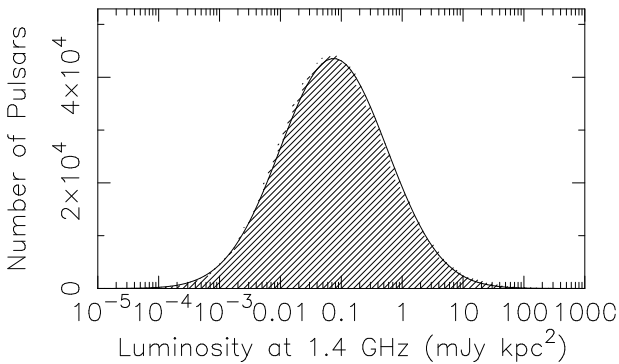}
\caption{
Underlying distribution of pulsar luminosities at 1.4 GHz in a Monte Carlo realization of our optimal model.
The sample consists of all synthetic pulsars that have not crossed the model death line.
The solid curve represents the best-fit log-normal distribution, with mean $\langle \log{L} \rangle=-1.1$ (\mbox{$L=0.07$ mJy kpc$^{2}$}) and standard deviation $\sigma_{\log{L}}=0.9$ in the logarithm to the base 10.
}
\label{L underlying}
\end{center}
\end{figure}

An obvious way of correlating the pulsar luminosity with age is to make it a function of the pulsar's period and period derivative.
When adjusting the model parameters for the $P-\dot{P}$ power-law model, we found that in order to simultaneously reproduce the distributions in observed periods and magnetic fields and the $P-\dot{P}$ diagram, the power-law exponents $\epsilon_{P}$ and $\epsilon_{\dot{P}}$ must be close to $-1.5$ and $0.5$, respectively.
The simulations did not allow us to constrain these parameters to a precision better than a few tenths and we adopted the round numbers.
To illustrate the wide range of pulsar luminosities consistent with the observed sample, we have plotted the underlying distribution of luminosities at 1.4 GHz in a MC realization of our optimal model in Figure \ref{L underlying}.
It is well fit by a log-normal distribution with mean $\langle \log{L} \rangle=-1.1$ (\mbox{$L=0.07$ mJy kpc$^{2}$}) and standard deviation $\sigma_{\log{L}}=0.9$ in the logarithm to the base 10.

We note two interesting facts.
First, since $B \propto \sqrt{P \dot{P}}$, the modelled luminosity contour lines (ignoring the corrections) are parallel to the death line (Eq. \ref{death line}).
Thus, in this model, the pulsar luminosity is a function of the distance from the death line in the $P-\dot{P}$ plane and they uniformly fade down as they approach it.
This explains why we do not observe a pile-up near the model death line.
In a simulation with the optimal parameters listed in Table \ref{pop model params}, but with the death line not applied and \mbox{$t_{max}=15$ Gyr}, $30\pm5$ pulsars are detected beyond the line, compared to only 5 in the real sample considered (c.f. Figure \ref{ppdot real}). 
This suggests that a complete radio emission shut-off near the modelled death line is nevertheless required, although it is inconclusive due to the unmodelled reduction of sensitivity to long-period pulsars (see section \ref{death line section}).
Second, the luminosity contour lines are furthermore coincident with those of the quantity $\sqrt{\dot{E}}$, where \mbox{$\dot{E}=4 \pi^{2} I \dot{P} / P^{3}$} (the ``spin-down luminosity") is the rate of loss of rotational kinetic energy \citep{lk05}.
This in turn is proportional to the potential drop available for particle acceleration in magnetosphere models \citep[e.g.,][]{1969ApJ...157..869G}, which may be related to the physics of radio emission \citep[see, e.g.,][who reached conclusions similar to ours]{1987A&A...171..152S}.

From their maximum likelihood analysis of \mbox{400 MHz} surveys, \cite{2002ApJ...568..289A} inferred a total physical luminosity also scaling approximately as $\sqrt{\dot{E}}$ (more precisely, as $P^{-1.3\pm0.3} \dot{P}^{0.4\pm0.1}$).
However, in their model, this corresponds to the integrated radiated power over Gaussian core and conal beams, whose size and relative contributions are functions of the pulsar's period.
The observed flux, then, is not in general proportional to $P^{-1.3} \dot{P}^{0.4}$, but depends on the viewing geometry. 
We considered the modification by \cite{2004ApJ...604..775G} of the \cite{2002ApJ...568..289A} luminosity model, in which spectral dependence was included to model the luminosity at \mbox{1.4 GHz}.
For the simplest case in which we observe an orthogonal rotator whose beam is confined to a plane containing the line of sight, the flux averaged over the spin period of the pulsar scales approximately as $P^{-0.8} \dot{P}^{0.4}$.
The flux contour lines then are no longer parallel to the death line and the modelled pulsar fluxes are thus quite different than ours.
Moreover, the flux variations due to beaming geometries span several orders of magnitude \citep[see, e.g., Figure 2 of][]{2002ApJ...568..289A}, whereas in our case, the typical correction factor due to the imposed dither is \mbox{$\sim10$}.

The simple pseudo-luminosity law that we have used may be interpreted as what would effectively result if we had geometrically modelled the pulsar beams as being flat and sharp-edged, so that the observed flux is independent of the impact angle, provided that the beam crosses our line of sight. 
The dithering would then mostly account for intrinsic variations about the standard candle law.
\cite{1998MNRAS.298..625T}, whose beaming model we have adopted, have indeed found in their analysis of polarization data that the fluxes of pulsars are consistent with being independent of the impact angle, contrary to the luminosity models of \cite{2002ApJ...568..289A} and \cite{2004ApJ...604..775G}.
This supports the hypothesis that emission within the beam boundary is patchy, with a random distribution of component locations \citep{1988MNRAS.234..477L, 1995JApA...16..107M}.

\subsection{Spin-Down Torque and Magnetic Field}
\label{spin-down torque magnetic field}
We have throughout assumed that the magnetic field of our synthetic pulsars was dipolar, constant, and neglected possible evolution of the magnetic inclination angle.
Since the observed data set seems to be well reproduced, our simulations suggest that no significant torque decay occurs in isolated pulsars over their lifetime as radio-loud sources (\mbox{$\sim100$ Myr}).
In particular, this suggests that magnetic field decay is not significant over this time scale in isolated neutron stars.

\cite{2002ApJ...565..482G} and \cite{2004ApJ...604..775G} performed simulations similar to ours and, in contrast, found evidence in favor of field decay on time scales as short as \mbox{$\sim2.5-5$ Myr}.
The main argument, in both cases, is that without field decay, a pile-up of simulated detections is seen near the death line in the $P-\dot{P}$ diagram.
Our results suggest that the need for field decay may be an artefact of the luminosity models they have assumed.
While the first study effectively assumed a dithered pseudo-luminosity law like ours and the second a seemingly different, more detailed geometric model based on that of \cite{2002ApJ...568..289A}, in both cases the dependence of the observed fluxes on $P$ and $\dot{P}$ is similar.
In fact, before dithering, the pseudo-luminosity law of \cite{2002ApJ...565..482G} scales as $P^{-1} P^{1/3}$, which is close to the flux dependence $\propto P^{-0.8}P^{0.4}$ in the case of the geometric model of \cite{2004ApJ...604..775G} derived in section \ref{p pdot power law}.
In both cases, the modelled luminosity contour lines are not parallel to the death line and the pulsars do not uniformly dim as they approach it.

\subsection{Distribution of Braking Indexes}
\label{braking index distribution}
As mentioned in section \ref{evolution rotational}, we have assumed in our simulations the braking index of each synthetic pulsar to be equal to 3, the value for pure magnetic dipole braking in a vacuum, despite the measured values (c.f. Table \ref{measured n}) being systematically lower.
To demonstrate that our results are in fact robust to this assumption, we have repeated our ``optimal" simulations, but with a normal distribution of braking indexes with mean 2.4 and standard deviation 0.6, the sample estimates for the measured values, and with an uniform distribution in the range \mbox{$[1.4, 3.0]$}, bracketing the measured values.
The corresponding $P-\dot{P}$ diagrams for typical Monte Carlo realizations of the models were found to be qualitatively very similar to the one obtained with all braking indexes set to 3 (Figure \ref{ppdot best}).
This is somewhat surprising, since the flow of the pulsars in the $P-\dot{P}$ diagram depends on the braking indexes, and reflects the dominant role of the luminosity dependence on $P$ and $\dot{P}$ in shaping the diagram observed in our simulations.

\begin{deluxetable*}{cccc}
\tablecaption{Galactic Birthrate and Number of Pulsars\label{birthrate}}
\tablewidth{0pt}

\tablehead{
\colhead{Birthrate} & 
\colhead{Potentially Observable Pulsars\tablenotemark{a}} & 
\colhead{All Radio-Loud Pulsars} 
\\
\colhead{(psr century$^{-1}$)} & 
\colhead{} &  
\colhead{}
}

\startdata

$2.8 \pm 0.1$ & $120,000 \pm 20,000$ & $1,200,000 \pm 300,000$

\enddata

\tablenotetext{a}{Radio-loud and beamed toward us.}

\tablecomments{The values given are averages over the 50 Monte Carlo realizations of the optimal model. The uncertainties are the corresponding standard deviations and do not include the uncertainty on the model parameters, which is difficult to reliably quantify. Varying the model parameters about the optimal values suggests that the true uncertainties exceed the formal ones by a factor $\sim5$.}

\end{deluxetable*}

\subsection{Birthrate and Number of Pulsars}
Table \ref{birthrate} lists the derived Galactic pulsar birthrate and number of pulsars (potentially observable and total) averaged over the 50 MC realizations of our optimal population synthesis model, calculated assuming the TM98 beaming model.
The formal uncertainties provided are the corresponding standard deviations and do not include the uncertainty on the model parameters, which is difficult to reliably quantify.
Varying the model parameters about the optimal values suggests that the true uncertainties exceed the formal ones by a factor $\sim5$.

The pulsar birthrate estimate, \mbox{$2.8 \pm 0.1$} pulsars per century, is higher than the one derived by \cite{2004ApJ...617L.139V} using a more model-independent pulsar current analysis of the PM data (also based on the NE2001 and TM98 models) of \mbox{1.9 $\pm$ 0.4} pulsars per century.
As \cite{2004ApJ...617L.139V} explain, their value is more strictly interpreted as a lower limit on the true value, so that our results are consistent.

Between 13\% and 25\% of core-collapse supernovae (Type Ib, Ic, and II) are expected to leave a black-hole remnant instead of a neutron star, depending on the lower mass limit for fall-back black hole formation, for main-sequence stars of solar metallicity \citep{2003ApJ...591..288H}.
Under the hypothesis that every core-collapse supernova produces either a pulsar or a black hole, the expected Galactic core-collapse supernova rate is then $\sim3.2-3.7$ supernovae per century. 
\cite{1999MNRAS.302..693D} reviews estimates based on extragalactic data.
The values range from $\sim1$ to $\sim4$ per century, assuming that about 85\% of supernovae are core-collapse \citep[e.g.,][]{1994ApJS...92..487T}.
Thus, the data are consistent with all neutron stars, except for a small fraction affected by accretion from a binary companion (X-ray binaries) or endowed with an extraordinarily strong magnetic field (magnetars), being born as radio pulsars, although not necessarily beamed toward us. 
Unfortunately, the remaining uncertainties prevent us from making a conclusive statement.

\section{CONCLUSION}
\label{conclusion}
Motivated by recent advances in pulsar astronomy (a large, homogeneous sample of detections by the Parkes and Swinburne Multibeam pulsar surveys; an updated model of the Galactic free electron density, NE2001; and new astrometric measurements), we have revisited the problem of the birth and evolution of isolated radio pulsars.

We started by estimating the pulsar birth velocity distribution directly from proper motion measurements of \cite{2002ApJ...571..906B, 2003AJ....126.3090B} in section \ref{kick vel}.
A single-component Maxwellian distribution appears to be inadequate due to the detection of a few very high-velocity objects.
However, we do not find evidence for multimodality of the velocity distribution, as alternative single-parameter models with heavier tails accommodate the observations equally well as a two-component Maxwellian.
The exact shape of the velocity distribution is not well constrained.
We adopted a model in which the absolute one-dimensional birth velocity components are exponentially distributed and with three-dimensional mean \mbox{$\langle v_{3D} \rangle = 380^{+40}_{-60}$ km s$^{-1}$}.

In section \ref{pop synth}, we used this velocity distribution as input to a more general Monte Carlo-based population synthesis.
We described parametric prescriptions for the birth properties (location, velocity, spin period, magnetic field, and radio luminosity) of the pulsars and their time evolution (spatial, rotational, and radio emission shut-off).
We then generated synthetic pulsar populations on which we performed simulations of the PM and SM surveys.
By comparing the observed samples in the simulations with the real detections, we determined ``optimal" model parameters.

The Galactocentric radial distribution of pulsar formation appears consistent with the functional form proposed by \cite{2004A&A...422..545Y}, which incorporates a deficit in surface density near the Galactic Center.
Although the \cite{2004A&A...422..545Y} distribution was derived for the present-day distribution of evolved pulsars, our simulations suggest that younger pulsars are preferentially detected, a bias not treated in the \cite{2004A&A...422..545Y}, and that it may in fact be a better approximation to the birth distribution.
This distribution is qualitatively consistent (peaking midway between the Galactic Center and the Sun) with that of massive main-sequence stars, the purported pulsar progenitors, although there is a difference of \mbox{$\sim1.5$ kpc} between the radii of peak density, possibly due to remaining imperfections in the free electron density model.
Spiral arm structure is required to reproduce the spatial distribution of pulsars.
We have modelled this structure using fixed spirals, but there are apparent deficiencies.
Proper modelling of the angular motion of the spirals with respect to the Sun may be needed for more accurate results.

Pulsar radio luminosities independent of the period and period derivative can safely be ruled out.
They lead to the detection of too many old synthetic pulsars, as indicated by the exceedingly large observed scale height and a clear pile-up of detections on the death line.
A model in which the radio (pseudo-)luminosity is, before dithering, proportional to $P^{-1.5} P^{0.5}$ (the square root of the spin-down luminosity) favors the detection of younger objects and appears consistent with the observations.
In this model, the undithered luminosity is proportional to the voltage drop available for particle acceleration in magnetosphere models, which may be related to the physics of the radio emission mechanism.
Also, the modelled fluxes are independent of the viewing geometry, provided that the pulsar beam crosses the line of sight, which supports the hypothesis of sharp-edged beams with random component locations.

We do not find evidence for significant torque decay (due to magnetic field decay or otherwise) over the lifetime of the pulsars as radio sources (\mbox{$\sim100$ Myr}).
The conflicting conclusion of \cite{2002ApJ...565..482G, 2004ApJ...604..775G}, who found evidence for magnetic field decay on time scales \mbox{$\lesssim5$ Myr}, is most likely due to different assumed (effective) dependences of the radio luminosity on the period and period derivative of the pulsar, which resulted in a pile-up of synthetic pulsars on the death line without simulated field decay.
Our preferred luminosity model neatly avoids such a pile-up, as the undithered luminosity contours are parallel to the death line, so that the pulsars uniformly dim as they approach it.
While it is unclear whether we have identified the correct radio pulsar luminosity law, this argues that conclusions regarding torque decay in isolated neutron stars based on simulation studies are strongly dependent on assumptions regarding the luminosities of the pulsars.
We have demonstrated that it is possible to avoid the conclusion of magnetic field decay using a very simple and natural luminosity model.

Our method does not precisely constrain the distribution of birth spin periods of the pulsars, because the period of a pulsar is asymptotically independent of its initial value.
However, in order to avoid an excess of simulated detections with periods \mbox{$\lesssim100$ ms}, many pulsars must be formed with greater periods.
We found that a Gaussian distribution with mean \mbox{300 ms} and standard deviation \mbox{150 ms} accommodates the observations well.
This suggests that characteristic ages, which assume that the initial period is negligible compared to the observed one, are generally poor age indicators, at least for young pulsars.

We estimate the Galactic pulsar birthrate to be $\sim2.8$ pulsars per century.
After accounting for a fraction of core-collapse supernovae forming black holes,
this value is consistent with all neutron stars, except for a small fraction in X-ray binaries and magnetars, being born as radio pulsars, although not necessarily beamed toward us.
Using the beaming model of \cite{1998MNRAS.298..625T}, the Galaxy is estimated to contain $\sim120,000$ potentially observable ordinary pulsars.

An important point to emphasize is that, as is also the case with previous work, we have merely provided evidence for consistency of the observational data with a particular scenario.
The value of our study lies in the fact that our proposed model has, arguably, minimal complexity and does not involve any controversial component.
We have also implemented the recent advances in pulsar astronomy in a generally self-consistent manner.
That very similar studies differing principally by assumptions about poorly constrainted aspects -- such as the radio luminosity of pulsars - led to conflicting conclusions -- notably regarding torque decay - is strongly suggestive that pulsar population simulations require further independent input before they can be used to draw definitive conclusions.
Perhaps the single most important breakthrough would be an independently verified luminosity model, as it is what ultimately determines the detectability of pulsars.
Until then, Ockham's razor should be applied when considering the results of such studies.
Meanwhile, the pulsar data set is poised to continue to be enhanced by the on-going Arecibo L-Band Feed Array (ALFA) pulsar survey, which is expected to detect as many as 1000 ordinary pulsars and 50 millisecond ones \citep{fgk04, cfl+05}.

\acknowledgments
We thank Z. Arzoumanian for providing us with an electronic version of the \cite{1981A&A...100..209H} sky background brightness temperature map, and D.~R. Lorimer and R.~N. Manchester for sharing with us details of the Parkes Multibeam pulsar survey completeness and data prior to public release.
We also thank D.~R. Lorimer for a careful reading of the manuscript and useful comments, the anonymous referee for a thorough and thoughtful report, and S.~M. Ransom and M.~S.~E. Roberts for useful discussions.
CAFG acknowledges support from the National Science and Engineering Research Council of Canada (NSERC) in the form of Undergraduate Student Research Awards and of a Postgraduate Scholarship, and from a John P. and Carol J. Merrill Graduate Fellowship held at Harvard University.
VMK is a Canada Research Chair and an NSERC Steacie Fellow.
This work has made extensive use of the NASA Abstract Database System and of a computer cluster purchased with funds of the Canada Foundation for Innovation.
Further support was provided by NSERC, the Fonds qu\'eb\'ecois de la recherche sur la nature et les technologies, and the Canadian Institute for Advanced Research.

\appendix

\section{APPENDIX 1: $S_{MIN}$ CALCULATION}
We describe the calculation of the minimum flux theoretically detectable from a radio pulsar ($S_{min}$) using the \cite{dss+84} formula.
First, we calculate the observed pulse width,
\begin{equation}
W_{e}=\sqrt{W^2+\tau_{samp}^2+\left(t_{samp} \frac{DM}{DM_0}\right)^2+\tau_{scatt}^2},
\end{equation}
where $W=DC \times P$ is the intrinsic pulse width, $t_{samp}$ is the sampling interval, and $\tau_{scatt}$ is the pulse broadening due to interstellar scattering \citep{1993MNRAS.263..403L}.
$DC$ is the duty cycle, i.e. the fraction of the time that the pulsar's flux is above 50\% of its maximum.
We take $DC=5\%$, a typical measured value (see, e.g., Lorimer et al. 2005\nocite{1995MNRAS.273..411L}), for all pulsars.
We do not model the observed scatter about this value or its potential correlation with the spin period.
We do, however, model the related period dependence of the beaming fraction the spin period as explained in section \ref{radiation beaming}.
$\tau_{samp}$ is the effective sampling interval, taking into account details of the system hardware such as anti-aliasing filters.
We take $\tau_{samp}=1.5 t_{samp}$.
The proportionality coefficient adopted is somewhat arbitrary, but chosen to be of order unity.
The simulations are insensitive to the exact value of this parameter.
$N_{ch}$ is the number of channels across the receiver bandwidth and $DM_{0}$ is the dispersion measure at which the smearing of the pulsar in one channel is equal to $t_{samp}$.
The latter is given by
\begin{equation}
DM_{0}=\frac{N_{ch} t_{samp} \nu^{3}}{8299\Delta \nu _{ch}},
\end{equation}
where $DM_{0}$ is in units of \mbox{pc cm$^{-3}$}, $t_{samp}$ is in s, and $\nu$ (the observing frequency) and $\Delta \nu$ (the receiver bandwidth) are in MHz \citep{joh90}.
Finally,
\begin{equation}
S_{min}=\delta_{beam} \frac{\beta \sigma (T_{rec}+T_{sky})}{G\sqrt{N_{p}\Delta \nu t_{int}}}\sqrt{\frac{W_{e}}{P-W_{e}}}
\end{equation}
where, as in section \ref{sensitivity modelling}, $T_{rec}$ is the receiver temperature on cold sky, $T_{sky}$ is the sky background temperature, $G$ is the antenna gain, $N_{p}$ is the number of polarizations, $\Delta \nu$ is the receiver bandwidth, $t_{int}$ is the integration time, $P$ is the pulse period, $W_e$ is the effective pulse width, $\sigma$ is the signal-to-noise threshold, and $\beta$ is a constant accounting for various system losses. 
The sky background temperature at 408 MHz is taken from an electronic version of the \cite{1981A&A...100..209H} map.
The value is scaled to 1.4 GHz assuming a spectral index $\alpha_{bg}=-2.8$ \citep{1987MNRAS.225..307L}.
The flux degradation factor due to a pulsar lying away from the center of a telescope beam, assuming a Gaussian power pattern, is
\begin{equation}
\delta_{beam}=\exp[-(r/f_{beam})^2]
\end{equation}
where $f_{beam}=FWHM/(2\sqrt{\ln 2})$ \citep{1993MNRAS.263..403L}.
In the simulations, we assume that each synthetic pulsar in the sky area covered by the PM and SM surveys lies at a random position within the half-power cone of a telescope beam.
The gain $G$ is taken to be the average of over individual beams of the Parkes Multibeam receiver, the values of which are reported by \cite{2001MNRAS.328...17M}.
The values of the parameters for the PM and SM surveys are given in Table \ref{survey params}.

\section{APPENDIX 2: COMPUTATION OF THE BIRTH RADIAL DISTRIBUTION OF PULSARS FROM AN EVOLVED MODEL}
In this appendix, we describe a method to estimate the birth Galactic radial distribution of pulsars corresponding to a prescribed model for the distribution of the evolved pulsars.
We then apply it to the distribution of evolved pulsars derived by \cite{2004A&A...422..545Y}.

We first note that the relationship between the birth and evolved spatial distributions of pulsars depends on the details of their evolution (not only spatial, but also rotational, as this determines the pulsars that survive as radio sources in the evolved population) and that one can be estimated from the other only after an evolution model has been fixed.
We thus perform this calculation only after having identified an ``optimal" population synthesis model (see section \ref{population synthesis results}).

Let $p(r_{ev})$ be a prescribed probability density function (PDF) for the evolved radial coordinate $r_{ev}$ of a pulsar, i.e. suppose that a pulsar has probability \mbox{$p(r_{ev})\textrm{d}r_{ev}$} of having an evolved radial coordinate in the range \textrm{$[r_{ev}, r_{ev}+\textrm{d}r_{ev})$}.
Given an evolution model, we can compute an ``evolution kernel" $K(r_{ev}, r_{birth})$ by Monte Carlo using the computer code described in section \ref{pop synth}.
Specifically, for equidistant radial distances $r_{birth}$ in the range \mbox{$[0, 50)$ kpc} (\mbox{$r_{birth}=0.25 + 0.5i$ kpc}; \mbox{$i=0, .., 99$}), we generate MC pulsars with birth radial coordinate equal to $r_{birth}$ and evolve them in time, until $10,000$ evolved radio-loud pulsars are reached.
We then construct a histogram (with bins centered on the selected $r_{birth}$ values and a bin width of 0.5 kpc) to estimate the distribution of evolved radial distances $r_{ev}$.
Interpolating between the discrete values of $r_{birth}$ and $r_{ev}$, we obtain a kernel $K(r_{ev}, r_{birth})$ defined on the square \mbox{$[0, 50)\times[0, 50)$ kpc} describing the conditional PDF that a pulsar born with radial distance $r_{birth}$ has radial distance $r_{ev}$ at the time of observation.
By the law of total probability, the PDF for the birth radial distance, $p(r_{birth})$, is related to $p(r_{ev})$ by
\begin{equation}
p(r_{ev})=\int_{0}^{\infty}p(r_{birth})K(r_{ev}, r_{birth})\textrm{d}r_{birth}.
\end{equation}
Given a birth radial distribution, it is thus straightforward compute the corresponding evolved distribution.
Here, the evolved distribution is prescribed.
We can estimate the birth distribution by choosing a parametric functional form, \mbox{$p(r_{birth};$ {\boldmath$\theta$}$)$}, depending on the vector parameter {\boldmath$\theta$}, and carrying out a least squares minimization between the prescribed evolved distribution, $p(r_{ev};presc)$, and the one corresponding to the trial birth distribution, $p(r_{ev};trial)$, evaluated at the bin centers.

We considered two functional forms for \mbox{$p(r_{birth};$ {\boldmath$\theta$}$)$}.
First, a functional form
\begin{equation}
p(r_{birth}; YK04) \propto r_{birth} \left( \frac{r_{birth}+R_{1}}{R_{\odot}+R_{1}} \right)^{a}
\exp{\left[ -b \left( \frac{r_{birth}-R_{\odot}}{R_{\odot}+R_{1}} \right) \right]},
\end{equation}
identical to the one form proposed by \cite{2004A&A...422..545Y} for the evolved distribution\footnote{This expression differs from the expression for the radial surface density $\rho$ given in equation \ref{radial distr eq} by a factor of $r_{birth}$ due to a polar effect near $r_{birth}=0$.}. Second, a simpler two-parameter ``displaced Gaussian"
\begin{equation}
p(r_{birth};Gauss) \propto \exp{\left[-\frac{(r_{birth}-R_{peak})^{2}}{2\sigma_{r_{birth}}^{2}} \right]}
\end{equation}
truncated to $r_{birth}\geq0$.
Both functional forms resulted in equal least sums of squares, but the fit using the $p(r_{birth}; YK04)$ function was found to be of lesser numerical stability, leading to degenerate and unexpectedly large values of the parameters in order to reproduce the requisite relatively narrow concentration of pulsars at radius \mbox{$\sim7$ kpc}. 
We thus adopted the displaced Gaussian fit ($R_{peak}=7.04$ and $\sigma_{r_{birth}}=1.83$). 
This distribution must not be confused with the Gaussian radial distribution discussed in section \ref{radial distribution}.
In the latter case, the radial distribution in \textit{surface density} is Gaussian and peaks at the Galactic Center.
Here, it is the radial \textit{PDF} $p(r_{birth};Gauss)$ that is Gaussian and the peak of the distribution is displaced from the GC.
Figure \ref{f birth ev} shows the best fit birth radial PDF.
To test that it satisfactorily reproduces the prescribed evolved distribution, we generated and evolved MC pulsars using our optimal population synthesis model and plotted the resulting evolved distribution against the prescribed model.
The agreement is qualitatively good.

\bibliography{references}

\begin{thebibliography}{151}
\expandafter\ifx\csname natexlab\endcsname\relax\def\natexlab#1{#1}\fi

\bibitem[{{Andersson}(1998)}]{1998ApJ...502..708A}
{Andersson}, N. 1998, \apj, 502, 708

\bibitem[{{Andersson} {et~al.}(1999){Andersson}, {Kokkotas}, \&
  {Schutz}}]{1999ApJ...510..846A}
{Andersson}, N., {Kokkotas}, K., \& {Schutz}, B.~F. 1999, \apj, 510, 846

\bibitem[{{Arras} {et~al.}(2003){Arras}, {Flanagan}, {Morsink}, {Schenk},
  {Teukolsky}, \& {Wasserman}}]{2003ApJ...591.1129A}
{Arras}, P., {Flanagan}, E.~E., {Morsink}, S.~M., {Schenk}, A.~K., {Teukolsky},
  S.~A., \& {Wasserman}, I. 2003, \apj, 591, 1129

\bibitem[{{Arzoumanian} {et~al.}(2002){Arzoumanian}, {Chernoff}, \&
  {Cordes}}]{2002ApJ...568..289A}
{Arzoumanian}, Z., {Chernoff}, D.~F., \& {Cordes}, J.~M. 2002, \apj, 568, 289

\bibitem[{{Baym} {et~al.}(1969){Baym}, {Pethick}, \&
  {Pines}}]{1969Natur.224..673B}
{Baym}, G., {Pethick}, C.~J., \& {Pines}, D. 1969, \nat, 224, 673

\bibitem[{{Bhattacharya} \& {van den Heuvel}(1991)}]{1991PhR...203....1B}
{Bhattacharya}, D., \& {van den Heuvel}, E.~P.~J. 1991, \physrep, 203, 1

\bibitem[{{Bhattacharya} {et~al.}(1992){Bhattacharya}, {Wijers}, {Hartman}, \&
  {Verbunt}}]{1992A&A...254..198B}
{Bhattacharya}, D., {Wijers}, R.~A.~M.~J., {Hartman}, J.~W., \& {Verbunt}, F.
  1992, \aap, 254, 198

\bibitem[{{Binney} \& {Merrifield}(1998)}]{1998gaas.book.....B}
{Binney}, J., \& {Merrifield}, M. 1998, {Galactic astronomy} (Princeton, NJ,
  Princeton University Press)

\bibitem[{{Binney} \& {Tremaine}(1987)}]{1987gady.book.....B}
{Binney}, J., \& {Tremaine}, S. 1987, {Galactic dynamics} (Princeton, NJ,
  Princeton University Press)

\bibitem[{{Blaauw}(1961)}]{1961BAN....15..265B}
{Blaauw}, A. 1961, \bain, 15, 265

\bibitem[{{Brink} {et~al.}(2004){Brink}, {Teukolsky}, \&
  {Wasserman}}]{2004PhRvD..70l1501B}
{Brink}, J., {Teukolsky}, S.~A., \& {Wasserman}, I. 2004, \prd, 70, 121501

\bibitem[{{Brisken} {et~al.}(2002){Brisken}, {Benson}, {Goss}, \&
  {Thorsett}}]{2002ApJ...571..906B}
{Brisken}, W.~F., {Benson}, J.~M., {Goss}, W.~M., \& {Thorsett}, S.~E. 2002,
  \apj, 571, 906

\bibitem[{{Brisken} {et~al.}(2003){Brisken}, {Fruchter}, {Goss}, {Herrnstein},
  \& {Thorsett}}]{2003AJ....126.3090B}
{Brisken}, W.~F., {Fruchter}, A.~S., {Goss}, W.~M., {Herrnstein}, R.~M., \&
  {Thorsett}, S.~E. 2003, \aj, 126, 3090

\bibitem[{{Bronfman} {et~al.}(2000){Bronfman}, {Casassus}, {May}, \&
  {Nyman}}]{2000A&A...358..521B}
{Bronfman}, L., {Casassus}, S., {May}, J., \& {Nyman}. 2000, \aap, 358, 521

\bibitem[{{Burton}(1971)}]{1971A&A....10...76B}
{Burton}, W.~B. 1971, \aap, 10, 76

\bibitem[{{Camilo} {et~al.}(2000{\natexlab{a}}){Camilo}, {Kaspi}, {Lyne},
  {Manchester}, {Bell}, {D'Amico}, {McKay}, \&
  {Crawford}}]{2000ApJ...541..367C}
{Camilo}, F., {Kaspi}, V.~M., {Lyne}, A.~G., {Manchester}, R.~N., {Bell},
  J.~F., {D'Amico}, N., {McKay}, N.~P.~F., \& {Crawford}, F.
  2000{\natexlab{a}}, \apj, 541, 367

\bibitem[{{Camilo} {et~al.}(2000{\natexlab{b}}){Camilo}, {Lorimer}, {Freire},
  {Lyne}, \& {Manchester}}]{2000ApJ...535..975C}
{Camilo}, F., {Lorimer}, D.~R., {Freire}, P., {Lyne}, A.~G., \& {Manchester},
  R.~N. 2000{\natexlab{b}}, \apj, 535, 975

\bibitem[{{Camilo} {et~al.}(2002){Camilo}, {Manchester}, {Gaensler}, {Lorimer},
  \& {Sarkissian}}]{2002ApJ...567L..71C}
{Camilo}, F., {Manchester}, R.~N., {Gaensler}, B.~M., {Lorimer}, D.~R., \&
  {Sarkissian}, J. 2002, \apjl, 567, L71

\bibitem[{{Camilo} {et~al.}(2005){Camilo}, {Ransom}, {Gaensler}, {Slane},
  {Lorimer}, {Reynolds}, {Manchester}, \& {Murray}}]{CRG+05}
{Camilo}, F., {Ransom}, S.~M., {Gaensler}, B.~M., {Slane}, P.~O., {Lorimer},
  D.~R., {Reynolds}, J., {Manchester}, R.~N., \& {Murray}, S.~S. 2005, \apj,
  accepted

\bibitem[{{Carlberg} \& {Innanen}(1987)}]{1987AJ.....94..666C}
{Carlberg}, R.~G., \& {Innanen}, K.~A. 1987, \aj, 94, 666

\bibitem[{{Case} \& {Bhattacharya}(1998)}]{1998ApJ...504..761C}
{Case}, G.~L., \& {Bhattacharya}, D. 1998, \apj, 504, 761

\bibitem[{{Chatterjee} \& {Cordes}(2002)}]{2002ApJ...575..407C}
{Chatterjee}, S., \& {Cordes}, J.~M. 2002, \apj, 575, 407

\bibitem[{{Chatterjee} \& {Cordes}(2004)}]{2004ApJ...600L..51C}
---. 2004, \apjl, 600, L51

\bibitem[{{Chatterjee} {et~al.}(2005){Chatterjee}, {Vlemmings}, {Brisken},
  {Lazio}, {Cordes}, {Goss}, {Thorsett}, {Fomalont}, {Lyne}, \&
  {Kramer}}]{2005ApJ...630L..61C}
{Chatterjee}, S., {Vlemmings}, W.~H.~T., {Brisken}, W.~F., {Lazio}, T.~J.~W.,
  {Cordes}, J.~M., {Goss}, W.~M., {Thorsett}, S.~E., {Fomalont}, E.~B., {Lyne},
  A.~G., \& {Kramer}, M. 2005, \apjl, 630, L61

\bibitem[{{Chen} \& {Ruderman}(1993)}]{1993ApJ...402..264C}
{Chen}, K., \& {Ruderman}, M. 1993, \apj, 402, 264

\bibitem[{{Chevalier} \& {Emmering}(1986)}]{1986ApJ...304..140C}
{Chevalier}, R.~A., \& {Emmering}, R.~T. 1986, \apj, 304, 140

\bibitem[{{Clifton} {et~al.}(1992){Clifton}, {Lyne}, {Jones}, {McKenna}, \&
  {Ashworth}}]{1992MNRAS.254..177C}
{Clifton}, T.~R., {Lyne}, A.~G., {Jones}, A.~W., {McKenna}, J., \& {Ashworth},
  M. 1992, \mnras, 254, 177

\bibitem[{{Cordes} \& {Chernoff}(1998)}]{1998ApJ...505..315C}
{Cordes}, J.~M., \& {Chernoff}, D.~F. 1998, \apj, 505, 315

\bibitem[{{Cordes} {et~al.}(2005){Cordes}, {Freire}, {Lorimer}, {Camilo},
  {Champion}, {Nice}, {Ramachandran}, {Hessels}, {Vlemmings}, {van Leeuwen},
  {Ransom}, {Bhat}, {Arzoumanian}, {McLaughlin}, {Kaspi}, {Kasian}, {Deneva},
  {Reid}, {Chatterjee}, {Han}, {Backer}, {Stairs}, {Deshpande}, \&
  {Faucher-Giguere}}]{cfl+05}
{Cordes}, J.~M., {Freire}, P.~C.~C., {Lorimer}, D.~R., {Camilo}, F.,
  {Champion}, D., {Nice}, D.~J., {Ramachandran}, R., {Hessels}, J.~W.~T.,
  {Vlemmings}, W., {van Leeuwen}, J., {Ransom}, S.~M., {Bhat}, N.~D.~R.,
  {Arzoumanian}, Z., {McLaughlin}, M., {Kaspi}, V.~M., {Kasian}, L., {Deneva},
  J.~S., {Reid}, B., {Chatterjee}, S., {Han}, J.~L., {Backer}, D.~C., {Stairs},
  I.~H., {Deshpande}, A.~A., \& {Faucher-Giguere}, C.-A. 2005, \apj, submitted

\bibitem[{{Cordes} \& {Lazio}(2002)}]{NE2001}
{Cordes}, J.~M., \& {Lazio}, T.~J.~W. 2002,
  {http://xxx.lanl.gov/abs/astro-ph/0207156}

\bibitem[{{Cordes} {et~al.}(1993){Cordes}, {Romani}, \&
  {Lundgren}}]{1993Natur.362..133C}
{Cordes}, J.~M., {Romani}, R.~W., \& {Lundgren}, S.~C. 1993, \nat, 362, 133

\bibitem[{{Cordes} {et~al.}(1991){Cordes}, {Ryan}, {Weisberg}, {Frail}, \&
  {Spangler}}]{1991Natur.354..121C}
{Cordes}, J.~M., {Ryan}, M., {Weisberg}, J.~M., {Frail}, D.~A., \& {Spangler},
  S.~R. 1991, \nat, 354, 121

\bibitem[{{Crawford}(2000)}]{2000PhDT.........1C}
{Crawford}, F. 2000, Ph.D.~Thesis

\bibitem[{{Crawford} {et~al.}(1998){Crawford}, {Kaspi}, {Manchester}, {Camilo},
  {Lyne}, \& {D'Amico}}]{1998MmSAI..69..951C}
{Crawford}, F., {Kaspi}, V.~M., {Manchester}, R.~N., {Camilo}, F., {Lyne},
  A.~G., \& {D'Amico}, N. 1998, Memorie della Societa Astronomica Italiana, 69,
  951

\bibitem[{D'Agostini(2003)}]{DAG03}
D'Agostini, G. 2003, Reports on Progress in Physics, 66, 1383

\bibitem[{{Damashek} {et~al.}(1978){Damashek}, {Taylor}, \&
  {Hulse}}]{1978ApJ...225L..31D}
{Damashek}, M., {Taylor}, J.~H., \& {Hulse}, R.~A. 1978, \apjl, 225, L31

\bibitem[{{Davies} {et~al.}(1977){Davies}, {Lyne}, \&
  {Seiradakis}}]{1977MNRAS.179..635D}
{Davies}, J.~G., {Lyne}, A.~G., \& {Seiradakis}, J.~H. 1977, \mnras, 179, 635

\bibitem[{{Deeter} {et~al.}(1999){Deeter}, {Nagase}, \&
  {Boynton}}]{1999ApJ...512..300D}
{Deeter}, J.~E., {Nagase}, F., \& {Boynton}, P.~E. 1999, \apj, 512, 300

\bibitem[{Dewey {et~al.}(1984)Dewey, Stokes, Segelstein, Taylor, \&
  Weisberg}]{dss+84}
Dewey, R.~J., Stokes, G.~H., Segelstein, D.~J., Taylor, J.~H., \& Weisberg,
  J.~M. 1984, in Millisecond Pulsars, ed. S.~Reynolds \& D.~Stinebring (NRAO :
  Green Bank), 234--240

\bibitem[{{Dewey} {et~al.}(1985){Dewey}, {Taylor}, {Weisberg}, \&
  {Stokes}}]{1985ApJ...294L..25D}
{Dewey}, R.~J., {Taylor}, J.~H., {Weisberg}, J.~M., \& {Stokes}, G.~H. 1985,
  \apjl, 294, L25

\bibitem[{{Dragicevich} {et~al.}(1999){Dragicevich}, {Blair}, \&
  {Burman}}]{1999MNRAS.302..693D}
{Dragicevich}, P.~M., {Blair}, D.~G., \& {Burman}, R.~R. 1999, \mnras, 302, 693

\bibitem[{{Edwards} {et~al.}(2001){Edwards}, {Bailes}, {van Straten}, \&
  {Britton}}]{2001MNRAS.326..358E}
{Edwards}, R.~T., {Bailes}, M., {van Straten}, W., \& {Britton}, M.~C. 2001,
  \mnras, 326, 358

\bibitem[{{Emmering} \& {Chevalier}(1989)}]{1989ApJ...345..931E}
{Emmering}, R.~T., \& {Chevalier}, R.~A. 1989, \apj, 345, 931

\bibitem[{{Endal} \& {Sofia}(1978)}]{1978ApJ...220..279E}
{Endal}, A.~S., \& {Sofia}, S. 1978, \apj, 220, 279

\bibitem[{{Faucher-Giguere} \& {Kaspi}(2004)}]{fgk04}
{Faucher-Giguere}, C.-A., \& {Kaspi}, V.~M. 2004,
  {http://astrosun2.astro.cornell.edu/$\sim$cordes/PALFA/alfa\_sims.pdf}

\bibitem[{{Feast} \& {Whitelock}(1997)}]{1997MNRAS.291..683F}
{Feast}, M., \& {Whitelock}, P. 1997, \mnras, 291, 683

\bibitem[{{Goldreich} \& {Julian}(1969)}]{1969ApJ...157..869G}
{Goldreich}, P., \& {Julian}, W.~H. 1969, \apj, 157, 869

\bibitem[{{Goldreich} \& {Reisenegger}(1992)}]{1992ApJ...395..250G}
{Goldreich}, P., \& {Reisenegger}, A. 1992, \apj, 395, 250

\bibitem[{{Gonthier} {et~al.}(2002){Gonthier}, {Ouellette}, {Berrier},
  {O'Brien}, \& {Harding}}]{2002ApJ...565..482G}
{Gonthier}, P.~L., {Ouellette}, M.~S., {Berrier}, J., {O'Brien}, S., \&
  {Harding}, A.~K. 2002, \apj, 565, 482

\bibitem[{{Gonthier} {et~al.}(2004){Gonthier}, {Van Guilder}, \&
  {Harding}}]{2004ApJ...604..775G}
{Gonthier}, P.~L., {Van Guilder}, R., \& {Harding}, A.~K. 2004, \apj, 604, 775

\bibitem[{{Gott} {et~al.}(1970){Gott}, {Gunn}, \&
  {Ostriker}}]{1970ApJ...160L..91G}
{Gott}, J.~R.~I., {Gunn}, J.~E., \& {Ostriker}, J.~P. 1970, \apjl, 160, L91+

\bibitem[{{Gotthelf} {et~al.}(2005){Gotthelf}, {Halpern}, \&
  {Seward}}]{2005ApJ...627..390G}
{Gotthelf}, E.~V., {Halpern}, J.~P., \& {Seward}, F.~D. 2005, \apj, 627, 390

\bibitem[{{Gouiffes} {et~al.}(1992){Gouiffes}, {Finley}, \&
  {Oegelman}}]{1992ApJ...394..581G}
{Gouiffes}, C., {Finley}, J.~P., \& {Oegelman}, H. 1992, \apj, 394, 581

\bibitem[{{Gregory} \& {Loredo}(1992)}]{1992ApJ...398..146G}
{Gregory}, P.~C., \& {Loredo}, T.~J. 1992, \apj, 398, 146

\bibitem[{{Gressman} {et~al.}(2002){Gressman}, {Lin}, {Suen}, {Stergioulas}, \&
  {Friedman}}]{2002PhRvD..66d1303G}
{Gressman}, P., {Lin}, L., {Suen}, W., {Stergioulas}, N., \& {Friedman}, J.~L.
  2002, \prd, 66, 041303

\bibitem[{{Gunn} \& {Ostriker}(1970)}]{1970ApJ...160..979G}
{Gunn}, J.~E., \& {Ostriker}, J.~P. 1970, \apj, 160, 979

\bibitem[{{Hansen} \& {Phinney}(1997)}]{1997MNRAS.291..569H}
{Hansen}, B.~M.~S., \& {Phinney}, E.~S. 1997, \mnras, 291, 569

\bibitem[{{Harrison} {et~al.}(1993){Harrison}, {Lyne}, \&
  {Anderson}}]{1993MNRAS.261..113H}
{Harrison}, P.~A., {Lyne}, A.~G., \& {Anderson}, B. 1993, \mnras, 261, 113

\bibitem[{{Hartman}(1997)}]{1997A&A...322..127H}
{Hartman}, J.~W. 1997, \aap, 322, 127

\bibitem[{{Hartman} {et~al.}(1997){Hartman}, {Bhattacharya}, {Wijers}, \&
  {Verbunt}}]{1997A&A...322..477H}
{Hartman}, J.~W., {Bhattacharya}, D., {Wijers}, R., \& {Verbunt}, F. 1997,
  \aap, 322, 477

\bibitem[{{Haslam} {et~al.}(1981){Haslam}, {Klein}, {Salter}, {Stoffel},
  {Wilson}, {Cleary}, {Cooke}, \& {Thomasson}}]{1981A&A...100..209H}
{Haslam}, C.~G.~T., {Klein}, U., {Salter}, C.~J., {Stoffel}, H., {Wilson},
  W.~E., {Cleary}, M.~N., {Cooke}, D.~J., \& {Thomasson}, P. 1981, \aap, 100,
  209

\bibitem[{{Heger} {et~al.}(2003){Heger}, {Fryer}, {Woosley}, {Langer}, \&
  {Hartmann}}]{2003ApJ...591..288H}
{Heger}, A., {Fryer}, C.~L., {Woosley}, S.~E., {Langer}, N., \& {Hartmann},
  D.~H. 2003, \apj, 591, 288

\bibitem[{{Heinke} {et~al.}(2005){Heinke}, {Grindlay}, {Edmonds}, {Cohn},
  {Lugger}, {Camilo}, {Bogdanov}, \& {Freire}}]{2005ApJ...625..796H}
{Heinke}, C.~O., {Grindlay}, J.~E., {Edmonds}, P.~D., {Cohn}, H.~N., {Lugger},
  P.~M., {Camilo}, F., {Bogdanov}, S., \& {Freire}, P.~C. 2005, \apj, 625, 796

\bibitem[{{Ho} \& {Lai}(2000)}]{2000ApJ...543..386H}
{Ho}, W.~C.~G., \& {Lai}, D. 2000, \apj, 543, 386

\bibitem[{{Hobbs} {et~al.}(2005){Hobbs}, {Lorimer}, {Lyne}, \&
  {Kramer}}]{2005MNRAS.tmp..475H}
{Hobbs}, G., {Lorimer}, D.~R., {Lyne}, A.~G., \& {Kramer}, M. 2005, \mnras, 475

\bibitem[{{Hulse} \& {Taylor}(1975)}]{1975ApJ...201L..55H}
{Hulse}, R.~A., \& {Taylor}, J.~H. 1975, \apjl, 201, L55

\bibitem[{{Iben} \& {Tutukov}(1996)}]{1996ApJ...456..738I}
{Iben}, I.~J., \& {Tutukov}, A.~V. 1996, \apj, 456, 738

\bibitem[{{Ivanova} {et~al.}(2005){Ivanova}, {Fregeau}, \& {Rasio}}]{IVA04}
{Ivanova}, N., {Fregeau}, J.~M., \& {Rasio}, F.~A. 2005, in Astronomical
  Society of the Pacific Conference Series, Vol. 328, Binary Radio Pulsars, ed.
  F.~A. {Rasio} \& I.~H. {Stairs}, 231--+

\bibitem[{{Johnston}(1990)}]{joh90}
{Johnston}, S. 1990, Ph.D.~Thesis

\bibitem[{{Johnston}(1994)}]{1994MNRAS.268..595J}
---. 1994, \mnras, 268, 595

\bibitem[{{Johnston} {et~al.}(1992){Johnston}, {Lyne}, {Manchester}, {Kniffen},
  {D'Amico}, {Lim}, \& {Ashworth}}]{1992MNRAS.255..401J}
{Johnston}, S., {Lyne}, A.~G., {Manchester}, R.~N., {Kniffen}, D.~A.,
  {D'Amico}, N., {Lim}, J., \& {Ashworth}, M. 1992, \mnras, 255, 401

\bibitem[{{Justel} {et~al.}(1997){Justel}, {Pe\~na}, \& {Zamar}}]{JUS97}
{Justel}, A., {Pe\~na}, D., \& {Zamar}, R. 1997, Statistics \& Probability
  Letters, 35, 251

\bibitem[{Kaspi {et~al.}(1996)Kaspi, Bailes, Manchester, Stappers, \&
  Bell}]{kbm+96}
Kaspi, V.~M., Bailes, M., Manchester, R.~N., Stappers, B.~W., \& Bell, J.~F.
  1996, Nature, 381, 584

\bibitem[{{Kaspi} {et~al.}(1994){Kaspi}, {Manchester}, {Siegman}, {Johnston},
  \& {Lyne}}]{1994ApJ...422L..83K}
{Kaspi}, V.~M., {Manchester}, R.~N., {Siegman}, B., {Johnston}, S., \& {Lyne},
  A.~G. 1994, \apjl, 422, L83

\bibitem[{{Kaspi} {et~al.}(2001){Kaspi}, {Roberts}, {Vasisht}, {Gotthelf},
  {Pivovaroff}, \& {Kawai}}]{2001ApJ...560..371K}
{Kaspi}, V.~M., {Roberts}, M.~E., {Vasisht}, G., {Gotthelf}, E.~V.,
  {Pivovaroff}, M., \& {Kawai}, N. 2001, \apj, 560, 371

\bibitem[{{Kramer}(1998)}]{1998ApJ...509..856K}
{Kramer}, M. 1998, \apj, 509, 856

\bibitem[{{Kramer} {et~al.}(2003{\natexlab{a}}){Kramer}, {Bell}, {Manchester},
  {Lyne}, {Camilo}, {Stairs}, {D'Amico}, {Kaspi}, {Hobbs}, {Morris},
  {Crawford}, {Possenti}, {Joshi}, {McLaughlin}, {Lorimer}, \&
  {Faulkner}}]{2003MNRAS.342.1299K}
{Kramer}, M., {Bell}, J.~F., {Manchester}, R.~N., {Lyne}, A.~G., {Camilo}, F.,
  {Stairs}, I.~H., {D'Amico}, N., {Kaspi}, V.~M., {Hobbs}, G., {Morris}, D.~J.,
  {Crawford}, F., {Possenti}, A., {Joshi}, B.~C., {McLaughlin}, M.~A.,
  {Lorimer}, D.~R., \& {Faulkner}, A.~J. 2003{\natexlab{a}}, \mnras, 342, 1299

\bibitem[{{Kramer} {et~al.}(2003{\natexlab{b}}){Kramer}, {Lyne}, {Hobbs}, {L{\"
  o}hmer}, {Carr}, {Jordan}, \& {Wolszczan}}]{2003ApJ...593L..31K}
{Kramer}, M., {Lyne}, A.~G., {Hobbs}, G., {L{\" o}hmer}, O., {Carr}, P.,
  {Jordan}, C., \& {Wolszczan}, A. 2003{\natexlab{b}}, \apjl, 593, L31

\bibitem[{{Kuijken} \& {Gilmore}(1989)}]{1989MNRAS.239..651K}
{Kuijken}, K., \& {Gilmore}, G. 1989, \mnras, 239, 651

\bibitem[{{Lai}(2003)}]{lai03}
{Lai}, D. 2003, in 3D Signatures of Stellar Explosion, a workshop honoring J.C.
  Wheeler's 60th Birthday

\bibitem[{{Lai} {et~al.}(1995){Lai}, {Bildsten}, \&
  {Kaspi}}]{1995ApJ...452..819L}
{Lai}, D., {Bildsten}, L., \& {Kaspi}, V.~M. 1995, \apj, 452, 819

\bibitem[{{Large}(1971)}]{1971IAUS...46..165L}
{Large}, M.~I. 1971, in IAU Symp. 46: The Crab Nebula, 165--+

\bibitem[{{Large} \& {Vaughan}(1971)}]{1971MNRAS.151..277L}
{Large}, M.~L., \& {Vaughan}, A.~E. 1971, \mnras, 151, 277

\bibitem[{{Lawson} {et~al.}(1987){Lawson}, {Mayer}, {Osborne}, \&
  {Parkinson}}]{1987MNRAS.225..307L}
{Lawson}, K.~D., {Mayer}, C.~J., {Osborne}, J.~L., \& {Parkinson}, M.~L. 1987,
  \mnras, 225, 307

\bibitem[{{Lequeux}(1979)}]{1979A&A....80...35L}
{Lequeux}, J. 1979, \aap, 80, 35

\bibitem[{{Lindblom} {et~al.}(1998){Lindblom}, {Owen}, \&
  {Morsink}}]{1998PhRvL..80.4843L}
{Lindblom}, L., {Owen}, B.~J., \& {Morsink}, S.~M. 1998, Physical Review
  Letters, 80, 4843

\bibitem[{{Livingstone} {et~al.}(2005{\natexlab{a}}){Livingstone}, {Kaspi}, \&
  {Gavriil}}]{2005ApJ...633.1095L}
{Livingstone}, M.~A., {Kaspi}, V.~M., \& {Gavriil}, F.~P. 2005{\natexlab{a}},
  \apj, 633, 1095

\bibitem[{{Livingstone} {et~al.}(2005{\natexlab{b}}){Livingstone}, {Kaspi},
  {Gavriil}, \& {Manchester}}]{2005ApJ...619.1046L}
{Livingstone}, M.~A., {Kaspi}, V.~M., {Gavriil}, F.~P., \& {Manchester}, R.~N.
  2005{\natexlab{b}}, \apj, 619, 1046

\bibitem[{{Lorimer}(2003)}]{lor03}
{Lorimer}, D.~R. 2003, in IAU Symposium No. 218: Young Neutron Stars and Their
  Environments, ed. F.~{Camilo} \& B.~M. {Gaensler}

\bibitem[{{Lorimer} {et~al.}(1993){Lorimer}, {Bailes}, {Dewey}, \&
  {Harrison}}]{1993MNRAS.263..403L}
{Lorimer}, D.~R., {Bailes}, M., {Dewey}, R.~J., \& {Harrison}, P.~A. 1993,
  \mnras, 263, 403

\bibitem[{{Lorimer} {et~al.}(1997){Lorimer}, {Bailes}, \&
  {Harrison}}]{1997MNRAS.289..592L}
{Lorimer}, D.~R., {Bailes}, M., \& {Harrison}, P.~A. 1997, \mnras, 289, 592

\bibitem[{{Lorimer} \& {Kramer}(2005)}]{lk05}
{Lorimer}, D.~R., \& {Kramer}, M. 2005, {Handbook of Pulsar Astronomy}
  (Cambridge University Press (Cambridge Observing Handbooks for Research
  Astronomers))

\bibitem[{{Lorimer} {et~al.}(1995){Lorimer}, {Yates}, {Lyne}, \&
  {Gould}}]{1995MNRAS.273..411L}
{Lorimer}, D.~R., {Yates}, J.~A., {Lyne}, A.~G., \& {Gould}, D.~M. 1995,
  \mnras, 273, 411

\bibitem[{{Lyne} {et~al.}(1982){Lyne}, {Anderson}, \&
  {Salter}}]{1982MNRAS.201..503L}
{Lyne}, A.~G., {Anderson}, B., \& {Salter}, M.~J. 1982, \mnras, 201, 503

\bibitem[{{Lyne} \& {Lorimer}(1994)}]{1994Natur.369..127L}
{Lyne}, A.~G., \& {Lorimer}, D.~R. 1994, \nat, 369, 127

\bibitem[{{Lyne} \& {Manchester}(1988)}]{1988MNRAS.234..477L}
{Lyne}, A.~G., \& {Manchester}, R.~N. 1988, \mnras, 234, 477

\bibitem[{{Lyne} {et~al.}(1998){Lyne}, {Manchester}, {Lorimer}, {Bailes},
  {D'Amico}, {Tauris}, {Johnston}, {Bell}, \& {Nicastro}}]{1998MNRAS.295..743L}
{Lyne}, A.~G., {Manchester}, R.~N., {Lorimer}, D.~R., {Bailes}, M., {D'Amico},
  N., {Tauris}, T.~M., {Johnston}, S., {Bell}, J.~F., \& {Nicastro}, L. 1998,
  \mnras, 295, 743

\bibitem[{{Lyne} {et~al.}(1985){Lyne}, {Manchester}, \&
  {Taylor}}]{1985MNRAS.213..613L}
{Lyne}, A.~G., {Manchester}, R.~N., \& {Taylor}, J.~H. 1985, \mnras, 213, 613

\bibitem[{{Lyne} {et~al.}(1993){Lyne}, {Pritchard}, \&
  {Graham-Smith}}]{1993MNRAS.265.1003L}
{Lyne}, A.~G., {Pritchard}, R.~S., \& {Graham-Smith}, F. 1993, \mnras, 265,
  1003

\bibitem[{{Lyne} {et~al.}(1996){Lyne}, {Pritchard}, {Graham-Smith}, \&
  {Camilo}}]{1996Natur.381..497L}
{Lyne}, A.~G., {Pritchard}, R.~S., {Graham-Smith}, F., \& {Camilo}, F. 1996,
  \nat, 381, 497

\bibitem[{{Lyne} {et~al.}(1988){Lyne}, {Pritchard}, \&
  {Smith}}]{1988MNRAS.233..667L}
{Lyne}, A.~G., {Pritchard}, R.~S., \& {Smith}, F.~G. 1988, \mnras, 233, 667

\bibitem[{{Lyne} {et~al.}(1975){Lyne}, {Ritchings}, \&
  {Smith}}]{1975MNRAS.171..579L}
{Lyne}, A.~G., {Ritchings}, R.~T., \& {Smith}, F.~G. 1975, \mnras, 171, 579

\bibitem[{{Manchester}(1995)}]{1995JApA...16..107M}
{Manchester}, R.~N. 1995, Journal of Astrophysics and Astronomy, 16, 107

\bibitem[{{Manchester} {et~al.}(2005){Manchester}, {Hobbs}, {Teoh}, \&
  {Hobbs}}]{2005AJ....129.1993M}
{Manchester}, R.~N., {Hobbs}, G.~B., {Teoh}, A., \& {Hobbs}, M. 2005, \aj, 129,
  1993

\bibitem[{{Manchester} {et~al.}(2001){Manchester}, {Lyne}, {Camilo}, {Bell},
  {Kaspi}, {D'Amico}, {McKay}, {Crawford}, {Stairs}, {Possenti}, {Kramer}, \&
  {Sheppard}}]{2001MNRAS.328...17M}
{Manchester}, R.~N., {Lyne}, A.~G., {Camilo}, F., {Bell}, J.~F., {Kaspi},
  V.~M., {D'Amico}, N., {McKay}, N.~P.~F., {Crawford}, F., {Stairs}, I.~H.,
  {Possenti}, A., {Kramer}, M., \& {Sheppard}, D.~C. 2001, \mnras, 328, 17

\bibitem[{{Manchester} {et~al.}(1996){Manchester}, {Lyne}, {D'Amico}, {Bailes},
  {Johnston}, {Lorimer}, {Harrison}, {Nicastro}, \&
  {Bell}}]{1996MNRAS.279.1235M}
{Manchester}, R.~N., {Lyne}, A.~G., {D'Amico}, N., {Bailes}, M., {Johnston},
  S., {Lorimer}, D.~R., {Harrison}, P.~A., {Nicastro}, L., \& {Bell}, J.~F.
  1996, \mnras, 279, 1235

\bibitem[{{Manchester} {et~al.}(1978){Manchester}, {Lyne}, {Taylor}, {Durdin},
  {Large}, \& {Little}}]{1978MNRAS.185..409M}
{Manchester}, R.~N., {Lyne}, A.~G., {Taylor}, J.~H., {Durdin}, J.~M., {Large},
  M.~I., \& {Little}, A.~G. 1978, \mnras, 185, 409

\bibitem[{{Manchester} \& {Taylor}(1977)}]{1977QB843.P8M36....}
{Manchester}, R.~N., \& {Taylor}, J.~H. 1977, {Pulsars} (San Francisco :
  W.~H.~Freeman, c1977.), 36--+

\bibitem[{{Maron} {et~al.}(2000){Maron}, {Kijak}, {Kramer}, \&
  {Wielebinski}}]{2000A&AS..147..195M}
{Maron}, O., {Kijak}, J., {Kramer}, M., \& {Wielebinski}, R. 2000, \aaps, 147,
  195

\bibitem[{{Marshall} {et~al.}(2004){Marshall}, {Gotthelf}, {Middleditch},
  {Wang}, \& {Zhang}}]{2004ApJ...603..682M}
{Marshall}, F.~E., {Gotthelf}, E.~V., {Middleditch}, J., {Wang}, Q.~D., \&
  {Zhang}, W. 2004, \apj, 603, 682

\bibitem[{{Marshall} {et~al.}(1998){Marshall}, {Gotthelf}, {Zhang},
  {Middleditch}, \& {Wang}}]{1998ApJ...499L.179M}
{Marshall}, F.~E., {Gotthelf}, E.~V., {Zhang}, W., {Middleditch}, J., \&
  {Wang}, Q.~D. 1998, \apjl, 499, L179+

\bibitem[{{Migliazzo} {et~al.}(2002){Migliazzo}, {Gaensler}, {Backer},
  {Stappers}, {van der Swaluw}, \& {Strom}}]{2002ApJ...567L.141M}
{Migliazzo}, J.~M., {Gaensler}, B.~M., {Backer}, D.~C., {Stappers}, B.~W., {van
  der Swaluw}, E., \& {Strom}, R.~G. 2002, \apjl, 567, L141

\bibitem[{{Mihalas} \& {Binney}(1981)}]{1981gask.book.....M}
{Mihalas}, D., \& {Binney}, J. 1981, {Galactic astronomy: Structure and
  kinematics /2nd edition/} (San Francisco, CA, W.~H.~Freeman and Co.,
  1981.~608 p.)

\bibitem[{{Murray} {et~al.}(2002){Murray}, {Slane}, {Seward}, {Ransom}, \&
  {Gaensler}}]{2002ApJ...568..226M}
{Murray}, S.~S., {Slane}, P.~O., {Seward}, F.~D., {Ransom}, S.~M., \&
  {Gaensler}, B.~M. 2002, \apj, 568, 226

\bibitem[{{Narayan}(1987)}]{1987ApJ...319..162N}
{Narayan}, R. 1987, \apj, 319, 162

\bibitem[{{Narayan} \& {Ostriker}(1990)}]{1990ApJ...352..222N}
{Narayan}, R., \& {Ostriker}, J.~P. 1990, \apj, 352, 222

\bibitem[{{Nice} {et~al.}(1995){Nice}, {Fruchter}, \&
  {Taylor}}]{1995ApJ...449..156N}
{Nice}, D.~J., {Fruchter}, A.~S., \& {Taylor}, J.~H. 1995, \apj, 449, 156

\bibitem[{{Ostriker} \& {Gunn}(1969)}]{1969ApJ...157.1395O}
{Ostriker}, J.~P., \& {Gunn}, J.~E. 1969, \apj, 157, 1395

\bibitem[{{Owen} {et~al.}(1998){Owen}, {Lindblom}, {Cutler}, {Schutz},
  {Vecchio}, \& {Andersson}}]{1998PhRvD..58h4020O}
{Owen}, B.~J., {Lindblom}, L., {Cutler}, C., {Schutz}, B.~F., {Vecchio}, A., \&
  {Andersson}, N. 1998, \prd, 58, 084020

\bibitem[{{Paczynski}(1990)}]{1990ApJ...348..485P}
{Paczynski}, B. 1990, \apj, 348, 485

\bibitem[{{Paczynski} \& {Stanek}(1998)}]{1998ApJ...494L.219P}
{Paczynski}, B., \& {Stanek}, K.~Z. 1998, \apjl, 494, L219+

\bibitem[{{Pfahl} {et~al.}(2002){Pfahl}, {Rappaport}, \&
  {Podsiadlowski}}]{2002ApJ...573..283P}
{Pfahl}, E., {Rappaport}, S., \& {Podsiadlowski}, P. 2002, \apj, 573, 283

\bibitem[{{Phinney} \& {Blandford}(1981)}]{1981MNRAS.194..137P}
{Phinney}, E.~S., \& {Blandford}, R.~D. 1981, \mnras, 194, 137

\bibitem[{{Press} {et~al.}(1992){Press}, {Teukolsky}, {Vetterling}, \&
  {Flannery}}]{1992nrca.book.....P}
{Press}, W.~H., {Teukolsky}, S.~A., {Vetterling}, W.~T., \& {Flannery}, B.~P.
  1992, {Numerical recipes in C. The art of scientific computing} (Cambridge:
  University Press, |c1992, 2nd ed.)

\bibitem[{{Proszynski} \& {Przybycien}(1984)}]{1984bens.work..151P}
{Proszynski}, M., \& {Przybycien}, D. 1984, in Birth and Evolution of Neutron
  Stars: Issues Raised by Millisecond Pulsars, 151--+

\bibitem[{{Ramachandran} \& {Deshpande}(1994)}]{1994JApA...15...69R}
{Ramachandran}, R., \& {Deshpande}, A.~A. 1994, Journal of Astrophysics and
  Astronomy, 15, 69

\bibitem[{{Regimbau} \& {de Freitas Pacheco}(2001)}]{2001A&A...374..182R}
{Regimbau}, T., \& {de Freitas Pacheco}, J.~A. 2001, \aap, 374, 182

\bibitem[{{Reynolds}(1985)}]{1985ApJ...291..152R}
{Reynolds}, S.~P. 1985, \apj, 291, 152

\bibitem[{{Romani} \& {Ng}(2003)}]{2003ApJ...585L..41R}
{Romani}, R.~W., \& {Ng}, C.-Y. 2003, \apjl, 585, L41

\bibitem[{{Ruderman} \& {Sutherland}(1975)}]{1975ApJ...196...51R}
{Ruderman}, M.~A., \& {Sutherland}, P.~G. 1975, \apj, 196, 51

\bibitem[{{Sayer} {et~al.}(1997){Sayer}, {Nice}, \&
  {Taylor}}]{1997ApJ...474..426S}
{Sayer}, R.~W., {Nice}, D.~J., \& {Taylor}, J.~H. 1997, \apj, 474, 426

\bibitem[{{Spitkovsky}(2004)}]{2004IAUS..218..357S}
{Spitkovsky}, A. 2004, in IAU Symposium, 357--+

\bibitem[{{Stairs} {et~al.}(2003){Stairs}, {Manchester}, {Lyne}, {Kramer},
  {Kaspi}, {Camilo}, \& {D'Amico}}]{2003ASPC..302...85S}
{Stairs}, I.~H., {Manchester}, R.~N., {Lyne}, A.~G., {Kramer}, M., {Kaspi},
  V.~M., {Camilo}, F., \& {D'Amico}, N. 2003, in ASP Conf. Ser. 302: Radio
  Pulsars, 85--+

\bibitem[{{Stokes} {et~al.}(1986){Stokes}, {Segelstein}, {Taylor}, \&
  {Dewey}}]{1986ApJ...311..694S}
{Stokes}, G.~H., {Segelstein}, D.~J., {Taylor}, J.~H., \& {Dewey}, R.~J. 1986,
  \apj, 311, 694

\bibitem[{{Stollman}(1987{\natexlab{a}})}]{1987A&A...178..143S}
{Stollman}, G.~M. 1987{\natexlab{a}}, \aap, 178, 143

\bibitem[{{Stollman}(1987{\natexlab{b}})}]{1987A&A...171..152S}
---. 1987{\natexlab{b}}, \aap, 171, 152

\bibitem[{{Sturrock}(1971)}]{1971ApJ...164..529S}
{Sturrock}, P.~A. 1971, \apj, 164, 529

\bibitem[{{Tammann} {et~al.}(1994){Tammann}, {Loeffler}, \&
  {Schroeder}}]{1994ApJS...92..487T}
{Tammann}, G.~A., {Loeffler}, W., \& {Schroeder}, A. 1994, \apjs, 92, 487

\bibitem[{{Tauris} \& {Manchester}(1998)}]{1998MNRAS.298..625T}
{Tauris}, T.~M., \& {Manchester}, R.~N. 1998, \mnras, 298, 625

\bibitem[{{Taylor} \& {Cordes}(1993)}]{1993ApJ...411..674T}
{Taylor}, J.~H., \& {Cordes}, J.~M. 1993, \apj, 411, 674

\bibitem[{{Taylor} \& {Manchester}(1977)}]{1977ApJ...215..885T}
{Taylor}, J.~H., \& {Manchester}, R.~N. 1977, \apj, 215, 885

\bibitem[{{Thorsett} {et~al.}(2005){Thorsett}, {Dewey}, \&
  {Stairs}}]{2005ApJ...619.1036T}
{Thorsett}, S.~E., {Dewey}, R.~J., \& {Stairs}, I.~H. 2005, \apj, 619, 1036

\bibitem[{{Torii} {et~al.}(1999){Torii}, {Tsunemi}, {Dotani}, {Mitsuda},
  {Kawai}, {Kinugasa}, {Saito}, \& {Shibata}}]{1999ApJ...523L..69T}
{Torii}, K., {Tsunemi}, H., {Dotani}, T., {Mitsuda}, K., {Kawai}, N.,
  {Kinugasa}, K., {Saito}, Y., \& {Shibata}, S. 1999, \apjl, 523, L69

\bibitem[{{Vivekanand} \& {Narayan}(1981)}]{1981JApA....2..315V}
{Vivekanand}, M., \& {Narayan}, R. 1981, Journal of Astrophysics and Astronomy,
  2, 315

\bibitem[{{Vranesevic} {et~al.}(2004){Vranesevic}, {Manchester}, {Lorimer},
  {Hobbs}, {Lyne}, {Kramer}, {Camilo}, {Stairs}, {Kaspi}, {D'Amico},
  {Possenti}, {Crawford}, {Faulkner}, \& {McLaughlin}}]{2004ApJ...617L.139V}
{Vranesevic}, N., {Manchester}, R.~N., {Lorimer}, D.~R., {Hobbs}, G.~B.,
  {Lyne}, A.~G., {Kramer}, M., {Camilo}, F., {Stairs}, I.~H., {Kaspi}, V.~M.,
  {D'Amico}, N., {Possenti}, A., {Crawford}, F., {Faulkner}, A.~J., \&
  {McLaughlin}, M.~A. 2004, \apjl, 617, L139

\bibitem[{{Wainscoat} {et~al.}(1992){Wainscoat}, {Cohen}, {Volk}, {Walker}, \&
  {Schwartz}}]{1992ApJS...83..111W}
{Wainscoat}, R.~J., {Cohen}, M., {Volk}, K., {Walker}, H.~J., \& {Schwartz},
  D.~E. 1992, \apjs, 83, 111

\bibitem[{{Wang} {et~al.}(2001){Wang}, {Howell}, {H{\" o}flich}, \&
  {Wheeler}}]{2001ApJ...550.1030W}
{Wang}, L., {Howell}, D.~A., {H{\" o}flich}, P., \& {Wheeler}, J.~C. 2001,
  \apj, 550, 1030

\bibitem[{{Webbink}(1985)}]{1985IAUS..113..541W}
{Webbink}, R.~F. 1985, in IAU Symp. 113: Dynamics of Star Clusters, 541--577

\bibitem[{{Wex} {et~al.}(2000){Wex}, {Kalogera}, \&
  {Kramer}}]{2000ApJ...528..401W}
{Wex}, N., {Kalogera}, V., \& {Kramer}, M. 2000, \apj, 528, 401

\bibitem[{{Willems} {et~al.}(2004){Willems}, {Kalogera}, \&
  {Henninger}}]{2004ApJ...616..414W}
{Willems}, B., {Kalogera}, V., \& {Henninger}, M. 2004, \apj, 616, 414

\bibitem[{{Yusifov} \& {K{\" u}{\c c}{\" u}k}(2004)}]{2004A&A...422..545Y}
{Yusifov}, I., \& {K{\" u}{\c c}{\" u}k}, I. 2004, \aap, 422, 545

\end{thebibliography}

\end{document}